# PHYSIK - DEPARTMENT

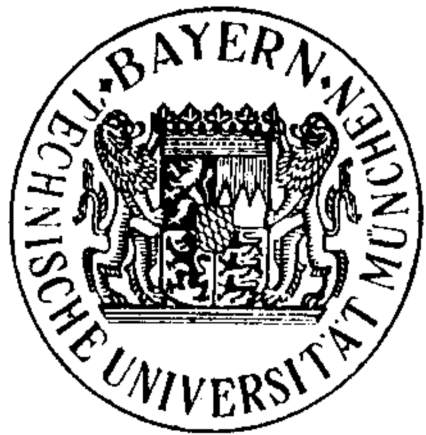

# TECHNISCHE UNIVERSITÄT

# MÜNCHEN

# Untersuchungen zur Verbesserung der Energieauflösung für Neutronenrückstreuspektrometer unter Verwendung von Idealkristallen mit geringer dynamischer Breite der Reflexionskurve

Diplomarbeit

von

Klaus-Dieter Liß

## Technische Universität München
## Fakultät für Physik

Institut E21, Professor Dr. W. Gläser

Juli 1990

# Inhaltsverzeichnis

# 1. Einführung

## 1.1. Allgemeines

Zur Untersuchung der kondensierten Materie besitzen Neutronen viele dazu notwendigen Eigenschaften in sich vereint: Ihre Energien sind in der Größenordnung der Anregungsenergien von Festkörpern, ihre Wellenlängen entsprechen denen atomarer Abstände und ihr magnetisches Dipolmoment gibt Einblick in magnetische Strukturen und deren Anregungen.

Insbesondere gewinnt in der Neutronenstreuung die hochauflösende Spektroskopie zur Aufklärung schwach inelastischer und quasielastischer Streuung, z.B. Messung von Tunnel- und Diffusionsmechanismen immer mehr an Bedeutung.Gängige Dreiachsen- und Flugzeitspektrometer erreichen Auflösungen von $\Delta E / E \approx 1 \cdot 10^{-3}$, was einem minimal meßbaren Energieübertrag von $\Delta E \approx 20~\mu eV$ entspricht [88B]. Mit hochauflösend bezeichnet man Instrumente die in einen Meßbereich von $\Delta E \approx 1~\mu eV$ und darunter vordringen. Zur Verwirklichung haben sich auf diesem Gebiet drei Instrumentgruppen ganz unterschiedlicher Meßmethoden herauskristallisiert, nämlich

a)   das Neutronen-(Resonanz)-Spinechospektrometer [88B], [89K], bei dem die Präzession des Neutronenspins als Maß für den Energieübertrag genommen wird,

b)   die Neutronen-Schwerkraftspektroskopie ultra kalter Neutronen, bei dem ballistische Flugbahnen der Neutronen zur Energieauswertung herangezogen werden, sowie

c)   das Rückstreuspektrometer, das Braggreflexe an Idealkristallen bei sehr großen Braggwinkeln verwendet.

## 1.2. Problemstellung

Für Neutronenexperimente aus den verschiedensten Gebieten der Festkörperphysik, z.B. Tunnelanregungen in Molekülkristallen oder quasielastische Streuungen in Festkörpern und Flüssigkeiten werden heutzutage Rückstreuspektrometer eingesetzt. Die derzeitig vorgegebene Auflösungsgrenze von $\Delta E = 300~neV$ erlaubt oft nur, wenige Spezialfälle physikalischer Modelle zu untersuchen.

Eine Verbesserung der Auflösung um einen Faktor 5 gäben dem Experimentator einen weiteren Horizont an Variationsmöglichkeiten für seine Untersuchungen:

Zum Beispiel sehen die Neutronen bei der Messung von Tunnelanregungen in Festkörpern nur die Bewegung von Wasserstoffgruppen (z. B. -CH$_3$ oder -NH $_3$), da andere Atome sich infolge ihrer größeren Masse viel langsamer bewegen und somit den Neutronen einen viel geringeren Energieübertrag mitgeben. Bei erhöhter Auflösung kann man gezielt Wasserstoff durch Deuterium ersetzen und somit den Einfluß der Masse auf die Bewegung studieren.

Betrachtet man Selbstdiffusionsmechanismen in der Metallphysik, so ist man nur auf eine Hand voll außerordentlich schnell diffundierender Systeme beschränkt, die alle eine kubisch raumzentrierte Kristallstruktur besitzen. Mit obiger Verbesserung des Instrumentes stünden dem Experimentator zum ersten Mal auch einige der langsamer diffundierenden Metalle mit kubisch flächenzentriertem Kristallgitter zur Messung bereit.

Die vorliegende Arbeit wurde im Rahmen eines Modernisierungsprogrammes am Institut Max von Laue - Paul Langevin (ILL) in Grenoble, Frankreich ausgeschrieben und soll experimentellen Aufschluß über eine Verbesserungsmöglichkeit der Auflösung von Rückstreuspektrometern geben [89L]. Speziell zur Untersuchung des vielversprechenden GaAs[200]-Reflexes ist eine Testbank in Rückstreugeometrie am ILL aufgebaut und verwendet worden.

# 2. Das Rückstreuspektrometer

## 2.1. Allgemeine Kristallspektrometer

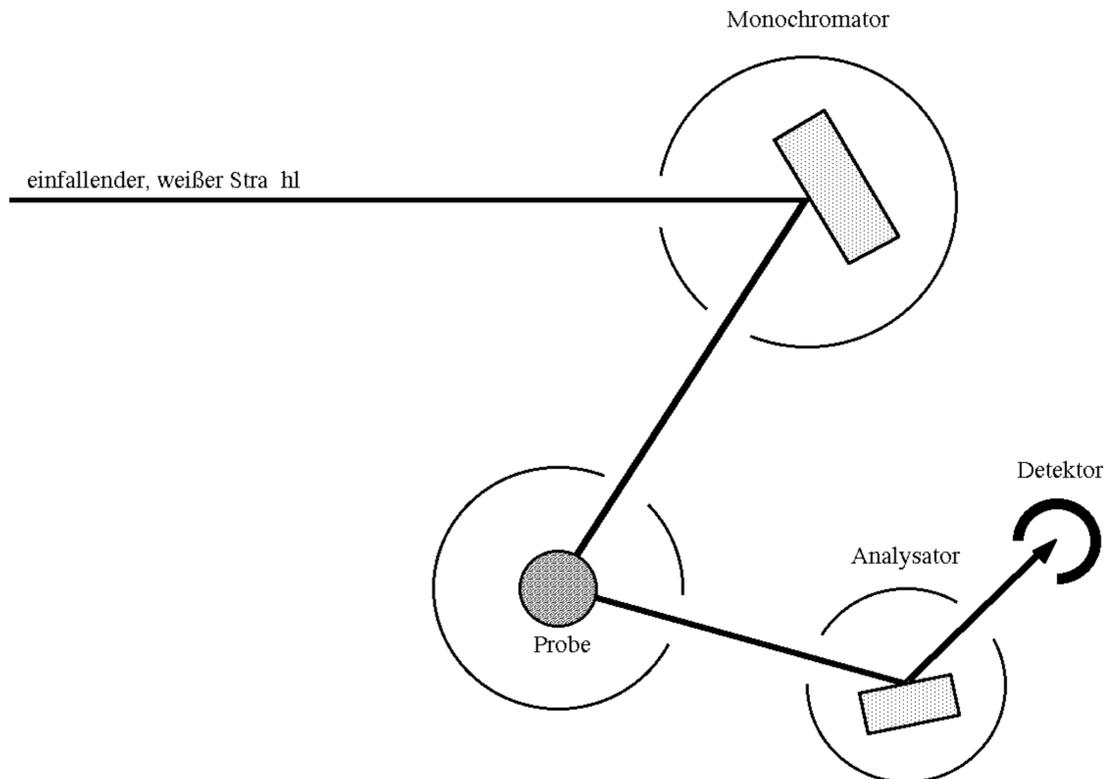

Abbildung 2.1:
Prinzip eines Dreiachsenspektrometers.

Wie das Dreiachsenspektrometer (Abbildung 2.1) besteht auch das Rückstreuspektrometer im wesentlichen aus einem Monochromatorkristall, Probe, Analysatorkristall und Detektor. Der Monochromator spezifiziert durch Braggreflexion die auf die Probe einfallende Wellenlänge bzw. Energie des Neutrons. An dieser wird es im allgemeinen inelastisch gestreut und ändert somit seine Wellenlänge (bzw. Energie), die dann wiederum durch Braggreflexion am Analysator gemessen wird. Aus dem Winkel zwischen einfallendem und gestreutem Strahl kann man gleichzeitig noch den Impulsübertrag ablesen.

Das Braggesetz zur Bestimmung der Wellenlänge λ lautet

$$\lambda = 2d\ \sin(\Theta)$$ (2.1)

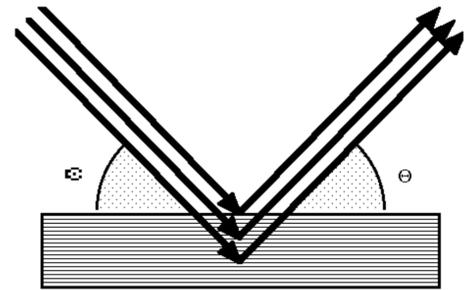

wobei $\Theta$ den Braggwinkel und d die Gitterkonstante für diesen Reflex bezeichnen (Abbildung 2.2). Bilden wir das totale Differential dieser Gleichung

Abbildubg 2.2:
Braggreflexion am Kristall. Der Braggwinkel ist zwischen dem einfallendem Strahl und den Gitterebenen zu messen.

$$\Delta\lambda = 2\ \sin(\Theta)\ \Delta d\ +\ 2\ d\ \cos(\Theta)\ \Delta\Theta$$ (2.2)

und dividieren es durch Gleichung (2.1), so erhalten wir

$$\frac{\Delta\lambda}{\lambda} = \frac{\Delta d}{d}\ +\ \cot(\Theta)\ \Delta\Theta\ .$$ (2.3)

Genauso erhält man aus der Beziehung

$$E = \frac{2\ \pi^2\ \hbar^2}{m\ \lambda^2}$$ (2.4)

(Wirkungsquantum $\hbar$, Neutronenmasse m)

$$\frac{\Delta E}{E} = -2\ \frac{\Delta\lambda}{\lambda}$$ (2.5)

und somit die Energieauflösung

$$\frac{\Delta E}{E}\ =\ -2\ \frac{\Delta d}{d}\ -\ 2\ \cot(\Theta)\ \Delta\Theta\ .$$ (2.6)

Sie wird also durch zwei Terme bestimmt, zum einen der Anteil $\Delta d / d$, der angibt, wie scharf die Gitterkonstante bestimmt ist und zum anderen der Teil proportional zu $\Delta\Theta$, der die Auflösung durch die Winkelungenauigkeitkeiten beschränkt, sei es durch Mosaizität des Kristalls oder durch die Winkeldivergenz des einfallenden Strahls.

Bei Dreiachsenmaschinen überwiegt der Term in $\Delta\Theta$ dem in $\Delta d$, so daß die Winkelabweichungen den führenden Beitrag zur Auflösung liefern. Wählt man allerdings einen Reflex mit einem Braggwinkel von $\Theta = \pi/2$, so verschwindet dessen Kotangens und wir erhalten aus obiger Gleichung

$$\boxed{\frac{\Delta E}{E} = -2\,\frac{\Delta d}{d} + O((\Delta\Theta)^2)} \qquad \text{für} \quad \Theta = \frac{\pi}{2}. \qquad (2.7)$$

In diesem Fall wird die Auflösung überwiegend durch die Definition der Gitterkonstanten bestimmt. Sie wird selbst bei der Reflexion an idealen Kristallen nicht beliebig gut, da die durch primäre Extinktion bedingte endliche Eindringtiefe der streuenden Welle in das Kristallvolumen im Impulsraum eine endliche Breite des Reflexes liefert. Dieses Phänomen wird im Kapitel über dynamische Beugung behandelt.

Ein Braggwinkel von $\Theta = \pi/2$ bedeutet geometrisch, daß die Achse des reflektierten Strahls mit der des einfallenden zusammenfällt; der Strahl wird in sich reflektiert, daher der Name Rückstreuspektrometer für dieses Instrument.

Beim Dreiachsenspektrometer konnten wir die auf die Probe einfallende Wellenlänge durch Änderung des Braggwinkels am Monochromator variieren. Am Rückstreuspektrometer hingegen ist diese einfache Methode nicht mehr anwendbar, da wir sofort unsere hohe Auflösung einbüßen würden. Wir müssen anstatt des Braggwinkels die Gitterkonstante d des reflektierenden Kristalls ändern, wozu sich zwei unterschiedlich Prinzipien durchgesetzt haben:

- Variation der Wellenlänge mittels thermischer Ausdehnung des Kristallgitters, sowie

- Variation der Wellenlänge mittels longitudinalem Dopplereffekt bei der Reflexion an einem in Richtung der Strahlachse bewegten Kristall.

Diese Verfahren werden in den Kapiteln 2.4 und 2.5 behandelt.

## 2.2. Grundprinzip des Rückstreuspektrometers

In der Anwendung finden wir die verschiedensten Varianten und Raffinessen des Rückstreuspektrometers. Hier soll ein typischer Aufbau, wie er auch in dieser Arbeit verwirklicht wurde, im wesentlichen besprochen werden. Auf Einzelheiten wie Datenerfassung oder Aufbereitung des Primärstrahls wird in den Kapiteln 4 und 5 dieser Arbeit eingegangen.

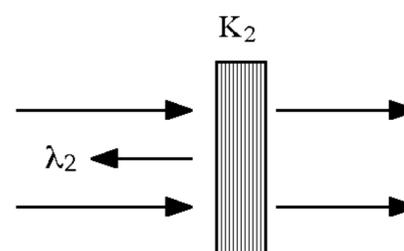

Durch Braggreflexion an einem Kristall $K_2$ unter einem Winkel von $\Theta = 90°$, also senkrechtem Einfall auf die Gitterebenen, wird aus dem weißen Primärstrahl[1] eine wohldefinierte Wellenlänge $\lambda_2$ in sich zurückgestreut (Abbildung 2.3). Die abgelenkten Neutronen treffen auf einen Detektor, während alle anderen ungehindert den Kristall durchdringen.

Abbildung 2.3:
Selektion der Wellenlänge durch Braggreflexion an dem Kristall $K_2$.

Da durch die Rückstreugeometrie auch für die einfallenden Neutronen der Detektor im Weg steht, muß dieser halbdurchlässig gemacht werden, das heißt, er sollte etwa 50% der auftreffenden Neutronen nachweisen, die andere Hälfte ungehindert hindurchlassen. Um noch unterscheiden zu können, aus welcher Richtung die nachgewiesenen Neutronen gekommen sind, ist es nötig den

---

[1] Ein weißer Strahl ist so definiert, daß er im ganzen Abtastintervall für jede Wellenlänge die gleiche Zählrate $R_W(\lambda)$ liefert. In diesem Fall sprechen wir auch von einem kontinuierlichen Spektrum:

$$R_W(\lambda) = \text{const.}$$

(2.8)

Strahl mit Hilfe eines Zerhackers zu pulsen, und die Detektorelektronik synchron zu den Strahlpaketen ein- und auszuschalten. In Abbildung 2.4 ist schematisch die Anordnung des Zerhackers und des Detektors, sowie ein einfallendes und ein reflektiertes Strahlpaket dargestellt.

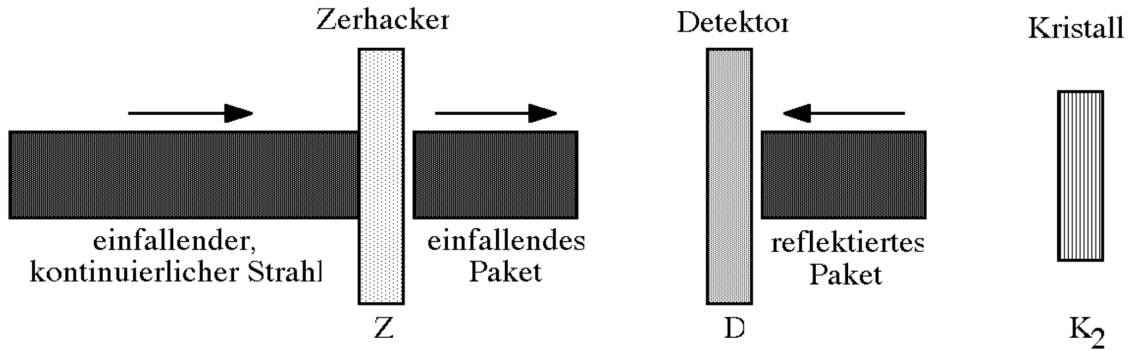

Abbildung 2.4:
Anordnung des Zerhackers und des halbdurchlässigen Detektors. Der kontinuierliche, einfallende Strahl wird in Pakete zerschnitten.

Verändern wir nun mit dem Gitterabstand an $K_2$ die reflektierte Wellenlänge $\lambda_2$, so kann man die Wellenlängenverteilung im einfallenden Strahl abtasten.

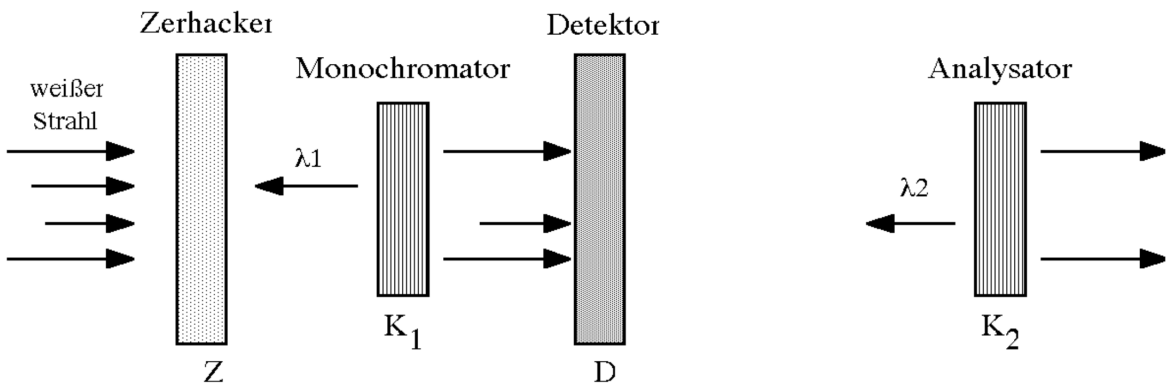

Abbildung 2.5:
Prinzipieller Aufbau eines Rückstreuspektrometers. Durch Braggreflexion am Monochromator $K_1$ wird aus dem einfallenden, weißen Strahl eine wohldefinierte Wellenlänge $\lambda_1$ ausgeblendet. Mit Änderung der Gitterkonstanten des Analysatorkristalls $K_2$ wird die resultierende Wellenlängenverteilung abgetastet.

Wird nun in den einfallenden Strahl ein weiterer Kristall $K_1$ gestellt (Abbildung 2.5), so blendet dieser wiederum durch Braggreflexion ein Wellenlängenband aus dem weißen Primärstrahl aus. Durch Änderung der Gitterkonstanten an $K_2$ können wir die re-sultierende Wellenlängenverteilung abtasten und finden so an der ausgeblendeten Stelle ein Loch im kontinuierlichen Spektrum, wie dies die Abbildung 2.6 andeutet. Wegen der Beziehung (2.7) wird die ausgeblendete Wellenlänge am schärfsten definiert, wenn beide Kristalle in Rückstreuung und damit parallel zueinander sind. Für viele Rückstreuspektrometer ist dies die Auflösungsfunktion. $K_1$ wird Monochromator, $K_2$ Analysator genannt.

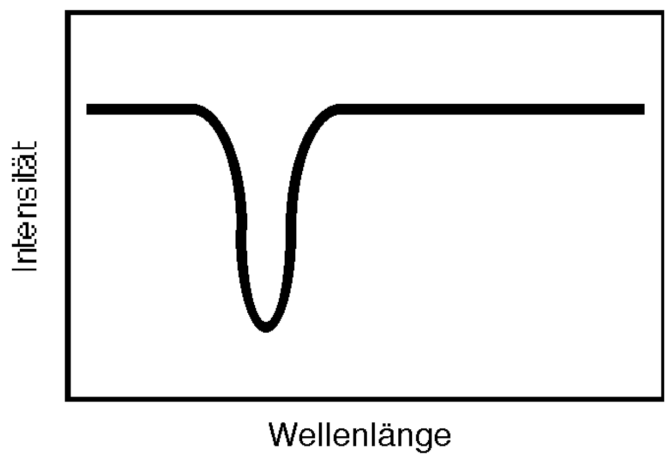

Abbildung 2.6:
Qualitativer Verlauf eines typischen Rückstreuspektrums.

Oft werden auch beide Kristalle in Reflexion betrieben, was dann anstatt eines Loches einen Berg liefert. Durch Einbringen einer Probe zwischen $K_1$ und $K_2$ kann die Verschiebung und Verformung des Berges untersucht werden, um Aufschluß über inelastische oder quasielastische Prozesse zu bekommen. Bestehende Rückstreuspektrometer sind mit Vieldetektorsystemen und großflächigen Analysatorplatten (bis zu einigen Quadratmetern) ausgestattet, um gleichzeitig in verschiedenen Raumrichtungen und somit bei verschiedenen Impulsüberträgen die Streuung zu untersuchen. Abbildung 2.7 zeigt z.B. den Aufbau des Rückstreuspektrometers IN10C am ILL.

Die beschriebene Anordnung bietet aber auch die Möglichkeit, Aussagen über die Kristalle $K_1$ oder $K_2$ selbst zu machen. Meßgrößen sind z.B. die Form, Breite, Tiefe, Position oder integrierte Reflektivität der Linie. Die Position, zum Beispiel, gibt direkten Aufschluß über die Gitterkonstante des Kristalls, die Halbwertsbreite deren Unschärfe, sei sie z.B. durch innere Spannungen oder Mosaizität hervorgerufen.

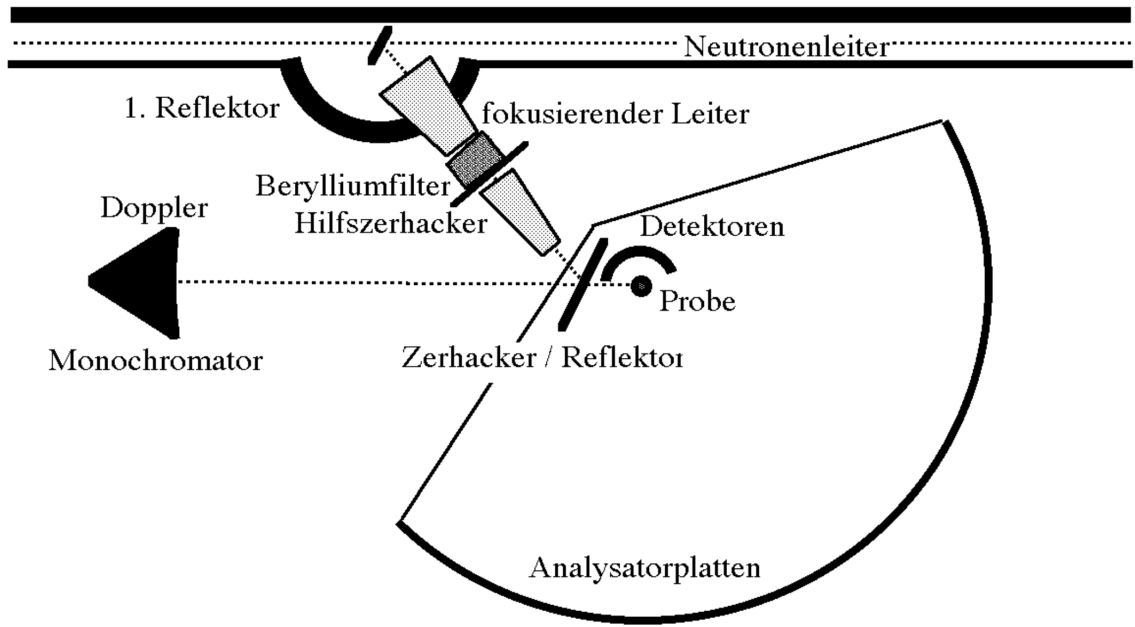

Abbildung 2.7:
Anordnung des im Bau befindlichen Rückstreuspektrometers IN10C am
Institut Laue Langevin in Grenoble. Die Neutronen werden hier mit zwei
Graphitreflektoren aus dem Leiter ausgeblendet. DieBauteile sind fokusie-
rend ausgelegt, um den Fluß auf die Probe zu erhöhen. Monochromator und
Analysator belegen eine Fläche von etwa 7 m$^2$. (Quelle: [88G]).

## 2.3. Wellenlängenauflösung

Die Auflösungsfunktion $A(\lambda)$ des Instrumentes ergibt sich durch
Faltung der Transmissionsfunktion $T_1(\lambda)$ von $K_1$ mit der Reflex-
ionsfunktion $R_2(\lambda)$ von $K_2$:

$$A(\lambda) = \int_{-\infty}^{\infty} T_1(\tau) \cdot R_2(\lambda - \tau) \, d\tau$$

(2.9)

Die beste Auflösung des Instruments ergibt sich unter
Verwendung idealer Kristalle. Die minimaleLinienbreite wird dann
von der dy namischen Streutheorie vorhergesagt und hängt stark
vom Material und dem verwendeten Reflex ab. In diesem Fall sind
$T_1(\lambda)$ und $R_2(\lambda)$ Ewaldkurven und in der Breite linear zum
Stukturfaktor $F^b_{hkl}$ des Reflexes. Die grundlegenden Beziehungen
hierzu werden in Kapitel 3.4 zusammengestellt.

## 2.4. Variation der Einfallsenergie mittels thermischer Gitterausdehnung

Eine der heute angewendeten Methoden zur Änderung der Einfallsenergie an Rückstreuspektrometern ist die Ausnützung thermischer Ausdehnung des Monochromatorkristalls.

Sei $\alpha(T)$ der thermische Ausdehnungskoeffizient des Kristalls bei der Temperatur T und bezüglich der Größe $L(T)$, eine seiner linearen Abmessungen, so folgt durch Integration von

$$dL = \alpha(T) \cdot L \cdot dT \qquad (2.10)$$

für deren Änderung $\Delta L$ mit $L_0 := L(T_0)$

$$\Delta L + L_0 := L(T) = L_0 \cdot e^{\int_{T_0}^{T} \alpha(\tau)\, d\tau} . \qquad (2.11)$$

oder relativ betrachtet

$$\frac{\Delta L}{L} = e^{\int_{T_0}^{T} \alpha(\tau)\, d\tau} - 1 . \qquad (2.12)$$

Oft wird in der Literatur dieser Wert auch direkt durch ein Polynom n-ten Grades approximiert. Diese Größen müssen allerdings mit Vorsicht genossen werden, da der hier benötigte mikroskopische Ausdehnungskoeffizient nicht immer gleich dem makroskopischen ist. In der Anwendung sollte daher eine Eichkurve der Neutronenenergie gegen die Temperatur aufgenommen werden.

Diese Methode wird am ILL am thermischen Rückstreuspektrometer IN13 angewandt, wobei der $CaF_2$ [422]-Monochromator einen Temperaturbereich von -196 °C bis 450 °C durchfahren kann, was einem Energiebereich von -125 µeV < δE < 300 µeV entspricht. Die Auflösung dieses Instruments liegt bei $\Delta E \approx 8$ µeV ($\lambda_0 = 2,23$ Å).

# 2.5. Variation der Einfallsenergie mittels Dopplereffekt

Die andere gängige Methode zur Energieänderung nutzt den Dopplereffekt bei der Reflexion an einem bewegten Monochromatorkristall (Abbildung 2.8).

Sei $v_0$ die Geschwindigkeit des an einem ruhenden Kristall reflektierten Neutrons, so ergibt sich bei der Reflexion an dem mit der Geschwindigkeit V parallel zur Strahlachse bewegten Monochromator für das Neutron

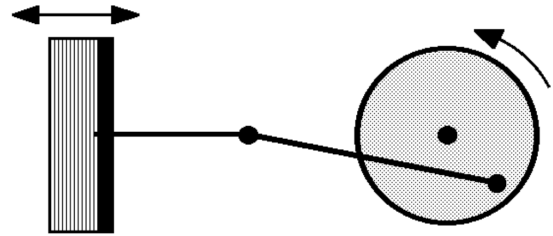

Abbildung 2.8:
Prinzip eines Dopplerantriebs

$$v = v_0 + V$$

(2.13)

Dabei ist das Vorzeichen von V positiv, wenn sich der Kristall parallel zum reflektierten Neutron bewegt, sonst negativ.

Es ist bemerkenswert, daß ein einzelnes Neutron, das den Mono - chromator mit dieser Geschwindigkeit verläßt vorher im weißen Strahl die Geschwindigkeit $v' = v_0 - V$ gehabt hat. Es erhält also, wie beim elastischen Stoß nicht anders erwartet, einen Geschwindigkeitszuwachs der doppelten Kristallgeschwindigkeit. Da die Braggreflexion und damit die Selektion dieses Neutrons aber im Kristallsystem stattfindet, addiert sich bei der Transformation ins Laborsystem nur einmal die Dopplergeschwindigkeit hinzu.

Diese Überlegung bedeutet auch, daß eine eventuell inhomogene Geschwindigkeitsverteilung des einfallenden Strahls am Monochromator um $v_0$ gespiegelt wird.

Die Beziehung

$$\frac{\Delta E}{E} = 2 \frac{\Delta v}{v}$$

(2.14)

stellt uns den Zusammenhang zur Neutronenenergie E her:

$$\boxed{\dfrac{\Delta E}{E_0} = 2\,\dfrac{V}{v_0}}$$

(2.15)

In der Praxis wird der Kristall näherungsweise sinusförmig hin- und herbewegt. Dazu ist dieser auf einen Kolben montiert, der durch eine Pleuelstange mit einem sich drehenden Rad verbunden ist (Abbildung 2.8). Die Näherung besteht darin, daß die Länge dieser Pleuelstange wesentlich größer als die Kurbellänge A ist.

Seien $\omega = 2\pi\,\nu$ die Kreisfrequenz und A die Amplitude der Schwingung, so ergibt sich für die Kristallgeschwindigkeit in Abhängigkeit von der Zeit t

$$V(t) = \omega\,A\,\cos(\omega\,t)$$

(2.16)

und somit ein Geschwindigkeitsintervall des Neutrons von

$$v \in [v_0 - V_{max},\ v_0 + V_{max}]$$

(2.17)

mit der Geschwindigkeitsamplitude

$$\boxed{V_{max} := \omega\,A = 2\pi\nu\,A}.$$

(2.18)

In Energien und Wellenlängen ausgedrückt bedeutet dies:

$$\boxed{\delta E_{max} = \pi\,\sqrt{8\,m\,E_0}\cdot\nu\,A}$$

(2.19)

$$\boxed{\delta\lambda_{max} = \dfrac{m\,\lambda_0{}^2}{\mathcal{L}}\cdot\nu\,A}$$

(2.20)

Diese Methode hat sich seit ca. 20 Jahren in verschiedenen Forschungseinrichtunegn etabliert und wird am ILL an den Rückstreuspektrometern IN10 und IN10C erfolgreich angewendet. Hier erreicht man z.B. mit einem Si [111]-Monochromator eine Auflösung von $\Delta E = 300$ neV in einem dynamischen Bereich von $\Delta E_{max} = 15\ \mu$eV. [88B].

# 2.6.    Form der Spektren und Normierung

Ziel von Messungen an Spektrometern ist es, Aussagen über Intensitäten in Abhängigkeit der Wellenlänge[2] zu machen. Im allgemeinen wird dem Spektrum dabei noch eine Abhängigkeit des zeitlichen Verlaufs der Messung aufgeprägt, die anschließend weg normiert werden muß, da sie für die Physik nicht von Interesse ist. Diese Abhängigkeit kommt daher, daß wir bei verschiedenen Wellenlängen verschieden lang messen und somit proportional zu dieser Zeit mehr oder weniger Neutronen für diesen Meßwert unseres Spektrums erhalten. Dies soll hier am oben angeführten Beispiel der Methode des Dopplereffekts erläutert werden:

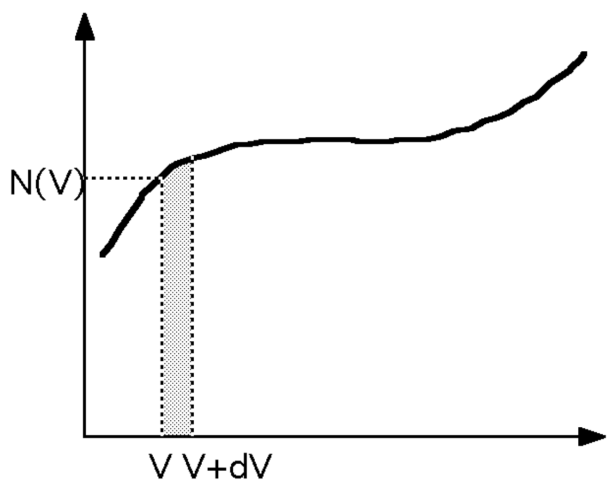

Abbildung 2.9:
Veranschaulichung von N(V) dV

Die Messung soll den direkten Zusammenhang zwischen Intensität und Kristallgeschwindigkeit wiedergeben. Diese läßt sich für kleine Geschwindigkeitsintervalle mittels linearer Beziehungen, analog zur Gleichung (2.5) in Wellenlängen oder Energien umrechnen. Gehen wir der Einfachheit halber zunächst wieder von einem gleichverteilten, weißen Primärstrahl aus, das heißt, die Zählrate ist von der Kristallgeschwindigkeit V(t) unabhängig:

$$R(V) = \text{const.} \qquad\qquad (2.21)$$

Dann ist die Zahl N(V) dV der Ereignisse im Intervall [V, V+dV] (Abbildung 2.9) mit R proportional zur Zeit dt, in der sich die Geschwindigkeit in diesem Intervall befindet:

$$N(V)\ dV = R\ dt$$

---

2  Genauso kann man die Abhängigkeiten von anderen Größen wie Energie, Betrag der Geschwindigkeit usw. erhalten, die alle über eine feste Beziehung mit der Wellenlänge verknüpft sind.

$\Rightarrow$

$$N(V)\ dV = R \frac{1}{\left|\dfrac{\partial V}{\partial t}\right|}\ dV$$

(2.22)

Dies ergibt somit

$$\boxed{N(V) = R \frac{1}{\left|\dfrac{\partial V}{\partial t}\right|}}$$ .

(2.23)

In unserem Fall ist die Zeitabhängigkeit der Geschwindigkeit durch

$$V(t) = V_{max} \cdot \cos(\omega\ t)$$

(2.24)

gegeben, was nach t aufgelöst

$$t = \frac{1}{\omega}\ a\cos(\frac{V}{V_{max}})$$

(2.25)

ergibt. Bilden wir nun die Ableitung von (2.24), so erhalten wir

$$\left|\frac{\partial V}{\partial t}\right| = V_{max}\ \omega\ \sin(\omega\ t)$$

$\Leftrightarrow$

$$\left|\frac{\partial V}{\partial t}\right| = V_{max}\ \omega\ \sin\left( a\cos\left(\frac{V}{V_{max}}\right)\right)$$

$\Leftrightarrow$

$$\boxed{\left|\frac{\partial V}{\partial t}\right| = V_{max}\ \omega\ \sqrt{1 - \left(\frac{V}{V_{max}}\right)^2}}$$ .

(2.26)

Den letzten Schritt kann man sich leicht am Einheitskreis überlegen. Wir erhalten also

$$N(V) = \frac{R}{V_{max}\,\omega}\,\sqrt{1 - \left(\frac{V}{V_{max}}\right)^2}\,.$$

(2.27)

Da V(t) und somit dessen Ableitung periodisch ist[3], wird die Form des Spektrums während einer Periodendauer $T_P$ bestimmt. Mißt man Zeiten t » $T_P$ mit

$$T_P := \frac{2\pi}{\omega}\cdot\frac{1}{2}\,,$$

(2.28)

so erhalten wir für die Zahl $N(V,t)$ der Ereignisse pro Geschwindigkeitsintervall

$$N(V,t) = N(V)\cdot\frac{t}{T_P}$$

(2.29)

und eingesetzt

$$N(V,t) = \frac{R\,t}{\pi\,V_{max}}\cdot\frac{1}{\sqrt{1 - \left(\frac{V}{V_{max}}\right)^2}}\,.$$

(2.30)

Man beachte, daß hier die Frequenzabhängigkeit noch in der Geschwindigkeisamplitude $V_{max} = \omega\,A$ enthalten ist. Dies bedeutet, daß bei V = 0 die Zählrate umgekehrt proportional zur Dopplerfrequenz ist.

Die Form des Spektrums entspricht also dem Kehrwert des Kreises mit Radius $V_{max}$ und ist in Abbildung 2.10 wiedergegeben.

---

3   Da in unsere Formeln nur der Betrag der Ableitung eingeht, ergibt sich in Gleichung (2.28) der Faktor 1/2.

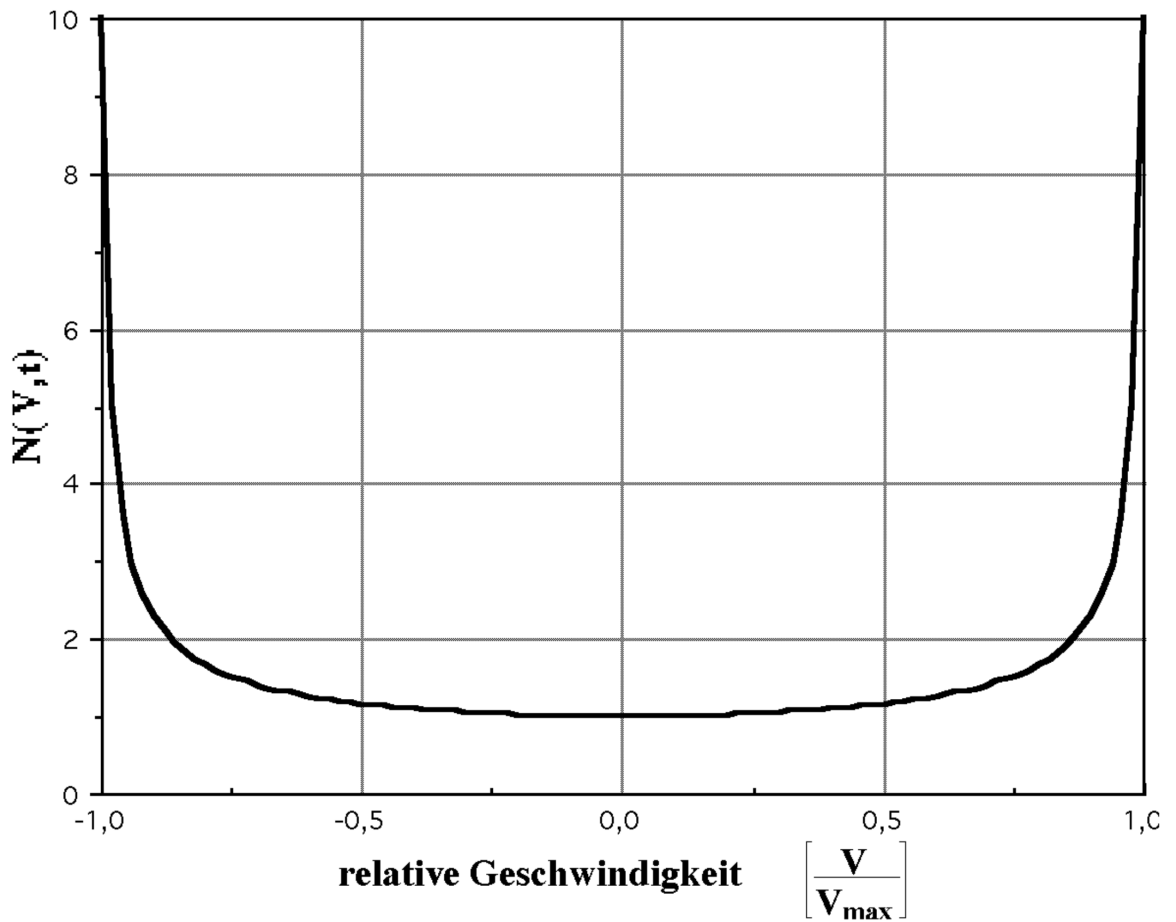

Abbildung 2.10:
Der Kehrwert des Einheitskreises ist das theoretische Monitorspektrum bei
Reflexion an einem sich sinusförmig hin- und herbewegenden Kristall.

Die Gesamtzahl aller Ereignisse im Spektrum zur Zeit t erhalten wir
durch Integration obiger Gleichung:

$$N(t) = \int_{-V_{max}}^{V_{max}} N(V,t) \, dV \qquad (2.31)$$

$\Leftrightarrow$

$$N(t) = \frac{R\,t}{\pi} \int_{-V_{max}}^{V_{max}} \frac{1}{\sqrt{V_{max}^2 - V^2}} \, dV$$

$\Leftrightarrow$

$$N(t) = \frac{R\,t}{\pi} \left[ asin\left(\frac{V}{V_{max}}\right) \right]_{-V_{max}}^{V_{max}}$$

$\Leftrightarrow$

$$\boxed{N(\text{t}) = R\,t}$$

(2.32)

Dies entspricht genau unserer Voraussetzung (2.21).

Experimentell wird ein weißes Spektrum, das sogenannte Monitorspektrum $N(V,t)$ mitgemessen, um damit das Meßergebnis zu normieren. Dies hat den Vorteil, daß man sich erstens die numerische Anpassung erspart und zweitens jeden beliebigen, nicht nur rein sinusförmigen Geschwindigkeitsverlauf wegnormieren kann.

# 3. Einblick in die dynamische Streutheorie

Die Wechselwirkung von Wellen mit Idealkristallen wird von der dynamischen Streutheorie beschrieben. Sie basiert auf der Lösung der Wellengleichung im periodischen Potential des Kristallgitters und ist auf alle Arten von Wellen wie z.B. Röntgen-, Elektronen- und Neutronenstrahlen anwendbar. In diesem Kapitel wollen wir uns auf das Neutron beschränken. Außerdem werden nur die grundlegenden Gleichungen zum Verständnis der Braggreflexion herausgestellt. Mehr oder weniger ausführliche Abhandlungen dieser Theorie finden sich in [76R], [78R], [45Z], [78D], [17E] und [14D]. Die Nomenklatur richtet sich nach der Arbeit von H. Rauch und D. Petrascheck [76R].

In der kinematischen Beugungstheorie wird die primäre Extinktion des auf den Kristall einfallenden Strahls vernachlässigt. Unter primärer Extinktion versteht man die Abschwächung der einfallenden Welle zugunsten der gestreuten Welle. Diese Näherung scheitert bei Kristalldicken, die vergleichbar oder größer als eine typische Eindrigtefe der Welle in das Kristallvolumen ist. In der dynamischen Theorie hingegen wird versucht, die Wellengleichung exakt zu lösen.

## 3.1. Grundgleichung für die Beugung am Kristall

Die Beschreibung von Neutronen genügt der Schrödingergleichung

$$\left\{ -\frac{\hbar^2}{2\,m}\,\Delta + V(\vec{r}) \right\} \Psi = E\,\Psi \quad . \tag{3.1}$$

Für langsame Neutronen (Wellenlänge « Kernradius) können wir die Wechselwirkung mit dem Kristall durch das Fermi-Pseudopotential

$$V(\vec{r}) = \frac{2\,\pi\,\hbar^2}{m} \cdot \sum_j b_c^j\,\delta(\vec{r} - \vec{r}_j) \tag{3.2}$$

ansetzen. Dabei bedeuten m die Masse des Neutrons, $b_j$ die kohä­rente, gebundene Streulänge für das j-te Atom, $\vec{r}_j$ der Ortsvektor des j-ten Gitteratoms, $\hbar$ das Plancksche Wirkungsquantum und

$$E = \frac{\hbar^2 k^2}{2\,m} \qquad (3.3)$$

die Energie des Neutrons mit dem Wellenvektor $k = |\vec{k}|$. Da $V(\vec{r})$ die Gitterperiodizität besitzt, kann das Wellenfeld $\Psi(\vec{r})$ als Eigenfunktion eines unendlichen Kristalls, die sogenannte Blochwelle geschrieben werden:

$$\Psi(\vec{r}) = e^{\,i\vec{K}\vec{r}} \cdot u(\vec{r}) \qquad (3.4)$$

$\vec{K}$ ist der Wellenvektor im Kristall. Das Wellenfeld ist also eine Ebene Welle, die mit einer periodischen Blochfunktion $u(\vec{r})$ modu­liert ist. Bei Transformation in den Fourierraum erhalten wir für das Potential

$$V(\vec{r}) = \sum_{\vec{G}} V(\vec{G}) \cdot e^{\,i\vec{G}\vec{r}} \qquad (3.5)$$

und für die Wellenfunktion entsprechend

$$\Psi(\vec{r}) = \sum_{\vec{G}} u(\vec{G}) \cdot e^{\,i(\vec{K}+\vec{G})\vec{r}} \qquad (3.6)$$

Dabei sind $\vec{G}$ reziproke Gittervektoren,

$$V(\vec{G}) = \frac{2\pi\,\hbar^2}{m\,V_z} \left| F_{hkl}^b \right| \qquad (3.7)$$

und $u(\vec{G})$ die Entwicklungskoeffizienten für das Potential bzw. für die Wellenfunktion. Der zu $\vec{G}$ gehörende Strukturfaktor ist durch

$$F_{hkl}^b = \sum_j e^{\,-i\vec{G}\vec{\rho}_j} \cdot b_c^j \qquad (3.8)$$

definiert. Hier wird über alle j Atome an den Positionen $\vec{\rho}_j$ der Einheitszelle summiert. Setzen wir die Fouriertransformationen (3.5) und (3.6) in die Schrödingergleichung (3.1) ein, so erhalten wir die Grundgleichung der dynamischen Beugungstheorie

$$\left\{ \frac{\hbar^2}{2\,m} \left| \vec{K} + \vec{G} \right|^2 - E \right\} u(\vec{G}) = - \sum_{\vec{G}'} V(\vec{G}-\vec{G}') \, u(\vec{G}')$$

(3.9)

Diese Gleichung beschreibt unendlich viele Blochwellen im Kristall, die sich in den Wellenvektoren gerade um die reziproken Gittervektoren unterscheiden. Selbst die einfallende Welle mit $\vec{G} = \vec{0}$ ist nicht besonders ausgezeichnet. Vom mathematischen Standpunkt her haben wir es mit einem System unendlich vieler homogenerer Gleichungen zu tun. Dieses ist nicht exakt lösbar.

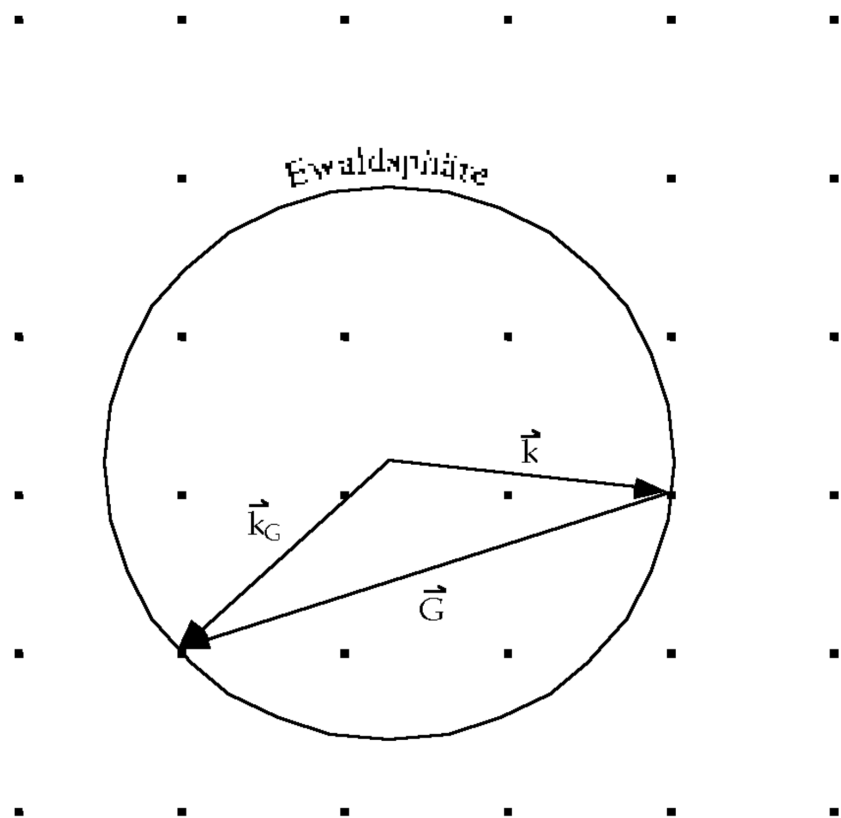

Abbildung 3.1:
Ewaldkonstruktion für die angeregten Braggreflexe im reziproken Gitter. Nur Gitterpunkte, die auf der Ewaldsphäre, einer Fläche konstanter Energie liegen erfüllen die Braggbedingung.

Zur Vereinfachung wird die Zahl der Wellen im Kristall einge schränkt. Nach der Ewaldschen Konstruktion des Braggesetzes tragen nur Wellen mit nichtverschwindender Amplitude $u(\vec{G})$ zur Interferenz bei, deren zugehörige reziproken Gitterpunkte auf der Ewaldsphäre, also einer Kugel im reziproken Gitter mit Radius $|\vec{K}|$ um den Ansatzpunkt von $\vec{k}$, liegen (Abbildung 3.1). Da die Bragg - bedingung die Lage der Beugungsmaxima recht gut beschreibt, werden in der dynamischen Theorie nur Wellen zugelassen, die zu einem reziproken Gittervektor "nahe" der Ewaldsphäre gehören.

## 3.2. Einstrahlnäherung

Außer dem Ursprung $\vec{0}$ liegt kein weiterer reziproker Gittervektor in der Nähe der Ewaldsphäre. Dann gibt es keinen abgebeugten Strahl, und es wird nur der einfallende Strahl in Vorwärtsrichtung beschrieben. Gleichung (3.9) vereinfacht sich zu

$$\left( \frac{\hbar^2 K_0^2}{2m} - E \right) u(\vec{0}) = -V(\vec{0})\, u(\vec{0}) \quad . \tag{3.10}$$

Diese Gleichung gibt uns den Zusammenhang zwischen der Wel lenzahl $K_0$ im Kristall und der Wellenzahl $k$ des einfallenden Strahls im Vakuum wieder. Das Verhältnis dieser Größen

$$n := \frac{K_0}{k} = 1 - \frac{1}{2}\frac{V(\vec{0})}{E} \qquad \text{für } \frac{V(\vec{0})}{E} \ll 1 \tag{3.11}$$

definiert einen mittleren Brechungsindex $n$ des Kristalls. Gleichung (3.7) eingesezt, erhalten wir

$$n = 1 - \frac{\lambda^2 N \overline{b}_c}{2\pi} \quad , \tag{3.12}$$

worin $N$ die Zahl der Streuzentren und $\overline{b}_c$ die mittlere Streulänge des Materials bedeuten.

### 3.2.1. Brechung

Mit Gleichung (3.12) ist eine Brechung des einfallenden Strahls beim Eintritt in den Kristall verbunden, die sich mit Hilfe von Dispersionsflächen beschreiben läßt. Dispersionsflächen eines Mediums sind Flächen konstanter Energie im Fourierraum, z.B. die Fermifläche bei Elektronen. Für freie Teilchen im Vakuum sowie bei der Einstrahlnäherung unserer Theorie sind dies Kugeloberflächen um den Nullpunkt des reziproken Raumes mit Radien k bzw. $K_0$.

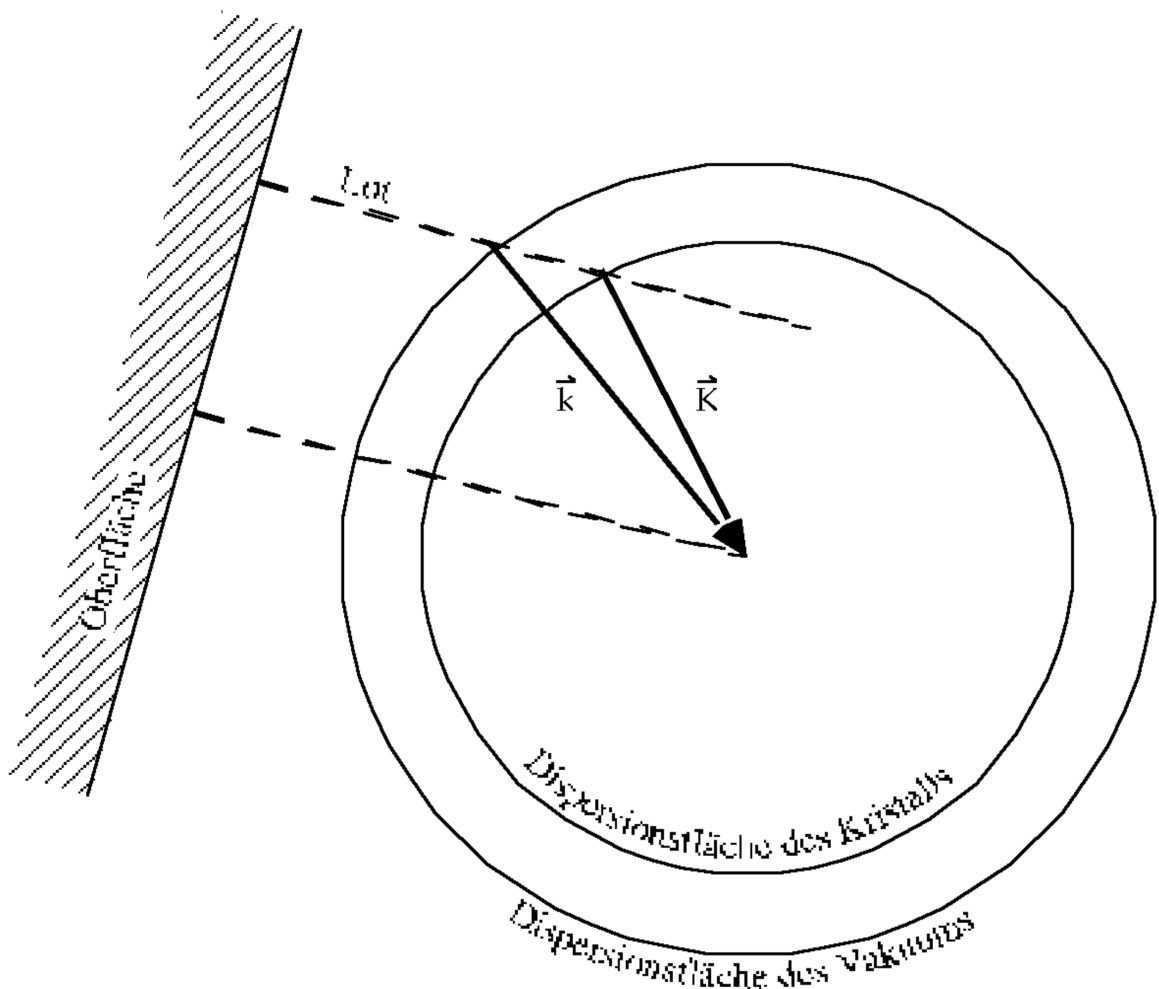

Abbildung 3.2:
Beschreibung der optischen Brechung mit Hilfe der Dispersionsflächen. Die beiden dargestellten Dispersionsflächen gehören zur gleichen Energie, einmal im Vakuum, einmal im Inneren des Kristalls. Wegen der Energieerhaltung kann sich ein Neutron nur auf diesen Sphären bewegen. Die Oberfläche setzt eine Randbedingung, nämlich daß die Parallelkomponente des Wellenvektors erhalten bleiben muß.

Die Kristalloberfläche definiert die Randbedingung, daß die zu ihr parallele Komponente k" des Wellenvektors beim Eintritt in das Medium erhalten bleiben muß:

$$K_0^{''} = k^{''} \qquad\qquad (3.13)$$

Dann läßt sich die Richtung von $\vec{K}_0$ wie in Abbildung 3.2 gezeigt konstruieren. Fällen wir das Lot vom Endpunkt von $\vec{k}$ auf die Kristalloberfläche, so gibt dessen Schnittpunkt mit der Dispersionsfläche für das Kristallvolumen den Endpunkt von $\vec{K}_0$.

### 3.2.2. Totalreflexion am optisch dünneren Medium

Trifft eine einfallende Welle unterhalb eines kritischen Winkels zur Oberfläche auf ein optisch dünneres Medium, so wird sie bekannt­lich totalreflektiert. Mit Hilfe der Dispersionsflächen im reziproken Raum ist diese Situation in Abbildung 3.3 dargestellt:

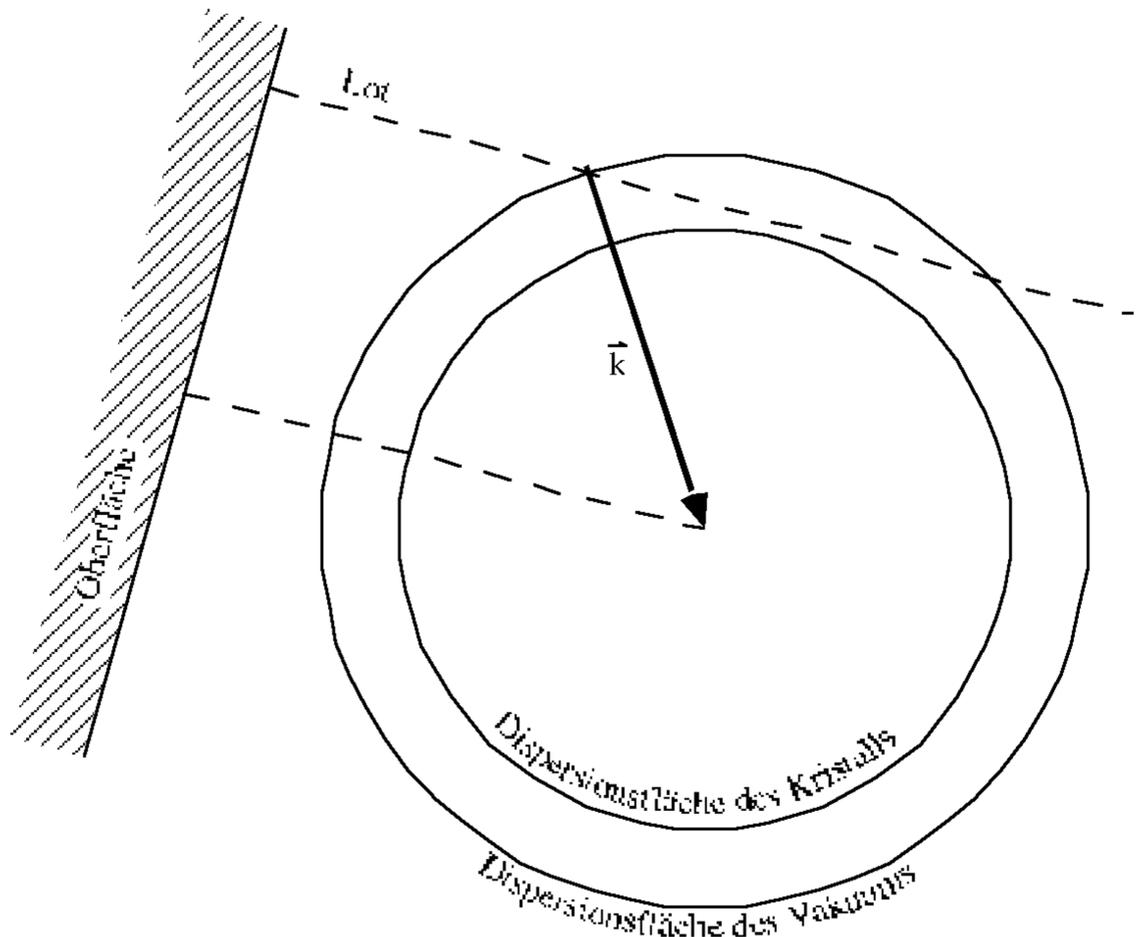

Abbildung 3.3:
Beschreibung der Totalreflexion mittels Dispersionsflächen. Der einfallende Strahl fällt unter einem kleinen Winkel auf die Oberfläche. Die Parallelkom­ponente des Wellenvektors ist größer als der maximal mögliche Wellenvektor im Kristall zu vorgegebener Energie. Im Kristall wird daher keine fortschreitende Welle angeregt, was zur Totalreflexion führt.

Fällen wir wieder das Lot vom Endpunkt von $\vec{k}$ auf die Oberfläche, so hat dieses keine Schnittpunkte mit der Dispersionsfläche des Kristallinneren. Die Randbedingung zur Erhaltung von k" kann nicht erfüllt werden. Im Kristall wird keine fortschreitende Welle angeregt; der einfallende Strahl wird totalreflektiert.

## 3.3. Zweistrahlnäherung

In diesem Fall liegt außer dem Nullpunkt noch ein weiterer Gitterpunkt $\vec{G}$ des reziproken Raumes nahe der Ewaldsphäre. Dieser Ansatz ist gerechtfertigt, wenn alle anderen reziproken Gitterpunkte sehr viel weiter als $\vec{0}$ und $\vec{G}$ von der Sphäre wegliegen. Die Gleichung (3.9) wird dann zu

$$\left\{ \frac{\hbar^2}{2\,m} \left|\vec{K}\right|^2 - E \right\} u(\vec{0}) = -V(\vec{0})\,u(\vec{0}) - V(-\vec{G})\,u(\vec{G})$$

$$\left\{ \frac{\hbar^2}{2\,m} \left|\vec{K}+\vec{G}\right|^2 - E \right\} u(\vec{G}) = -V(\vec{G})\,u(\vec{0}) - V(\vec{0})\,u(\vec{G})$$

$$(3.14)$$

Definieren wir die Anregungsfehler

$$\varepsilon := \frac{\left|\vec{K}\right|}{k} - 1 \;\ll 1$$

$$(3.15a)$$

und

$$\varepsilon_G := \frac{\left|\vec{K}+\vec{G}\right|}{k} - 1 \;\ll 1$$

$$(3.15b)$$

als die relativen Unterschiede der Wellenzahlen im Vakuum und im Kristall so läßt sich das Gleichungssystem (3.14) in der Form

$$\left\{ 2\varepsilon + \frac{V(\vec{0})}{E} \right\} u(\vec{0}) \;+\; \frac{V(-\vec{G})}{E}\,u(\vec{G}) \;=\; 0$$

$$\frac{V(\vec{G})}{E}\,u(\vec{0}) \;+\; \left\{ 2\varepsilon_G + \frac{V(\vec{0})}{E} \right\} u(\vec{G}) \;=\; 0$$

$$(3.16)$$

schreiben. Dieses homogene Gleichungssystem at nur dann eine nichttriviale Lösung, wenn dessen Säkulardeterminante verschwindet:

$$\det\begin{bmatrix} \left\{2\varepsilon + \dfrac{V(\vec{0})}{E}\right\} & \dfrac{V(-\vec{G})}{E} \\[3mm] \dfrac{V(\vec{G})}{E} & \left\{2\varepsilon_G + \dfrac{V(\vec{0})}{E}\right\} \end{bmatrix} = 0 \qquad\qquad (3.17)$$

Dies führt zu einer quadratischen Gleichung in $\varepsilon$.

Physikalisch betrachtet, geben die Anregungsfehler $\varepsilon$ und $\varepsilon_G$ die relativen, radialen Abstände der Dispersionsfläche $S_M$ im Kristall zu den Vakuumdispersionsflächen $S_0$ um $\vec{0}$ und $\vec{G}$ mit Radius $k$ wieder. Man kann sie auch als die Koordinaten von $S_M$ in dem krummlinigen Koordinatensystem $S_0$ betrachten. Die quadratische Gleichung (3.17) führt zu einer hyperbolischen Aufspaltung von $S_M$ an den Schnittpunkten der Kugeln mit Radius $|\vec{K}|$ um $\vec{0}$ und Radius $|\vec{K}+\vec{G}|$ um $\vec{G}$. Dies ist in Abbildung 3.4 übertrieben dargestellt. Letztere Kugeln waren die Dispersionsflächen der Einstrahlnäherung.

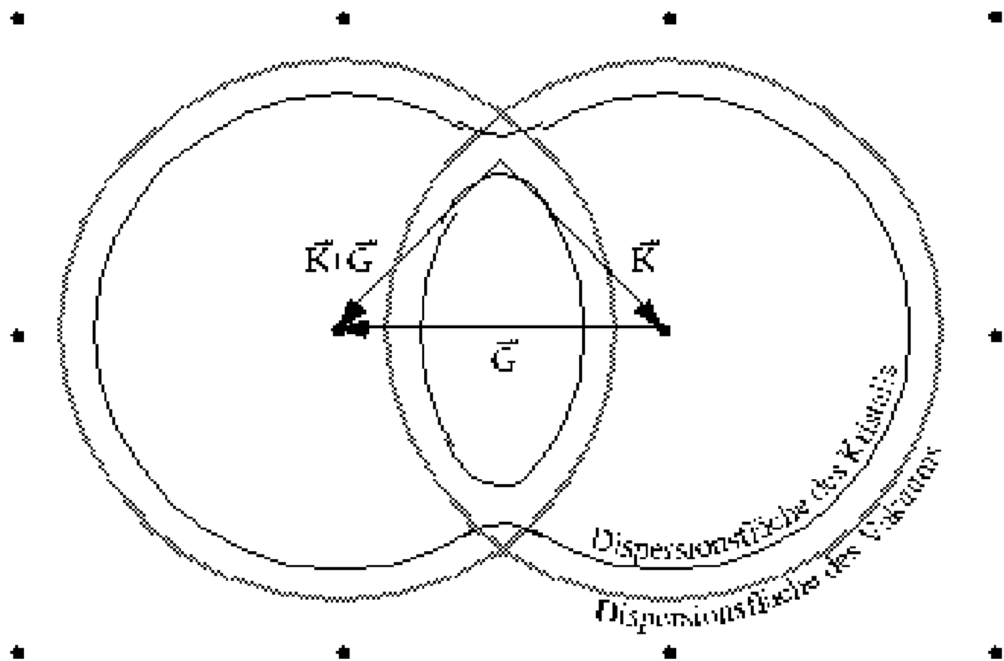

Abbildung 3.4:
Aufspaltung der Dispersionsflächen an den Grenzen der Brillouinzonen im reziproken Gitter.

### 3.3.1. Pendellösungsoszillationen in Laue-geometrie

Eine Vergrößerung der Aufspaltung der Dispersionsfläche ist in Abbildung 3.5 wiedergegeben. Analog zur Konstruktion in Abbildung 3.2 erhalten wir die im Kristall angeregten Wellenvektoren durch die Schnittpunkte des Lots vom Endpunkt von $\vec{k}$ auf die Kristalloberfläche mit den Dispersionsflächen. Dabei gibt es je zwei angeregte Wellen, $\vec{K}_{01}$ und $\vec{K}_{02}$ zu $\vec{0}$ gehörend, sowie $\vec{K}_{G1}$ und $\vec{K}_{G2}$ zu $\vec{G}$ gehörend. Diese entsprechen den zwei unabhängigen Lösungen einer Differenzialgleichung zweiter Ordnung, wie die Schrödingergleichung eine darstellt.

Die Wellenzahlen $K_{01}$ und $K_{02}$ bzw $K_{G1}$ und $K_{G2}$ sind in ihrem Betrag leicht unterschiedlich. Treffen die Wellen bei einem Kristall endlicher Dicke auf eine Oberfläche (Laue-Geometrie), so interferieren sie hier je nach Phasenlage konstruktiv oder destruktiv. Diese örtlichen Schwebungen sind die bekannten Pendellösungsoszillationen, die man erhält, wenn man die Kristalldicke variiert.

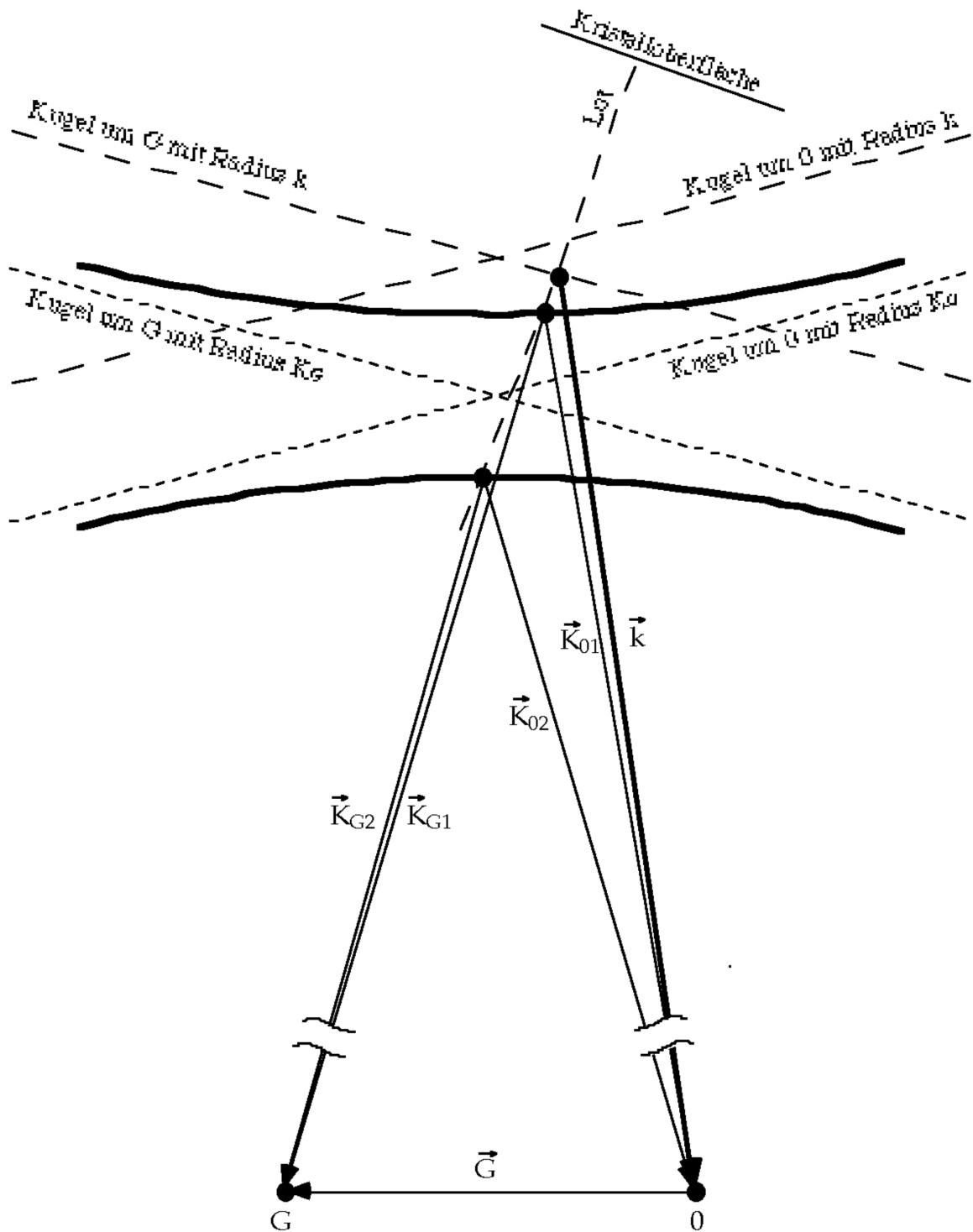

**Abbildung 3.5:**
Auschnitt im Bereich der Aufspaltung der Dispersionsflächen im reziproken Gitter. Die extreme Vergrößerung erlaubt es nicht, die Wellenvektoren Maßstabgerecht darzustellen. Die Konstruktion liefert uns je zwei Wellenvektoren $K_{01}$ und $K_{02}$ zu 0 gehörend, sowie $K_{G1}$ und $K_{G2}$ zu G gehörend.

### 3.3.2. Totalreflexion in symmetrischer Bragg-stellung

Im sogenannten symmetrischen Braggfall liegen die Gitterebenen des reflektierenden Kristalls parallel zu seiner Oberfläche. Betrachten wir diese Situation wieder an den Dispersionsflächen im reziproken Raum (Abbildung 3.6), so finden wir einfallende Wellenvektoren, für die es keine Schnittpunkte des Lots zur Oberfläche mit der aufgespaltenen Dispersionsfläche $S_M$ gibt. Das bedeutet, in diesem Bereich gibt es Totalreflexion, analog zum Kapitel 3.2.2. Drehen wir langsam die Richtung von $\vec{k}$, so kommen wir an einen Punkt, ab dem die Schnittpunkte wieder einsetzen und Wellen im Kristall angeregt werden. Diese verlieren bei weiterer Drehung sehr schnell an Intensität, da wir uns dann zu weit von der Ewald sphäre entfernen.

Dieser Totalreflexionsbereich wird Braggreflex genannt. Seine Breite hängt rein vom Fermi-Pseudopotential des verwendeten Reflexes und vom Strukturfaktor des Reflexes ab und kann nicht unterschritten werden. Das vollkomene Analogon hierfür ist die elektronische Bandlücke an den Grenzen der Brillouinzone.

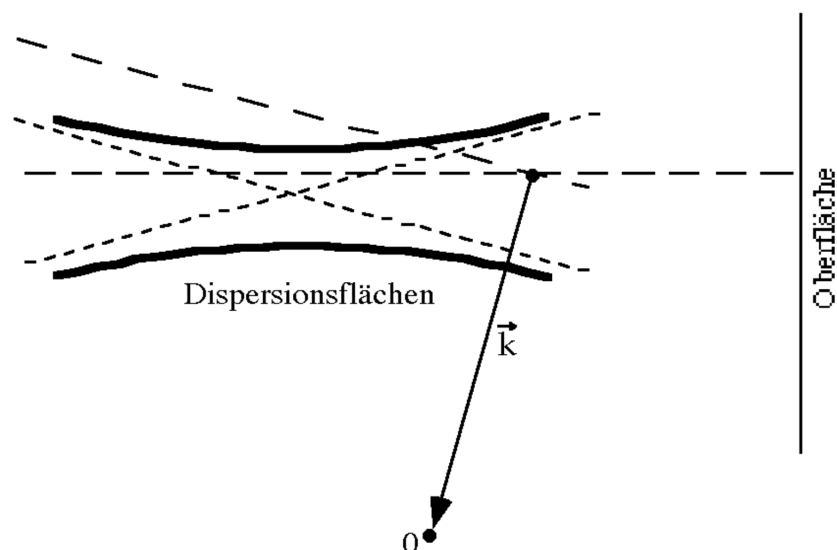

Abbildung 3.6:
In Bragg-Geometrie finden sich einfallenbde Wellenvektoren, deren Parallel-komponente im Kristall verboten ist und somit zu einem Bereich der Totalre-flexion führen.

# 3.4.    Formeln zur Energieauflösung

In diesem Abschnitt sollen die wichtigsten Ergebnisse der dynamischen Theorie zur Beschreibung der zu messenden Bragglinien in Rückstreugeometrie (siehe Kapitel 2) aufgelistet werden.

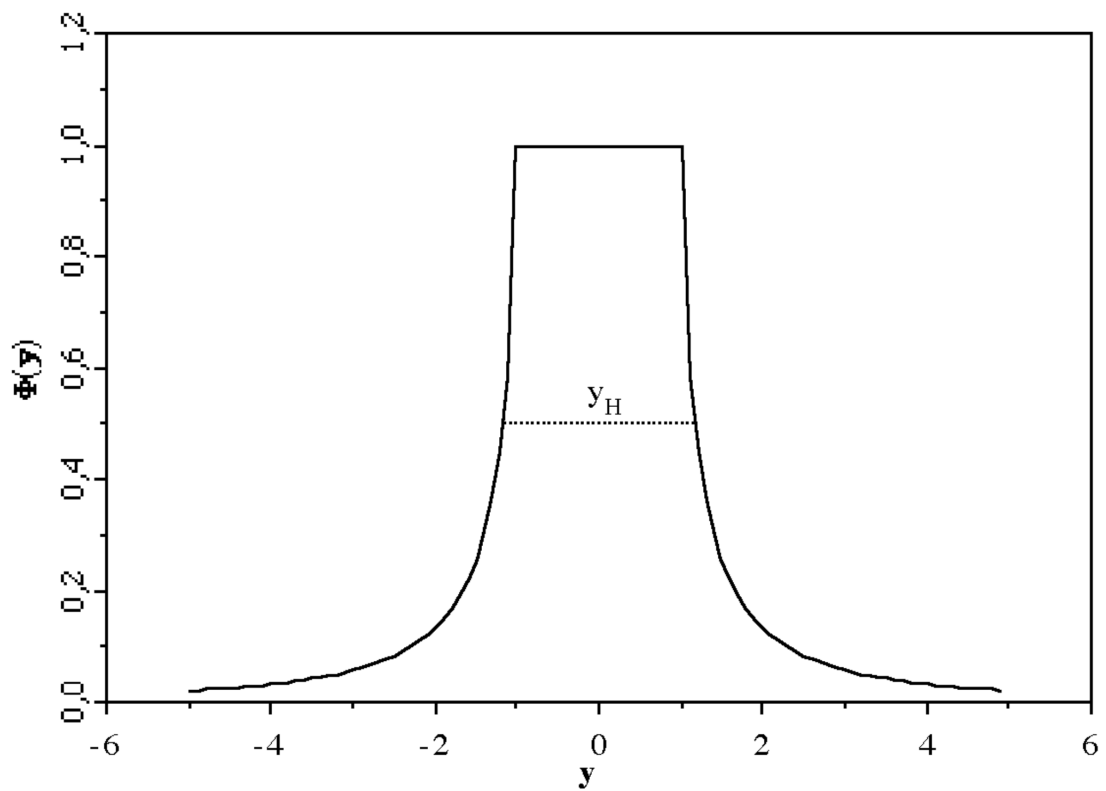

Abbildung 3.7:

Graph der Ewaldfunktion $\Phi(y)$ in Abhängigkeit des dimensionslosen Parameters y.

Die Reflektivität eines Braggreflexes in Abhängigkeit eines dimensionslosen Parameters y wird durch die Ewaldfunktion (Abbildung 3.7)

$$\phi(y) = \begin{cases} 1 & |y| \leq 1 \\ 1 - \sqrt{\dfrac{y^2 - 1}{|y|}} & |y| > 1 \end{cases}$$

(3.18)

beschrieben. Dabei ist y eine Linearkombination der Größe, in deren Abhängigkeit die Reflektivität gemessen wird, in unserem Fall der Neutronengeschwindigkeit v oder der Energie E.

$$y = \frac{v - v_0}{v_0} \cdot \frac{2\pi \, V_z}{\lambda^2 \left| F_{hkl}^b \right|}$$

$$y = \frac{E - E_0}{E_0} \cdot \frac{\pi \, V_z}{\lambda^2 \left| F_{hkl}^b \right|}$$

(3.19)

Hierin bedeuten $|F_{hkl}^b|$ der Betrag des die Steulängen $b_c$ enthaltenden Strukturfaktors, $V_z$ das Volumen einer Einheitszelle sowie $\lambda$ die Wellenlänge.
Die volle Halbwertsbreite der Ewaldkurve ist durch

$$y_H = \frac{4}{\sqrt{3}}$$

(3.20)

gegeben. Dies in die Formeln (3.19) eingesetzt und aufgelöst gibt auf der Geschwindigkeits- und Energieskala relative Breiten von

$$\frac{\Delta E_H}{E_0} = \frac{4}{\sqrt{3}} \cdot \frac{\lambda^2 \left| F_{hkl}^b \right|}{\pi \, V_z}$$

$$\frac{\Delta v_H}{v_0} = \frac{4}{\sqrt{3}} \cdot \frac{\lambda^2 \left| F_{hkl}^b \right|}{2\pi \, V_z}$$

(3.21)

Die integrierte Reflektivität $R_y$ in dimensionslosen Einheiten ist durch

$$R_y = \pi$$

(3.22)

gegeben. In Energien bzw. Geschwindigkeiten ausgedrückt gibt dies[4]

$$R_E = R_y \, \frac{\Delta E_H}{y_H} = \frac{\sqrt{3}}{4} \, \pi \, \Delta E_H$$

(3.23)

---

[4] So kompliziert die Funktion $\phi(y)$ auch aussieht, läßt sie sich vom mathematischen Standpkt her genauso wie eine Lorentz- oder Gaußfunktion durch genau drei Parameter beschreiben, nämlich Position, Breite und Höhe. Eine Umrechnung dieser Parameter in andere Einheiten entspricht nur einer Verschiebung, Stauchung oder Streckung.

$$R_v = R_y \, \frac{\Delta v_H}{y_H} = \frac{\sqrt{3}}{4} \, \pi \, \Delta v_H \quad .$$

Die bisher angeführten Größen und Funktionen beziehen sich alle auf die Reflexion an einem einzigen Kristall. Wie in Kapitel 2.3 bereits erwähnt, werden in der Doppelkristallanordnung zwei solcher Ewaldkurven miteinander gefaltet. Die tatsächlich meßbare Linienbreite $\Delta E / E_0$ wird um einen Faltungsfaktor f von

$$f = \frac{\pi}{2\pi - 4} \approx 1{,}376 \qquad\qquad (3.24)$$

verbreitert, die Linientiefe hingegen um diesen Faktor von 1 auf $\tau$ verkleinert :

$$\frac{\Delta E}{E_0} = f \cdot \frac{\Delta E_H}{E_0} \qquad\qquad (3.25)$$

$$\tau = \frac{1}{f} \qquad\qquad (3.26)$$

Diese Formeln geben die theoretischen Werte in Tabelle 6.1 zur Datenauswertung wieder.

# 4. Die Meßapparatur

## 4.1. Die Elektronik

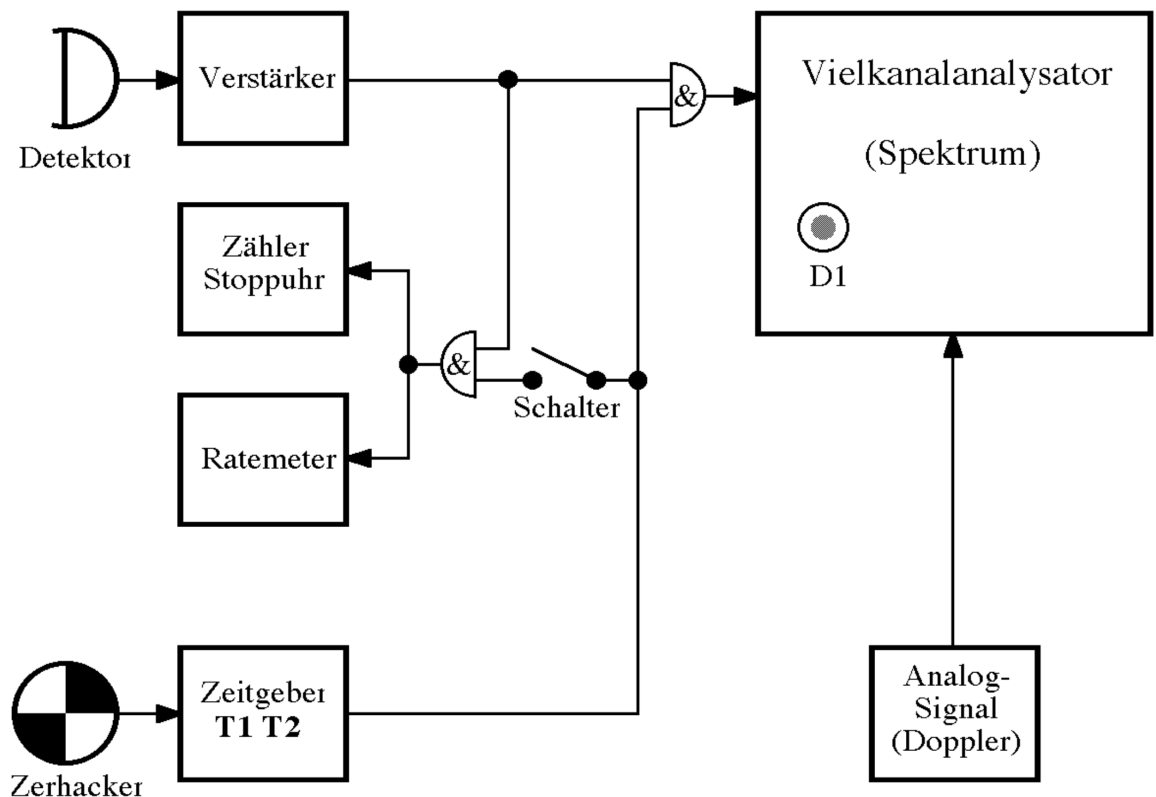

Abbildung 4.1:
Schema der elektronischen Bauteile.

Die Elektronik besteht im wesentlichen aus dem Detektorteil, dem Zeitgebermodul sowie einem Vielkanalanalysator (Abbildung 4.1).

Die Detektorelektronik dient vor allem zur digitalen Aufbereitung der Zählrohrimpulse. Zusätzlich sind noch ein einfacher Zähler mit gekoppelter Stoppuhr sowie ein Ratemeter integriert.

Der Zeitgeber stellt die Zeifenster zur Ein- und Ausblendung des Detekors für die Zähler zur Verfügung (Abbildung 4.2). Das vom Zerhacker gelieferte Startsignal löst eine Wartezeit $T_1$ aus, nach deren Ablauf für eine Öffnungszeit $T_2$ der Detektor eingeschaltet wird.

Der Vielkanalanalysator besteht aus zwei mit wahlweise 128 oder 256 Kanälen ausgestatteten Zählern. Dabei hängt die für beide Zähler paralle Kanalzuweisung vonder augenblicklich angelegten Spannung am Analogeingang dieses Gerätes ab. Diese Zuweisung kann durch eine einstellbare Zeit $D_1$ verzögert werden.

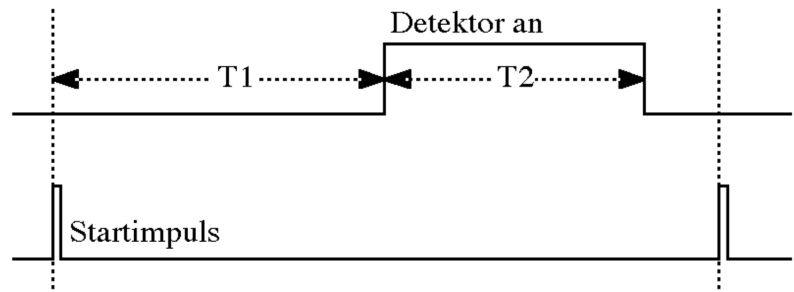

Abbildung 4.2:
Signalverlauf am Zeitgeber

Für die Dauer der Messungen stand ein Prototyp der IN10-Elek - tronik des ILL zur Verfügung. Diese wurde im Rahmen dieser Arbeit an die individuellen Bedürfnisse der Meßbank angepaßt und lauffähig gemacht.

In unserem Fall wird eine zur Dopplergeschwindigkeit proportionale Spannung an den Analogeingang gelegt. An den ersten Zählereingang wird das Detektorsignal, an den zweiten ein Monitorsignal zur Normierung gelegt. Die Verzögerung dient zur Elimination der Neutronenflugzeit vom Dopplerantrib zum Detektor.

Über einen VME-Bus sind die einzelnen Komponenten mit einem kleinen OS9-Rechner verbunden. Zeitgeber und Vielkanalanalysator können von ihm gesteuert und ausgelesen werden. Die Meßspektren werden über eine Terminalleitung automatisch auf das VAX-Verbundnetz im Haus übertragen, worauf mit Hilfe umfangreicher Bibliotheken Auswerte- und Graphikprogramme erstellt werden konnten.

## 4.2.    Der mechanische Aufbau

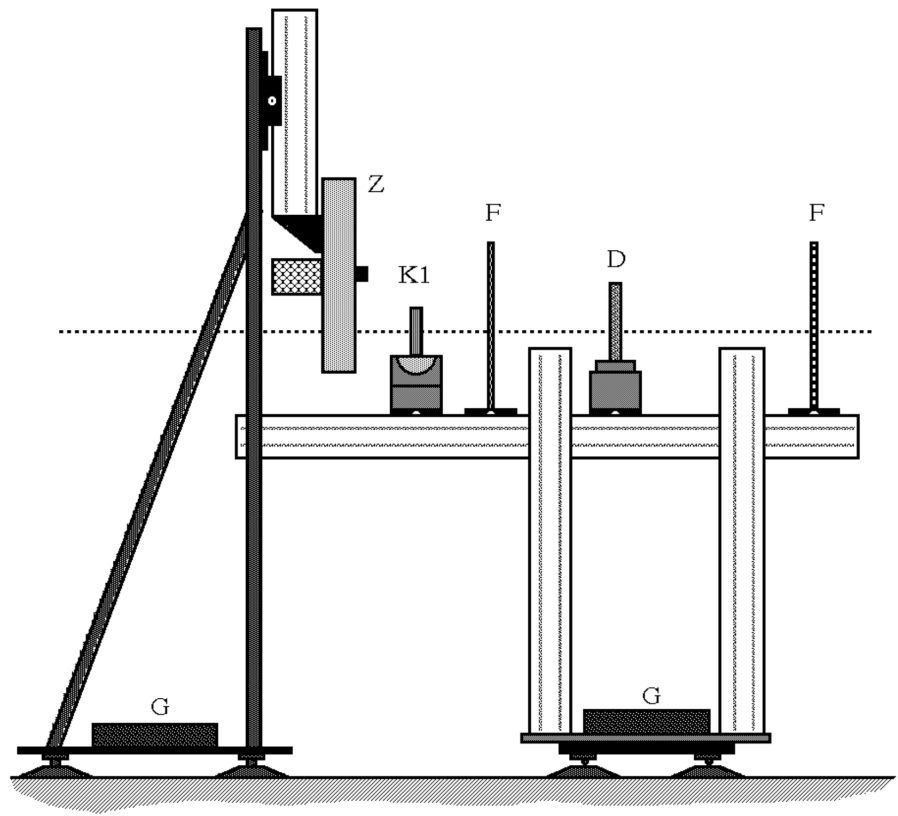

Abbildung 4.3:
Im rechten Teil ist die 2 m lange optische Bank dargestellt. Auf ihr stehen
der Detektor **D**, der Kristall **K₁**, als auch Hilfsmittel wie z.B. Fadenkreuze
**F**. Der linke, eigenständige Rahmen trägt den Zerhacker **Z** um Vibrationen
am Kristall zu vermeiden. Gewichte **G** beschweren die Anlage.

Die Messung soll nach dem Prinzip verlaufen, wie es in Kapitel
(2.2) erläutert wurde. Dazu wurde eine optische Bank aufgebaut
(Siehe Abbildung 4.3), die parallel zur Strahlachse einjustiert ist um
auf ihr die Goniometer mit der Kristallhalterung für $K_1$, den
Detektor sowie verschiedene Justiereinrichtungen zu installieren.
Um Vibrationen zu vermeiden wird für den Zerhacker eine
selbsttragende Vorrichtung gebaut, die über die optische Bank ge-
stellt wird. Beide Teile sind in der Höhe verstellbar um das Instru-
ment an verschiedenen Strahlrohren aufstellen zu können. Zur
einfachen Handhabung ist die ganze Apparatur auf Luftkis-
senfüßen aufgebaut. Dies bietet die Möglichkeit, das Instrument
mit Gewichten zu beschweren und damit gegen leichte Anstöße
während der Justierarbeiten unempfindlich zu machen. Zur opti-
schen Ausrichtung der Bank steht ein Laser bereit, der sowohl auf
als auch neben die Bank gestellt werden kann. Außerdem stehen ein Kadmiumfadenkreuz, eine Neutronenkamera, zwei feste, nur in der Strahlachse verstellbare optische Fadenkreuze für den Laser, eine regulierbare Blende sowie zwei Handgoniometer zur Verfügung. Das Probengoniometer ist mit Schrittmotoren ausgestattet.

## 4.2.1. Der Zerhacker

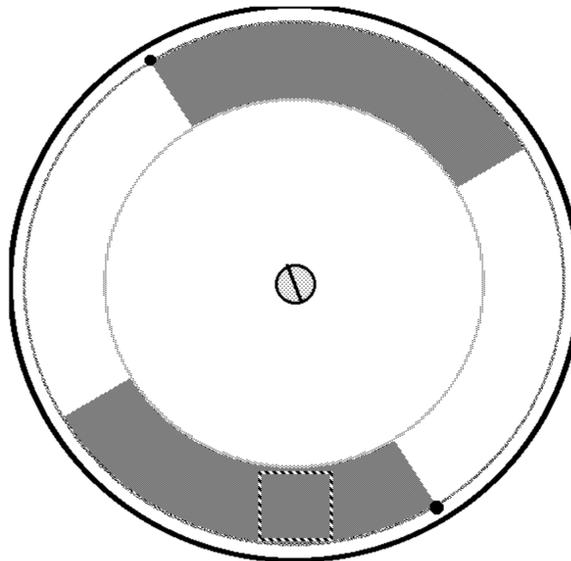

Abbildung 4.4:
Die grau ausgelegten Flächenstellen die Kadmiumbelegung auf der Zerhakkerscheibe dar. Der Querschnitt des Neutronenstrahls ist im unteren Teil angedeutet. Zwei magnetische Impulsaufnehmer, angedeutet durch zwei schwarze Punkte sorgen für die Synchronisation mit der Elektronik.

Der Zerhacker besteht im wesentlichen aus einer parallel zur Strahlachse rotierenden Aluminiumscheibe, deren zwei gegenüberliegende Viertelkreissegmente mit 1 mm dickem Kadmiumblech beklebt sind (Abbildung 4.4). Dabei wurde diese große Dicke verwendet, um auch für höhere Ordnungen der Wellenlänge ausreichende Absorption zu gewährleisten:
Sei $T(x,\lambda)$ das Transmissionsverhältnis der Wellenlänge $\lambda$ in der Tiefe x des Absorbers, dann gilt:

$$T(x,\lambda) = e^{-\sigma_A(\lambda)\,N\,x} \quad \text{mit} \quad N = \frac{N_A\,\rho}{A} \qquad (4.1)$$

Dabei bedeuten $\sigma_A$ den Absorptionsquerschnitt, N die Teilchendichte, $N_A$ die Avogadrokonstante, $\rho$ die Dichte und A die relative Atommasse des Absorbers. Mit den Werten für Kadmium

$$N_A = 6{,}0225 \cdot 10^{23} \, \frac{1}{\text{mol}}$$

$$\rho = 8{,}65 \, \frac{g}{cm^3}$$

$$A = 112{,}40 \, \frac{g}{\text{mol}}$$

erhalten wir für die Teilchendichte:

$$N = 4{,}63 \cdot 10^{22} \, \frac{1}{cm^3}$$

Die Absorptionsquerschnitte werden in der Literatur [84S] für $\lambda_0 = 1{,}80$ Å angegeben. Da sie in unserem Energiebereich linear mit der Wellenlänge gehen, sind sie bezüglich

$$\boxed{\sigma_A(\lambda) = \frac{\lambda}{\lambda_0} \, \sigma_A(\lambda_0)} \tag{4.2}$$

umzurechnen. Für Kadmium ist $\sigma_A(\lambda_0) = 2520$ barn. Für die Gal-liumarsenid-Rückstreuwellenlänge $\lambda = 5{,}65$Å und deren höhe ren Ordnungen erhalten wir die folgende Tabelle:

| | $\frac{\sigma(\lambda_0)}{\text{barn}}$ | $T_{0{,}5\,mm}$ | $T_{1\,mm}$ |
|---|---|---|---|
| $\lambda$ / 1 | 7910 | $1{,}12 \cdot 10^{-8}$ | $1{,}24 \cdot 10^{-16}$ |
| $\lambda$ / 2 | 3955 | $1{,}06 \cdot 10^{-4}$ | $1{,}12 \cdot 10^{-8}$ |
| $\lambda$ / 3 | 2637 | $2{,}23 \cdot 10^{-3}$ | $4{,}99 \cdot 10^{-6}$ |
| $\lambda$ / 4 | 1978 | $1{,}03 \cdot 10^{-2}$ | $1{,}06 \cdot 10^{-4}$ |
| $\lambda$ / 5 | 1582 | $2{,}57 \cdot 10^{-2}$ | $6{,}59 \cdot 10^{-4}$ |

Da ohne Berylliumfilter die 4. Ordnung im einfallenden Strahl noch erheblich vorhanden ist, wurden die Kreissegmente des Zerhackers mit 1 mm Kadmium belegt. Als Klebstoff dient ein Zweikomponenten-Kunstharz (Araldite®), das 10 min lang bei 180°C ausgehärtet wird und somit eine Endfestigkeit von 3000 N / cm$^2$ erreicht. Der Ze rhacker wurde ausgewuchtet und kann problemlos bis zu 9000 Umdrehungen pro Minute betrieben werden.

## 4.2.2.  Der Detektor

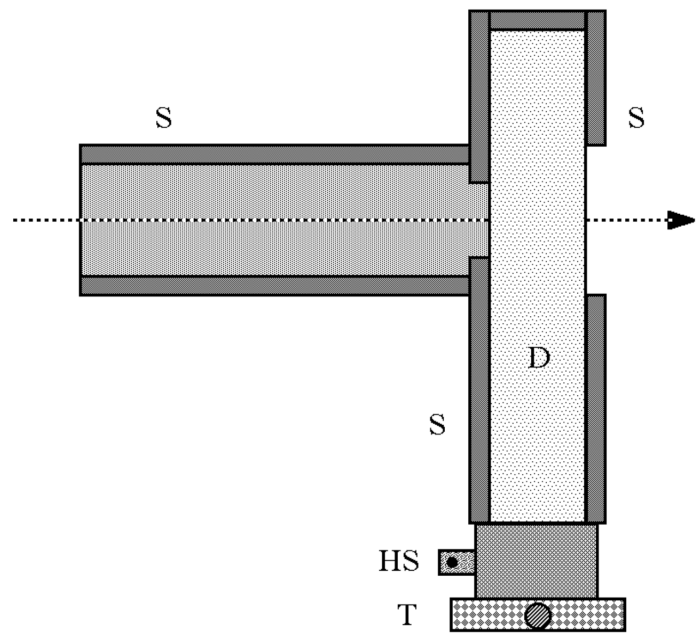

Abbildung 4.5:
Querschnitt des Detektors **D** im Maßstab 1:2. Um Untergrund zu vermeiden ist er von einer Abschirmung **S** umgeben. Die Größe der Ein- und Austritts­fenster sind ider Strahlgeometrie angepaßt. Des weiteren sind der Hochspan­nungsanschluß **HS** sowie ein Translationstisch **T** angedeutet.

Zum Nachweis der Neutronen wird ein mit $^3$He gefülltes Zählrohr verwendet. Die Reaktion $^3$He (n,p) $^3$T erzeugt aus den Neutronen geladene Teilchen, die durch Stoßionisation eine Ladungswolke und mittels der am Geiger-Müller-Zählrohr angelegten Hoch­spannung einen Stromimpuls erzeugen. Der Betrieb findet im elek­trischen Auslösebereich statt.

Das Besondere an dem hier verwendeten Detektor ist seine Halb­durchlässigkeit. Dafür hat das quaderförmige Detektorgehäuse je ein Neutronenfenster an zwei gegenüberliegenden Seiten. Die an­deren Seiten sind mit neutronenabsorbierendem Borkarbid gegen Untergrund abgeschirmt (Abbildung 4.5).

Um seine Transmission und Nachweiswahrscheinlichkeit über die ganze Detektorfläche zu ermitteln, wird dieser auf einen x-y-Translationsschlitten des Zweiachsendiffraktometers T13 gestellt und mit einem ca. 6 mm breiten Neutronenstrahl abgetastet. Die Betriebsspannung liegt im Bereich von 1900 V bis 2000 V. Die Meßergebnisse sind in Abbildung 4.6 dargestellt, woraus sich die empfindliche Fläche des Detektors als ziemlich homogen erweist. Seine Transmission bei einer Wellenlänge von $\lambda_D = 2{,}04$ Å ergibt sich aus

$$T_D(\lambda_D) = \frac{N_{13}}{N_D + N_{13}} \tag{4.3a}$$

zu

$$T_D(\lambda_D) = 0{,}71 \tag{4.3b}$$

und seine Nachweiswahrscheinlichkeit aus

$$W_D(\lambda_D) = \frac{N_D}{N_D + N_{13}} \tag{4.4a}$$

zu

$$W_D(\lambda_D) = 0{,}29 \ . \tag{4.4b}$$

Die Größen $N_D = 11500$ und $N_{13} = 27800$ sind die Zählraten des untersuchten, bzw des T13-Detektors.

Wir wollen Transmission und Absorption auf die uns interessierende Wellenlänge von $\lambda_0 = 5{,}65$ Å umrechnen. Dabei wird näherungsweise davon ausgegangen, daß Verluste, z.B. durch Streuung an den Detektorfenstern, unabhängig von der Wellenlänge sind. Da der Neutronennachweis auf Absorption beruht, ist die Transmission wieder durch ein Exponentialgesetz der Form (4.1) gegeben. Die Wellenlängenabhängigkeit steckt gemäß Gleichung (4.2) im Absorptionsquerschnitt der $^3$He-Kerne. Beide Gleichungen kombiniert ergeben

$$T_D(\lambda) = \left(T_D(\lambda_D)\right)^{\lambda/\lambda_D} \ . \tag{4.5}$$

Mit den Werten (4.3b) und (4.4b) ergibt dies

$$T_D(\lambda_0) = 0{,}48 \tag{4.6}$$

$$W_D(\lambda_0) = 0{,}52 \ . \tag{4.7}$$

Der hier untersuchte Detektor erfüllt also die gewünschten Anforderungen für Halbdurchlässigkeit und halbe Nachweiswahrscheinlichkeit.

Zur Messung in der Rückstreuanordnung wurden die ungenutzten Detektorflächen wieder mit Borkarbid abgeschirmt und dabei der Strahlgeometrie angepaßt (Siehe Abbildung 4.5).

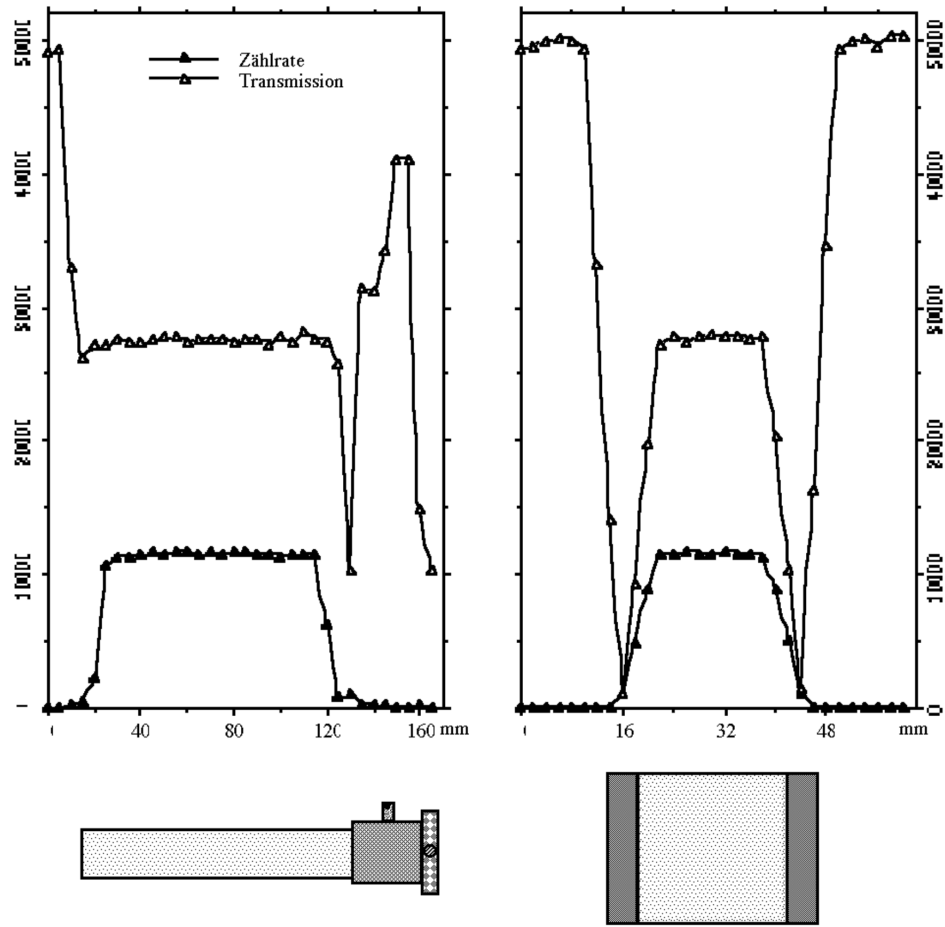

Abbildung 4.6:
Meßergebnisse beim Abtasten des Detektors mit einem Neutronenstrahl, links in Längs-, rechts in Querrichtung. Unter den Diagrammen ist jeweils im Maßstab der Abszisse der Detektor im Längs- bzw. Querschnitt gezeigt. Die Schrittweite zwischen zwei Meßpunkten beträgt links 5 mm, rechts 2 mm. Die tiefen Einschnitte in der Transmissionskurve in der Querrichtung rühren von der Borkarbidabschirmung.

## 4.3.    Strahlposition und Meßplatz

Die Testbank wurde für die Dauer der Messungen am Strahlrohr des sich im Aufbau befindenden Instrumentes IN10C in der neuen Leiterhalle des ILL installiert. Aus dem auf die vertikale kalte Quelle weisenden Neutronenleiter wird mit Hilfe eines Graphitreflektors ein geeignetes Wellenlängenband ausgeblendet und am

Ende eines sich verjüngenden, mit Superspiegeln beschichteten Neutronenleiters zur Verfügung gestellt. Optional können wir einen Berylliumfilter inden Strahlengang bringen, der alle Neutro- nen unterhalb seiner Abschneidewellenlänge von 3,6 Å unterdrückt.

## 4.3.1.  Der Berylliumfilter

Die Transmission des für das IN10C konstruierten Filters wurde zum ersten Mal im Rahmen dieser Arbeit vermessen und mit der Theorie verglichen. Dies soll hier kurz erläutert werden.

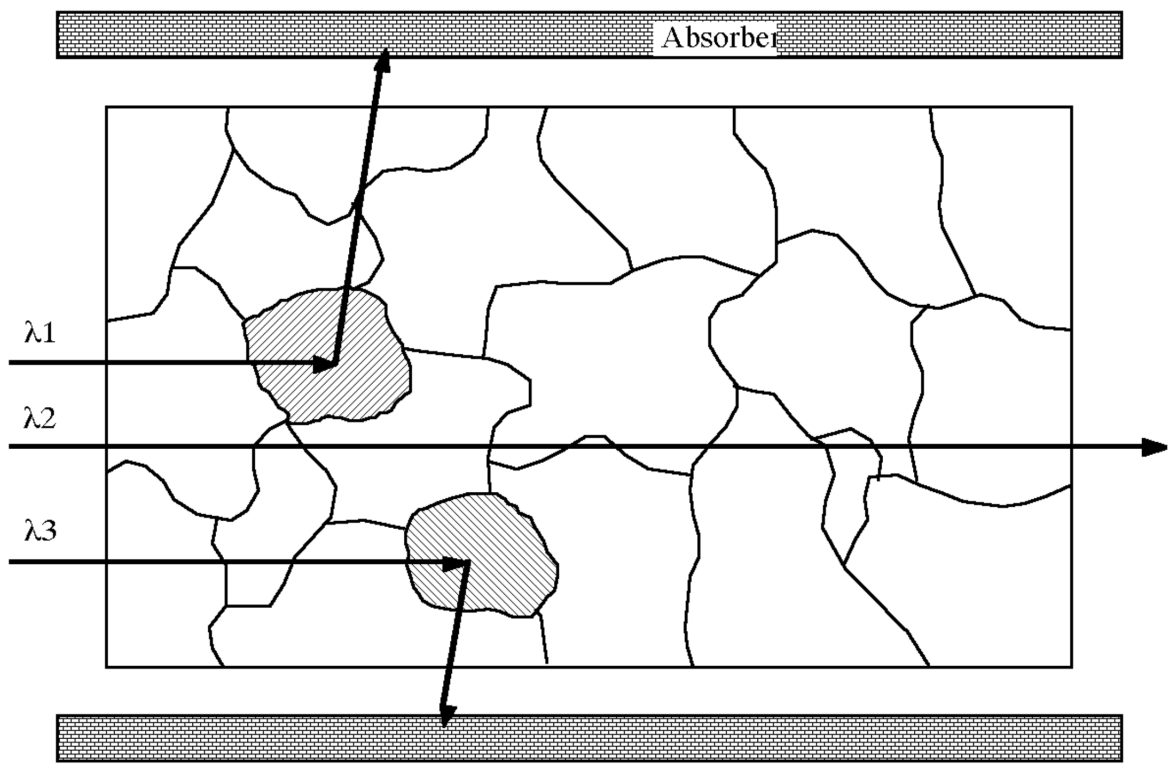

Abbildung 4.7:
Prinzip eines polykristallinen Berylliumfilters. Wellenlängen unterhalb einer Abschneidewellenlänge werdenan einer geeigneten Stelle Braggreflektiert, die anderen gehen unabgelenkt hindurch.

Beryllium wird als polykristallinerFilter verwendet. Solche Subs- tanzen nutzen die statistische Verteilung der Kristallite, indem sich jede einfallende Wellenlänge $\lambda$ ein richtig orientiertes Kriställchen zur Braggreflexion heraussucht (Abbildung 4.7). Das Neutron än- dert so seine Richtung und trifft auf einen den Filter umgebenden Absorber, hier [6]Li und Gd. Oberhalb einer Abschneidewellenlänge $\lambda_A$ gibt es keinen Braggreflex mehr. Bei Raumtemperatur steigt an dieser Kante der totale Wirkungsquerschnitt, der alle Arten von Streuung und Absorption umfaßt, zu kleineren Wellenlängen hin sprunghaft um eine Größenordnung an (Abbildung 4.8). Kühlen wir den Filter auf die Temperatur des flüssigen Stickstoffs, so sind dies aufgrund der wesentlich kleineren Phononenstreuung sogar zwei Zehnerpotenzen. Man beachte, daß der Wirkungsquerschnitt exponentiell in die Transmissiondes Filters eingeht; wir haben also eine erhebliche Unterdrückung unterhalb der Abschneidewellenlänge.

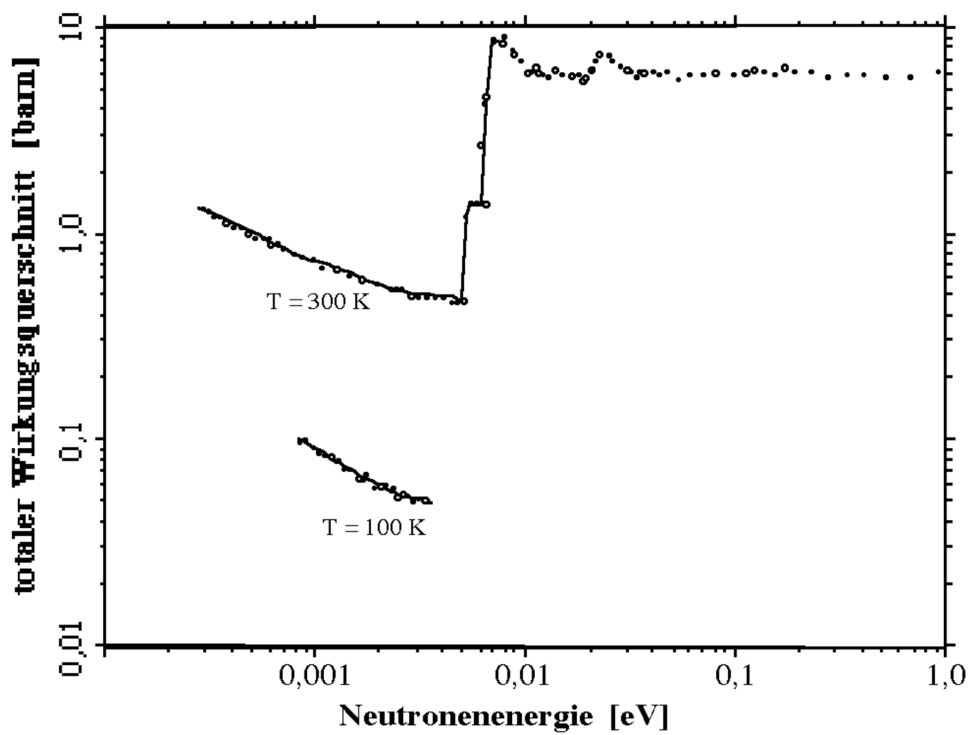

Abbildung 4.8:
Totaler Wirkungsquerschnitt für polykristallines Beryllium beiTemperaturen von 100K und 300K. Der Einsatz von Braggreflexionen bei einer Neutronenenergie von ca 50 meV führt zu einem sprunghaften Anstieg des Wirkungquerschnitts. (Quelle: [55H].

Der 12 cm lange Berylliumklotz ist auf Eingangs- und Ausgangsseite mit je einem 0,1 cm dickenVakuumfenster aus Aluminium umgeben, das mit in die Rechnung einbezogen werden soll: Die Transmission $T_S(x,\lambda,T)$ der Wellenlänge $\lambda$ durch die Substanz S der Dicke x und bei der Temperatur T ist analog zu Formel (4.1) durch

$$T_S(x,\lambda,T) = e^{-\sigma_{tot}^S(T) \cdot N_S \cdot x} \quad \text{mit} \quad N_S = \frac{N_A \, \rho_S}{A_S}$$

(4.8)

gegeben. Für die Wellenlänge $\lambda = 6{,}4$ Å und den Größen

$x_{Al} = 0{,}2$ cm $\qquad\qquad x_{Be} = 12$ cm

$\rho_{Al} = 2{,}7 \ \dfrac{g}{cm^3}$ $\qquad\qquad \rho_{Be} = 1{,}8477 \ \dfrac{g}{cm^3}$

$A_{Al} = 26{,}982 \ \dfrac{g}{mol}$ $\qquad\qquad A_{Be} = 9{,}0122 \ \dfrac{g}{mol}$

$\sigma_{Al}(300K) = 1{,}05 \cdot 10^{-24}$ cm$^2$ $\qquad \sigma_{Be}(300K) = 0{,}55 \cdot 10^{-24}$ cm$^2$

$\sigma_{Al}(100K) = 0{,}88 \cdot 10^{-24}$ cm$^2$ $\qquad \sigma_{Be}(100K) = 0{,}06 \cdot 10^{-24}$ cm$^2$

erhalten wir

$T_{Al}(300K) = 0{,}99$ $\qquad\qquad T_{Be}(300K) = 0{,}44$

$T_{Al}(100K) = 0{,}99$ $\qquad\qquad T_{Be}(100K) = 0{,}91$

Die Gesamttransmission $T_{tot}$ für den warmen und den kalten Zustand ergibt sich aus der Multiplikation

$$T_{tot}(T) = T_{Al}(T) \cdot T_{Be}(T) \qquad\qquad (4.9)$$

also

$T_{tot}(300K) = 0{,}44$

$T_{tot}(100K) = 0{,}90 \qquad\qquad\qquad\qquad (4.10)$

Die gerechneten Endergebnisse stimmen gut mit der in den Abbildungen 4.10 und 4.11 wiedergegebenen Messungen überein, vor allem wenn man berücksichtigt, daß die Temperaturen von 300 K bzw. 100 K, für die die Wirkungsquerschnitte zur Verfügung stehen, nur angenähert gelten.

Zur Trennung der verschiedenen Ordnungen wurde eine Flugzeitmessung gemacht (Abbildung 4.9). Ein Zerhacker mit sehr kurzer Öffnungszeit triggert eine Sägezahnspannung, die an dem Analogeingang unseres Vielkanalanalysators angelegt wird. Somit ist die Nummer des augenblicklich angesprochenen Kanals zur Registrierung eines eventuell ankommenden Neutrons proportional zu seiner Flugzeit zwischen Zerhacker und Detektor.

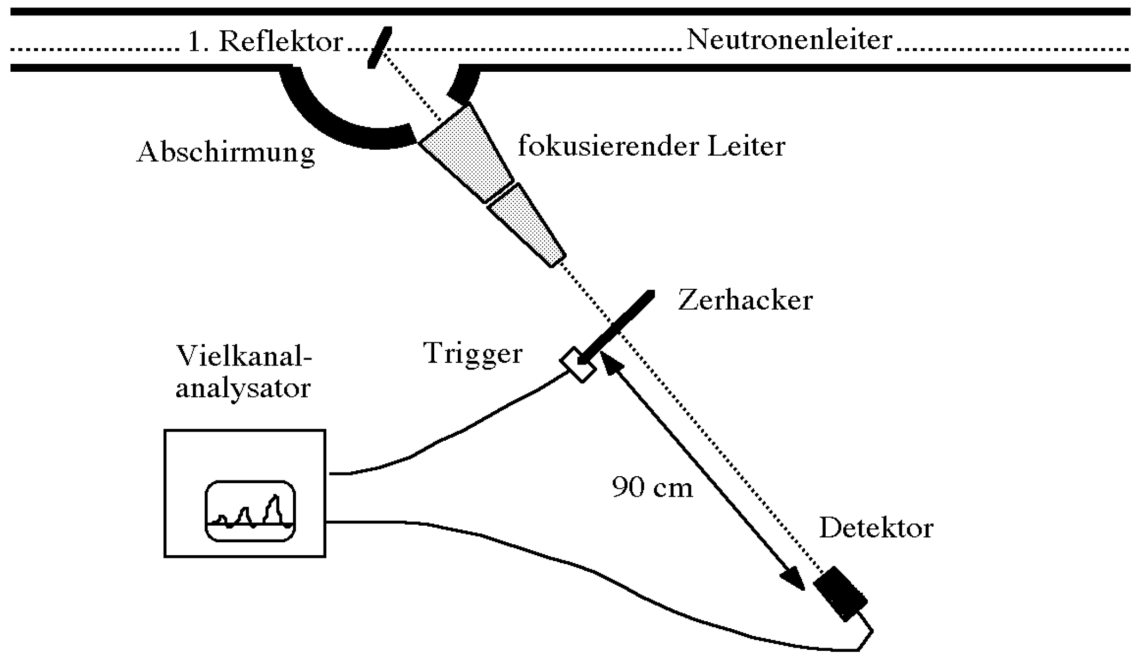

Abbildung 4.9:
Anordnung zur Flugzeitmessung (Quelle [88G]).

In den nachfolgenden Tabellen und Graphen sind die Ergebnisse Zusammengestellt.

Tabelle 4.1:  Zählrate in Abhängigkeit der Ordnung von $\lambda$ = 6,30 Å ohne Filter, mit Filter warm und kalt.

| Ordnung | ohne Filter n/min | Filter warm n/min | Filter kalt n/min |
|---|---|---|---|
| 1 | 12835 | 8144 | 11088 |
| 2 | 7727 | 0,57 | 0,73 |
| 3 | 1715 | 1,26 | 1,29 |
| 4 | 196 | 0,29 | 0,21 |
| 5 | 26 | 0,56 | % |

Tabelle 4.2:   Transmission des Filters warm und kalt in Abhängigkeit derOrd-
nung von $\lambda$ = 6,30 Å.

| Ordnung | Filter warm $10^{-4}$ | Filter kalt $10^{-4}$ |
|---------|------------|-----------|
| 1 | 6300 | 8600 |
| 2 | 0,74 | 0,94 |
| 3 | 7,3 | 7,5 |
| 4 | 14 | 11 |
| 5 | 220 | % |

Tabelle 4.3:   Intensitätsverhältnisse der höheren Ordnungen zur 1. Ordnung
jeweils ohne Filter, mit Filter warm und kalt.

| Ordnung | ohne Filter | Filter warm | Filter kalt |
|---------|-------------|-------------|-------------|
| 1 | 1 | 1 | 1 |
| 2 | 0,602 | 0,00007 | 0,00007 |
| 3 | 0,134 | 0,00015 | 0,00012 |
| 4 | 0,015 | 0,00004 | 0,00002 |
| 5 | 0,002 | 0,00007 | % |

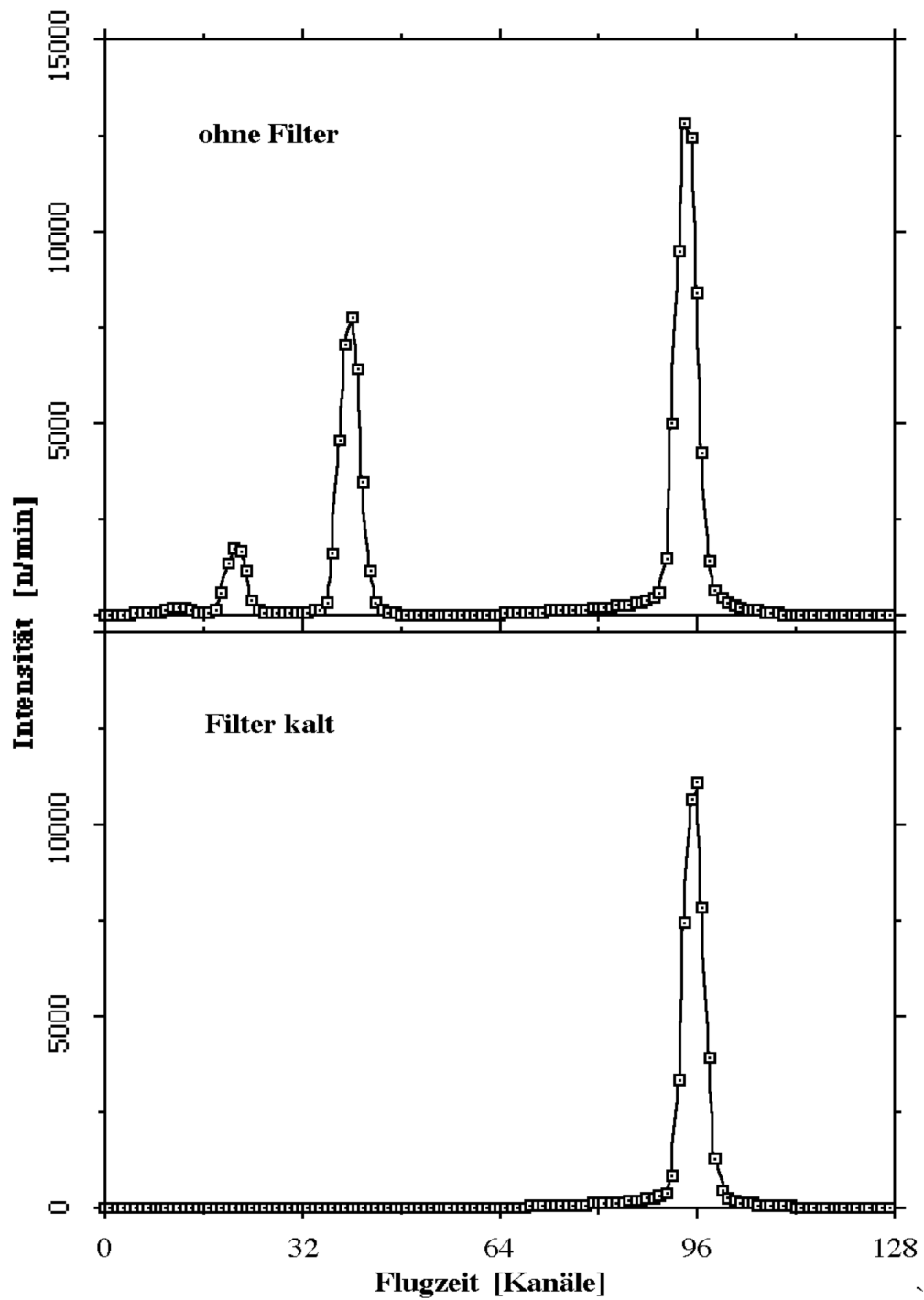

Abbildung 4.10:

Flugzeitspektren der Transmission am Berylliumfilter. Die Intensität in Neutronen pro Minute ist gegen die Flugzeit aufgetragen. Das erste Diagramm zeigt die Intensitätsverteilung ohne Filter. Das rechte Maximum rührt von $\lambda_0$ her, während nach links zu der Reihe nach die höheren Ordnungen bis zur fünften auftreten. Die numerischen Werte der Maxima sind in der Tabelle 4.1 angegeben. Im zweiten Diagramm ist die Intensitätsverteilung mit abgekühltem Berylliumfilter wiedergegeben. Es läßt sich nur noch die erste Ordnung erkennen.

Der Abstand zwischen Zerhacker und Detektor beträgt 90 cm, die Zerhackerfrequenz 317 Hz, die Öffnungszeit für ein Neutronenpaket 18 µs.

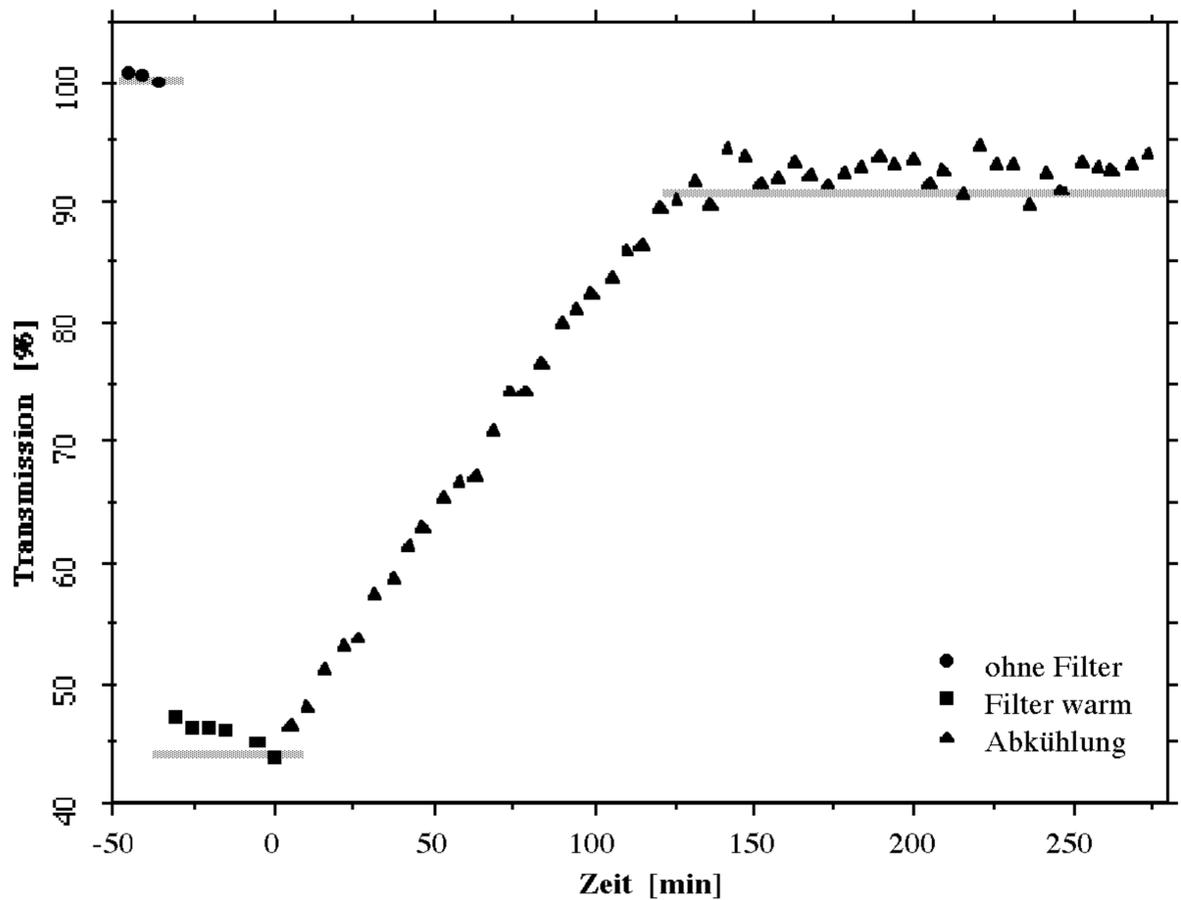

Abbildubg 4.11:
Abkühlung des Berylliumfilters. Die Transmission der 1. Ordnung ist gegen
die Zeit aufgetragen. Bei t = 0 wird flüssiger Stickstoff in den Voratsbehäl-
ter gefüllt und jede Viertelstunde ein neuses Spektrum gemessen. Als Refe-
renz dient der dritte Meßwert ohne Filter. Die horizontalen Linien deuten die
theoretischen Werte an.

# 5. Vorbereitung der Messungen

Aufgrund der Neuinstallation und der damit mangelnden Erfahrung am IN10c muß die Wellenlänge bei vorgegebener Reflektorstellung gemessen werden. Anschließend wird der Reflektor auf die gewünschte Wellenlänge eingestellt und diese durch eine neue Messung überprüft. Ist sie ausreichend justiert, so können wir die optische Bank und den Dopplerantrieb ausrichten. Bevor wir nun zur Justierung der Kristalle kommen müssen die Zeitfenster für den Detektor abgetastet und gesetzt werden.

## 5.1. Messung der Wellenlänge

### 5.1.1. Methode und Meßdaten

Die Wellenlänge wird durch Messung des Winkels bei Braggreflexion an einem bekannten Kristall bestimmt (siehe Abbildung 5.1). Dieser wird auf einem Rotationstisch so eingestellt, daß der Braggreflex in dem großflächigen Detektor $D_1$ sein Maximum erreicht. Die Winkelstellung des Rotationstisches wird abgelesen. Dann drehen wir den Kristall so, daß der Reflex symmetrisch zur Achse des einfallenden Strahls in den Detektor $D_2$ gelangt. Die Differenz der beiden Winkelstellungen gibt uns den Drehwinkel $\alpha$, der über die Beziehung

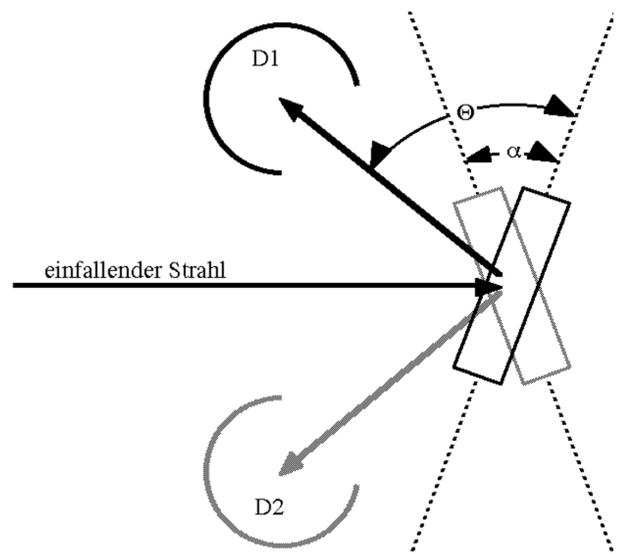

Abbildung 5.1:
Messung des Braggwinkels $\Theta$ zur Bestimmung der Wellenlänge. Erläuterungen im Text.

$$\alpha = 180° - 2\,\Theta \qquad (5.1)$$

mit dem Braggwinkel $\Theta$ verknüpft ist. Damit lautet das Braggesetz in $\alpha$ ausgedrückt

$$\lambda = 2\,\mathrm{d}\cdot\cos\!\left(\frac{\alpha}{2}\right).$$ (5.2)

Um den Drehwinkel $\alpha$ genau zu ermitteln wird bei den beiden Maximalpositionen je eine Winkelverteilung aufgenommen.

Die Messungen wurden zunächst mit dem [002]-Reflex eines Graphitkristalls mit 30' Mosaizität durchgeführt (Abbildung 5.2). Diese führt zu einer relativ ungenauen Bestimmung von $\alpha$, wozu die Messung mit dem [111]-Reflex eines perfekten Germaniumkristalls wiederholt wurde. Die beiden Reflexe weisen jetzt eine deutliche Winkelverteilung auf (Abbildung 5.3): Im Zentrum des linken Reflexes befindet sich ein schmales Hauptmaximum (1). Im Winkelabstand von $\gamma_l = 1{,}5°$ wird es auf jeder Seite von einem Nebenmaximum (2) und (3) umgeben. Der rechte Reflex besteht aus zwei schmalen Maxima (2') und (3') im Abstand von $2\gamma_r = 2{,}4°$. In der Mitte zwischen ihnen sitzt ein sehr breites, flaches Maximum (1').

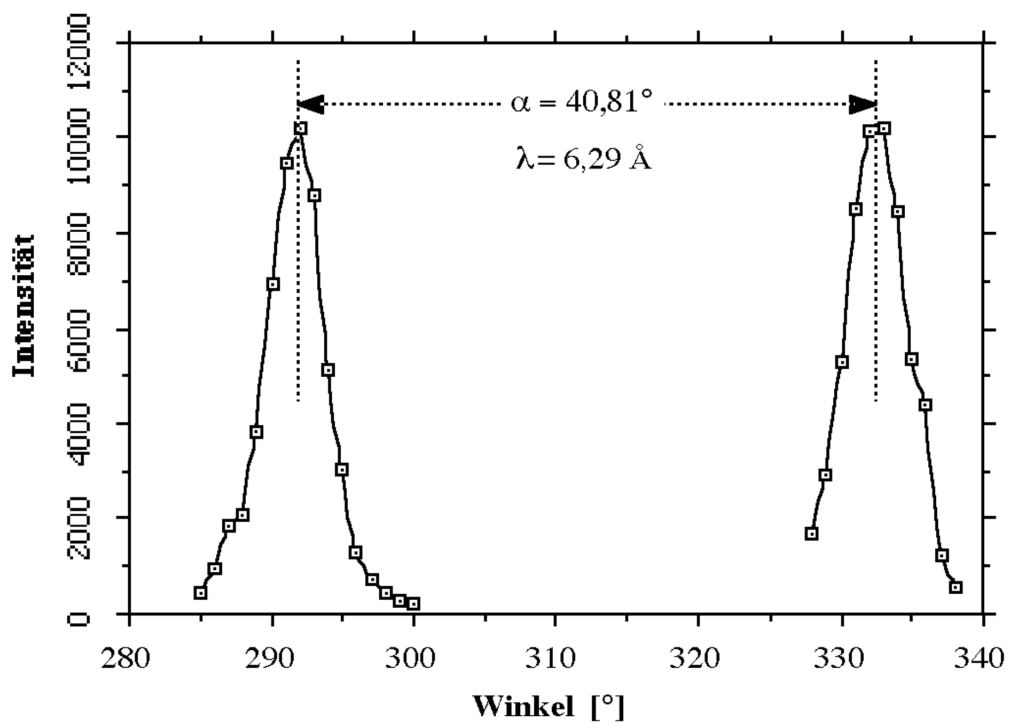

Abbildung 5.2:
Messung der Wellenlänge mit dem Graphit [002]-Reflex (d = 3,354 Å).

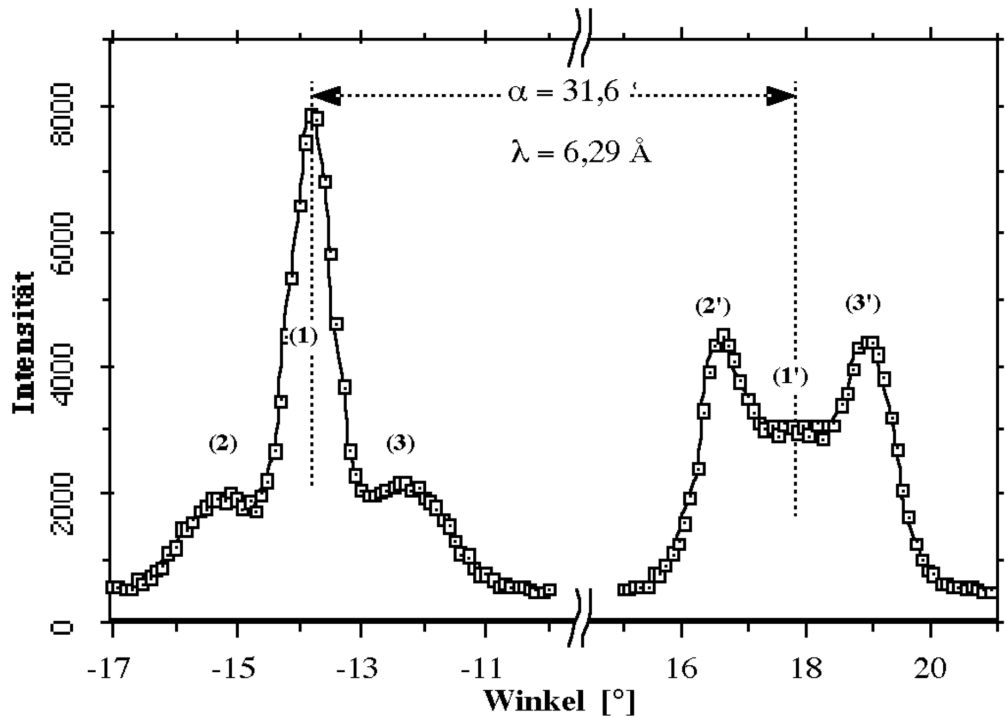

Abbildung 5.3:
Messung derselben Wellenlänge wie in Abbildung 5.2 mit dem Ge [111] - Reflex (d = 3,26659 Å). Es ist sehr gut eine Überstruktur der Winkelverteilung erkennbar, die von der speziellen Optik des IN10C-Strahlrohrs herrührt. Die eingestellte Wellenlänge ergibt sich zu 6,29 Å.

## 5.1.2. Interpretation der Winkelverteilung

## 5.1.2.1. Fokusierende und defokusierende Doppelkristallanordnungen

Bei divergenten Strahlen muß je nach Geometrie der Anordnung bei aufeinanderfolgender Reflexion an zwei Kristallen in zwei Fälle unterschieden werden:

Die fokusierende Anordnung besitzt Z-förmige Geometrie (Abbildung 5.4). Betrachten wir zwei repräsentative Teilstrahlen des weißen, divergenten, einfallenden Strahls. Der eine trifft unter einem Winkel $\Theta_1$, der andere unter $\Theta_2$ auf den ersten Kristall. Durch Braggreflexion werden unterschiedlich Wellenlängen $\lambda_1$ bzw. $\lambda_2$ selektiert. Steht der zweite Kristall parallel zum ersten, so treffen beide Wellenlängen $\lambda_1$ und $\lambda_2$ unter den zugehörigen Braggwinkeln $\Theta_1$ bzw. $\Theta_2$ auf die Gitterebenen und werden beide in Richtung des Detektors reflektiert. Bei Verdrehung des Kristalls aus der Parallelstellung heraus ist die Braggbedingung für keinen der einfallenden Strahlen mehr erfüllt. Wir verlieren daher sehr schnell an Intensität. Letztere gegen den Drehwinkel aufgetragen liefert uns ein schmales, hohes Maximum. Die so gewonnene Kurve gibt Aufschluß über Kristalleigenschaften, wie z.B. Mosaizität.

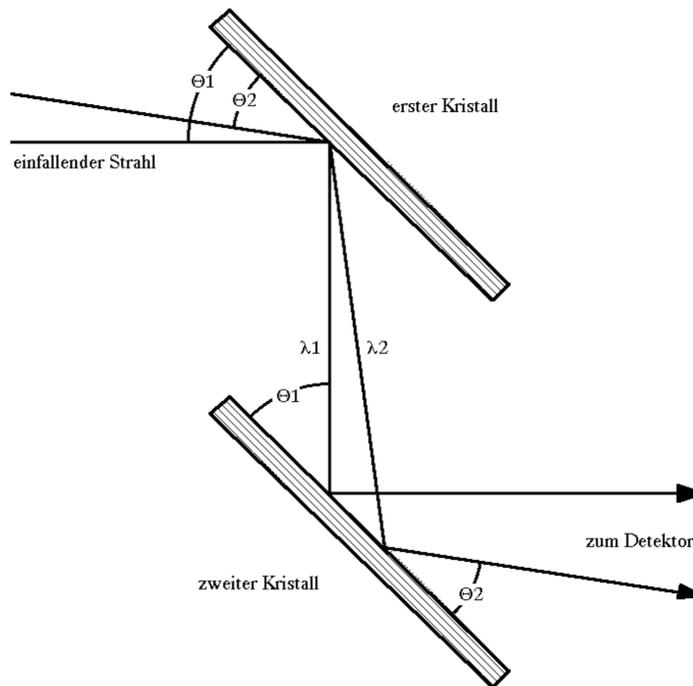

Abbildung 5.4:
Fokusierende Geometrie der Doppelkristallanordnung. Stehen beide Kristalle parallel zueinander, so gelangen Teilstrahlen aus dem ganzen Divergenzbereich des einfallenden Strahls in den Detektor.

Die defokusierende Anordnung ist in U-förmiger Geometrie gebaut (Abbildung 5.5). Wie in der fokusierenden Anordnung werden am ersten Kristall, durch die verschiedenen Einfallswinkel $\Theta_1$ und $\Theta_2$ bedingt, unterschiedliche Wellenlängen $\lambda_1$ und $\lambda_2$ selektiert. Stellen wir den zweiten Kristall so, daß z.B. für $\lambda_1$ die Braggbedingung erfüllt ist, und diese Wellenlänge in Richtung Detektor reflektiert wird. Dann ist im Gegensatz zur fokusierenden Anordnung für $\lambda_2$ der Einfallswinkel $\Phi$ auf die Gitterebenen verschieden von $\Theta_2$. Wir erhalten also keine Reflexion für diese Wellenlänge. Der Kristall muß verdreht werden, damit $\lambda_2$ auf Kosten von $\lambda_1$ die Braggbedingung erfüllen kann. Wir erhalten somit ein breites Maximum, wenn wir die Intensität wieder gegen den Drehwinkel auftragen. Diese Geo-metrie eignet sich, die Eigenschaften des Strahls, z.B. die Diver-
genz zu untersuchen.

Die integrierte Reflektivität der so aufgenommenen Kurven beider
Anordnungen ist gleich. Deswegen erhalten wir im fokusierenden
Fall ein scharfes, hohes Maximum, während es in der defokusie-
renden Geometrie breit und flach wird.

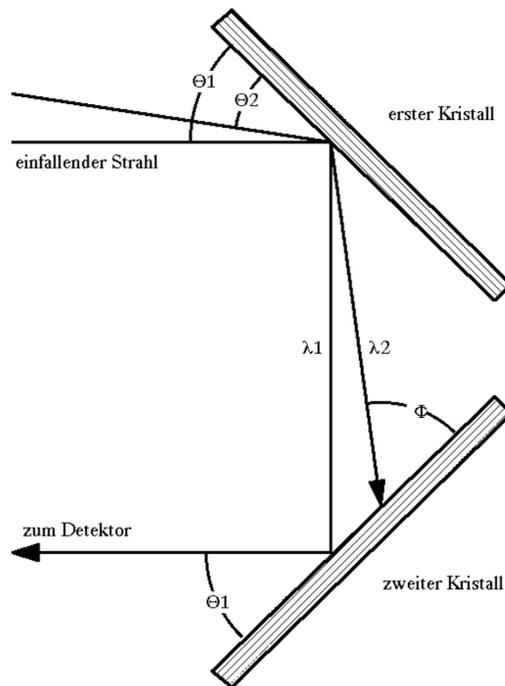

Abbildung 5.5:
Defokusierende Geometrie der Doppelkristallanordnung. Der zweite Kristall
ist so eingestellt, daß $\lambda_1$ durch Braggreflexion in den Detektor gelangt. Für
die andere Wellenlänge $\lambda_2$ ist die Braggbedingung nicht mehr erfüllt.

## 5.1.2.2. Erklärung der gemessenen Winkel-verteilung

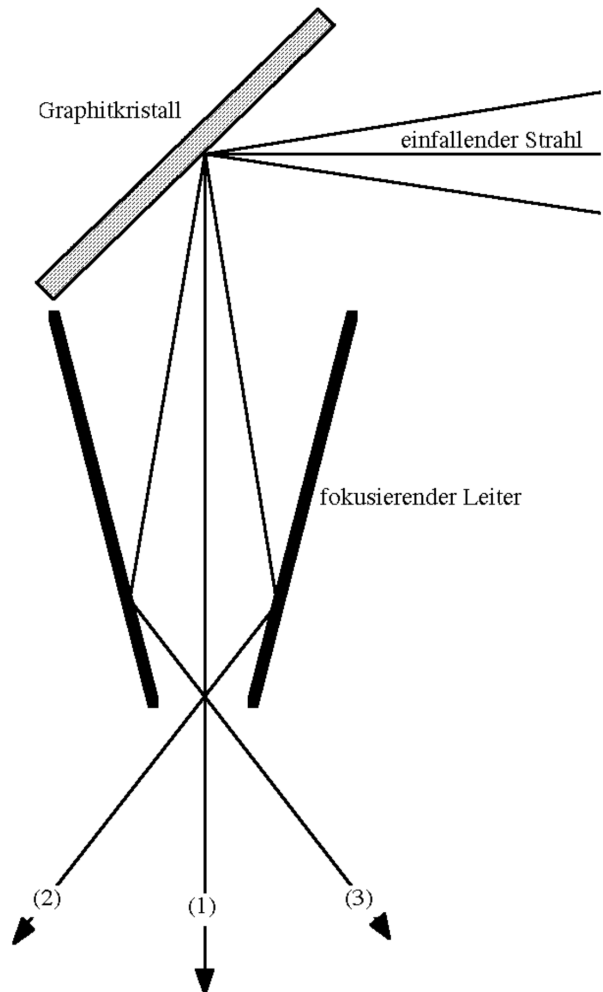

Abbildung 5.6:
Strahlengang durch den fokusierenden Neutronenleiter. Der Strahl(1) geht ungehindert durch den Leiter, während (2) und (3) je einmal reflektiert werden.

Die gemessenen Spektren (Abbildungen 5.3 und 5.8) ergeben sich aus der speziellen Optik des zur Verfügung stehenden Strahlrohrs. Am IN10C befindet sich hinter dem Reflektor $R_1$ ein sich verjüngender Neutronenleiter (Abbildung 5.6). Dieser soll den Strahl in der Probenregion fokusieren. Einzelheiten dazu sind in der Diplomarbeit von P. Goppelt [88G] beschrieben.

In Bezug auf Abbildung 5.3 können wir das mittlere Maximum (1) bzw. (1') dem direkten Strahl durch den fokusierenden Leiter zuordnen. Im linken Reflex (1) ist er in fokusierender Anordnung gemessen, also schmal und hoch, während er im rechten Reflex (1') defokusiert, also breit und flach ist. Die Nebenmaxima werden den einmal an den Wänden des Neutronenleiters reflektierten Strahl-gängen zugeschrieben, wie in Abbildung 5.6 angedeutet. Die Re-flexion vertauscht dabei die fokusierende mit der defokusierenden Kristallanordnung. Deswegen werden die Nebenmaxima (1) und (3) im linken Reflex breit und flach, während im rechten Reflex (1') und (3') schmal und hoch werden.

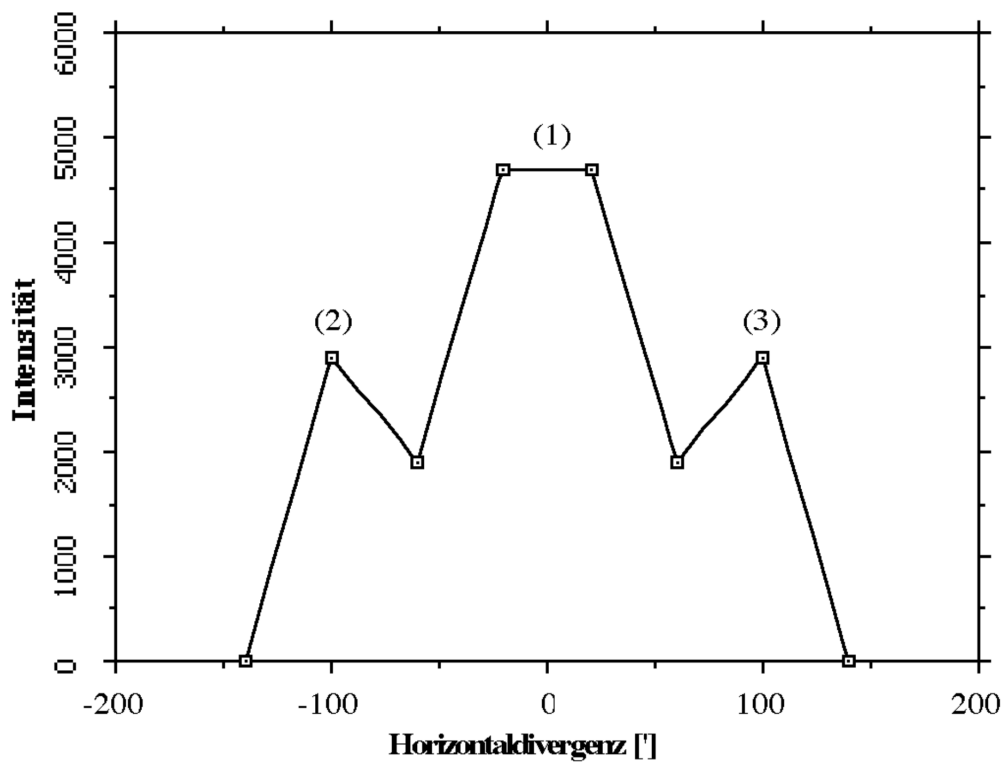

Abbildung 5.7:
Berechnete horizontaleStrahldivergenz beim Durchgang durch den fokusie renden Neutronenleiter (Quelle [00M]).

Die gemessene Winkelverteilung stimmt gut mit der von A. Magerl [00M] berechneten überein (Abbildung 5.7). Die folgende Tabelle erhält den gerechneten sowie die gemessenen Werte zwischen den Strahlen (1) und (2).

| berechnet | Reflex links | Reflex rechts |
|-----------|--------------|---------------|
| 1,7°      | 1,5°         | 1,2°          |

Der kleine Unterschied zwischen dem linken und rechten Reflex ist auf einen kleinen Wellenlängenunterschied der verschiedenen Strahlen zurückzuführen.

Die nachfolgende Tabelle erhält, gemäß der Gleichung (5.2) ausgewertet, die Wellenlängen der verschiedenen Maxima. Nachdem die erste Messung von Abbildung 5.3 ausgewertet war, konnte die Wellenlänge auf den gewünschten Wert für die Galliumarsenidmessungen eingestellt werden. Abbildung 5.8x zeigt eine Messung nach der Umstellug. Dieses Ergebnis ist ebenfalls in der folgenden Tabelle aufgelistet.

|  | Strahl | $\alpha$ [°] | $\lambda$ [Å] |
|---|---|---|---|
|  | 2 | 31,9 | 6,28 |
| vorher | **1** | **31,6** | **6,29** |
|  | 3 | 31,3 | 6,29 |
|  | 2 | 58,8 | 5,69 |
| nacher | **1** | **58,9** | **5,69** |
|  | 3 | 59,0 | 5,69 |

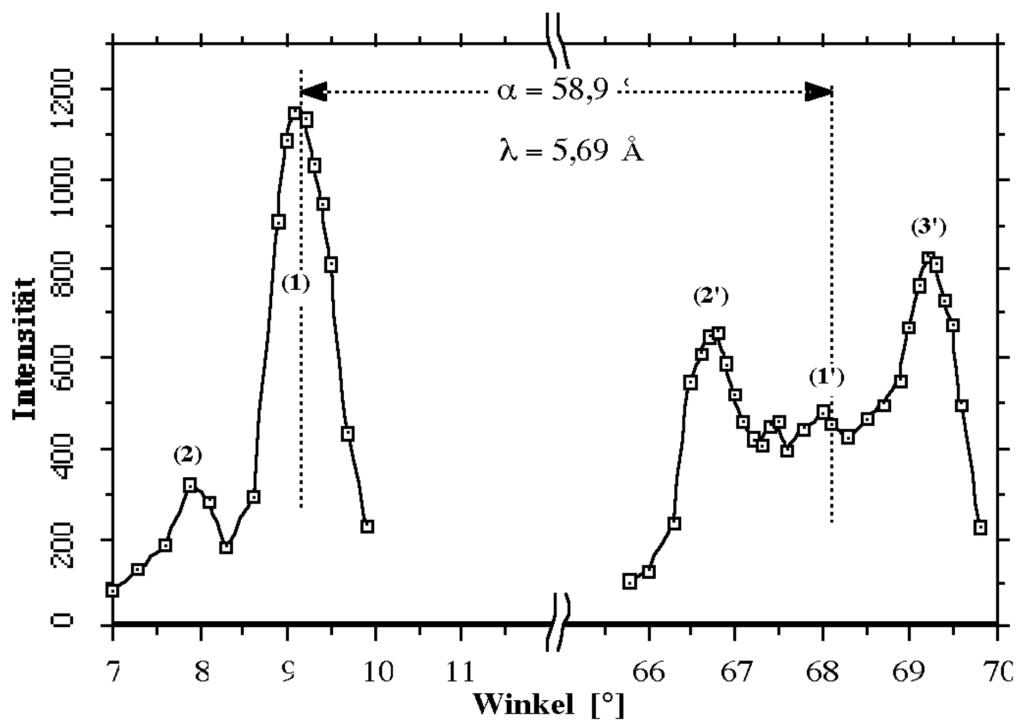

Abbildung 5.8:
Messung der Wellenlänge mittels Ge [111], nachdem sie auf die Rückstreu­wellenlänge an GaAs [200] eingestellt wurde. Qualitativ haben wir wieder die gleiche Form wie in der Abbildung 5.3. Die eingestellte Wellenlänge be­trägt 5,69 Å.

# 5.2.  Ausrichtung der optischen Bank

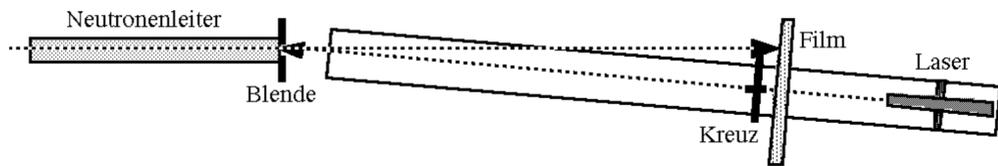

Abbildung 5.9:
Verfahren zur Einjustierung der Bank paralell zur Strahlachse.Erläuterungen im Text.

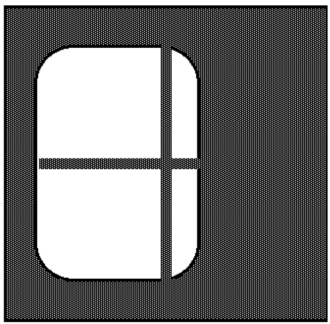

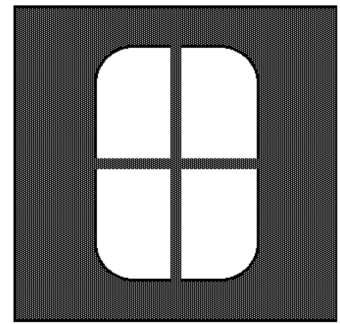

Ein paralell zur optischen Bank justierter Laserstrahl wird mit dieser so gedreht, daß er auf die Austrittsblende des Neutronenleiters trifft. Mit einem neutronenempfindlichen Filmwird der Schatten eines im Laserstrahl zentrierten Kadmium kreuzes aufgenommen

Abbildung 5.10:
Belichtung des Films bei dejustierter Bank.

Abbildung 5.11:
Belichtung des Films bei justierter Bank.

(Abbildungen 5.10 und 5.11). Die Bank wird dann so lange um das Austrittsfenster gedreht, bis der Schatten auf dem Bild im Zentrum des Neutronenstrahls sitzt.

# 5.3.  Zeitfenstereinstellung

Im Idealfall soll die Frequenz des im Kapitel 2.2 beschriebenen Zerhackers so eingestellt sein, daß die räumliche Länge des Neutronenpulses gerade dem doppelten Abstand D zwischen Zerhacker und Detektor entspricht. In diesem Fall treffen die ersten, von $K_2$ reflektierten Neutronen gerade dann auf den Detektor, wenn die letzten desselben Pakets ihn gerade auf dem Hinflug verlassen haben. Der zeitliche Verlauf ist in Abbildung 5.12 wiedergegeben. Bei bekannter Neutronengeschwindigkeit $v_0$ läßt sich diese örtliche Paketlänge in eine zeitliche umrechnen, aus der wir die Zerhackerfrequenz $v_z$ erhalten. Beachten wir, daß die Öffnungszeit genauso lange wie seine Verschlußzeit ist, so erhalten wir

$$v_z = \frac{v_0}{4\,D} \qquad . \qquad (5.3)$$

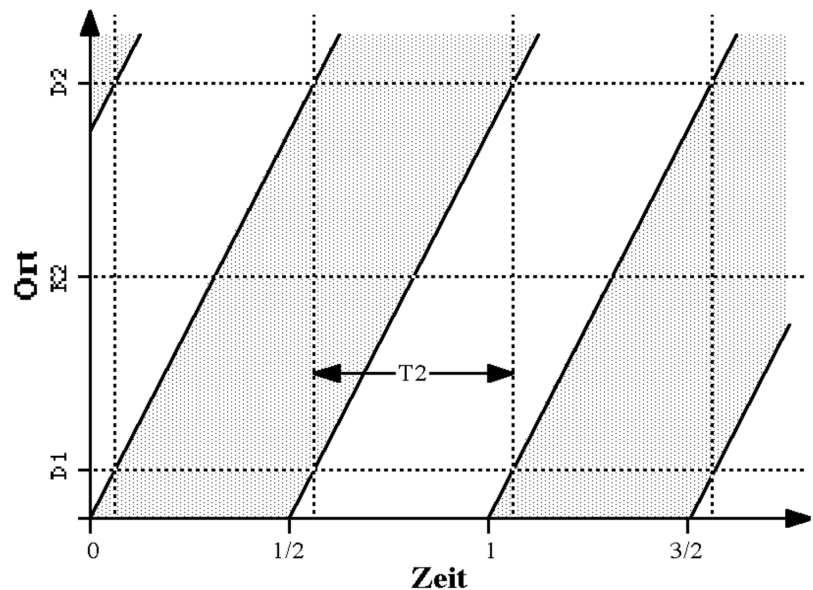

Abbildung 5.12:
Anpassung der räumlichen Impulslänge an den Abstand Detektor - Kristall.
Die Neutronen passieren den Detektor $D_1$ werden an $K_2$ reflektiert um in
den Detektor $D_2$ zu fallen. Während der Zeit $T_2$ fallen nur die reflektierten
Neutronen in den Detektor.

Nachdem die Zerhackerfrequenz abgestimmt ist, müssen wir das
Zeitfenster setzen, während dessen die Neutronen gezählt
werden sollen. Die Wartezeit $T_1$ hängt im wesentlichen von der
Phasenlage des Signalaufnehmers sowie vom Abstand zwischen
Zerhackers und Detektor ab. Die Öffnungszeit $T_2$ des Zeitfensters
soll so eingestellt werden, daß wir keine primär einfallenden Neu-
tronen erfaßt werden. Der Zeitverlauf der Intensität wird experi-
mentell bestimmt: Dazu schließen wir das zum Analysatorkristall $K_2$
gerichtete Austrittsfenster des Detektors um nur die einfallenden
Neutronen zu erfassen. Die Öffnungszeit $T_2$ dient uns hier als
Auflösungsfunktion und wird daher wesentlich kleiner als die Pe-
riodendauer des Zerhackers gewählt, typischerweise 2 bis 4 %
davon. Mit Variation der Wartezeit $T_1$ können wir nun die ganze
Periode abtasten. In Abbildung 5.13 ist der Zeitverlauf für die
Abtastung dargestellt. Das Ergebnis wird in Abbildung 5.14
gezeigt, woraus sich sofort die richtigen Werte für die Einstellung
von $T_1$ und $T_2$ ablesen lassen.

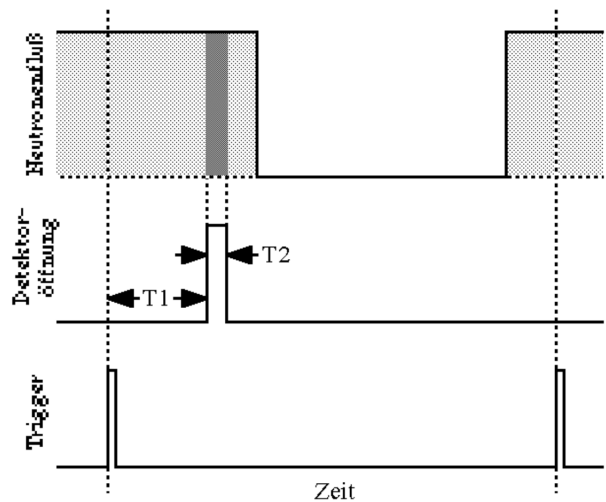

Abbildung 5.13:
Erläuterung zur Messung des zeitlichen Verlaufs des Neutronenflusses hinter dem Zerhacker. Gezählt werden die Neutronen während der Zeit $T_2$, also alle, die in die grau hinterlegte Fläche fallen. Diese Meßwerte geben als Funktion von $T_1$ den Zeitverlauf wieder.

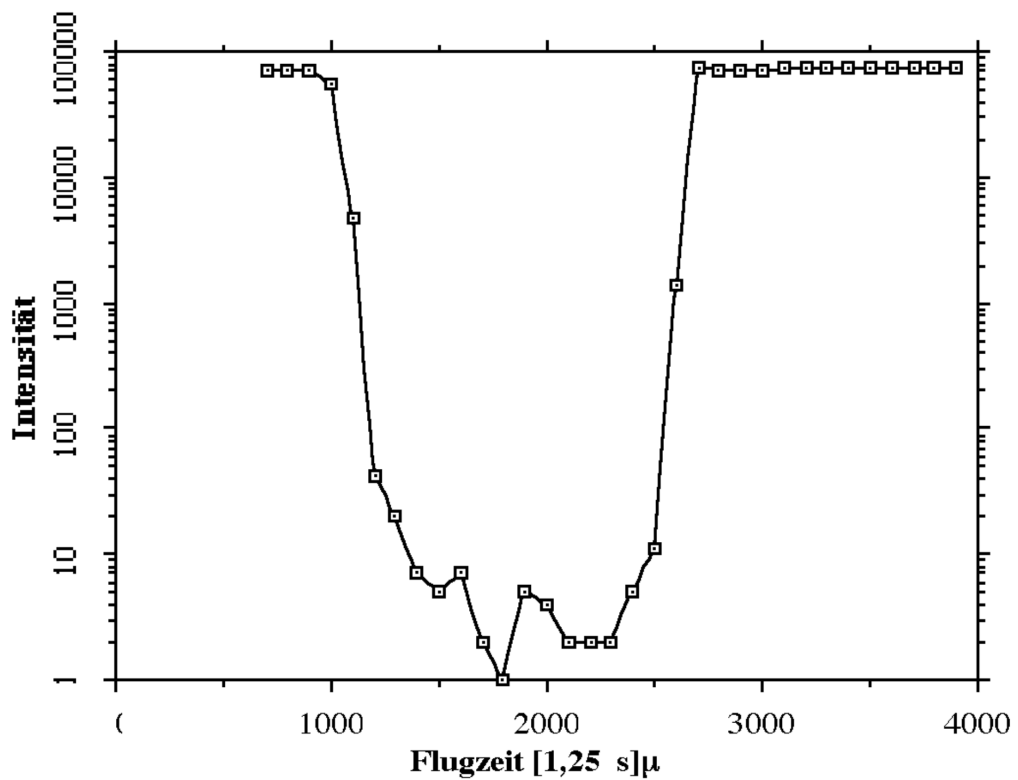

Abbildung 5.14:
Zeitlicher Intensitätsverlauf während einer Zerhackerperiode. Die Intensität ist logarythmisch gegen die Flugzeit in Einheiten von 1,25 μs aufgetragen. Die Zerhackerfrequenz beträgt 245 Hz, der Abstand Detektor - Kristall 72 cm und die Wellenlänge 5,69 Å. Das Zeitfenster wurde daraufhin auf 1800 bis 2400 in obigen Einheiten gesetzt.

# 5.4. Ausrichtung der Kristalle

Zunächst wird der Analysatorkristall $K_2$ justiert, da dieser auf der relativ unhandlichen Dopplermaschiene befestigt ist. Haben wir diesen Reflex gefunden, so kann man den Monochromator $K_1$ auf ein Goniometer setzen und ihn parallel zu $K_2$ ausrichten. Die Justierung jedes Kristalls geschieht zunächst mit Hilfe eines Lasers und wird anschließend neutronenoptisch nachgestellt.

Es soll hier eine Abschätzung gegeben werden, mit welcher Genauigkeit die Kristalle justiert werden müssen. Diese werden mit Meßdaten für die Ausrichtung verglichen.

## 5.4.1. Ausrichtung des Analysatorkristalls

Der Analysatorkristall $K_2$ alleine ist relativ unsensibel gegenüber Verkippungen gegen die Strahlachse. Dreht man den Kristall im Strahl und trägt die Intensität gegen den Drehwinkel auf, so ist die Breite $\Delta\Phi$ der so gemessenen Kurve ausschließlich durch die Geometrie bestimmt. Dies ist in Abbildung 5.15 veranschaulicht. Seien b die Breite des Detektorfensters und D der Abstand Kristall - Detektor, so erwarten wir für b « D

$$\Delta\Phi = \frac{b}{2\,D} \quad . \tag{5.4}$$

Um diesen Winkel muß der Kristall $K_2$ gedreht werden, um einen dünnen, parallelen Neutronenstrahl von einem Detektorrand zum anderen wandern zu lassen. Der Faktor 2 im Nenner berücksichtigt, daß sich der an $K_2$ reflektierte Strahl doppelt so schnell dreht, wie der Kristall selbst.

Analysatorkristall                                    Detektorfenster

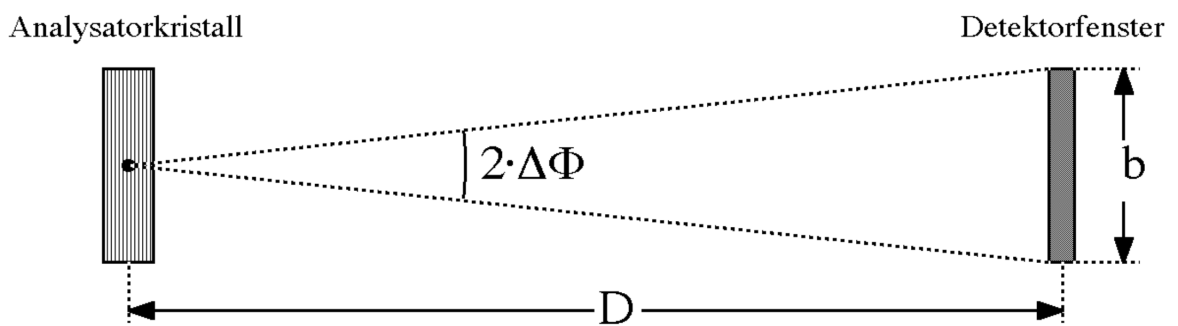

Abbildung 5.15:

Der Kristall kann um den Winkel $\Delta\Phi$ gedreht werden, um den reflektierten Strahl vom einen Ende des Detektorfensters zum anderen laufen zu lassen.

Der gemessene Drehwinkel von ca. $\Delta\Phi \approx 0{,}8°$ stimmt sehr gut mit dem theoretischen Wert von $0{,}84°$ überein. Dabei sind für $D = 72$ cm und $b = 2{,}1$ cm eingesetzt worden. Die Meßkurve ist in Abbildung 5.16 wiedergegeben.

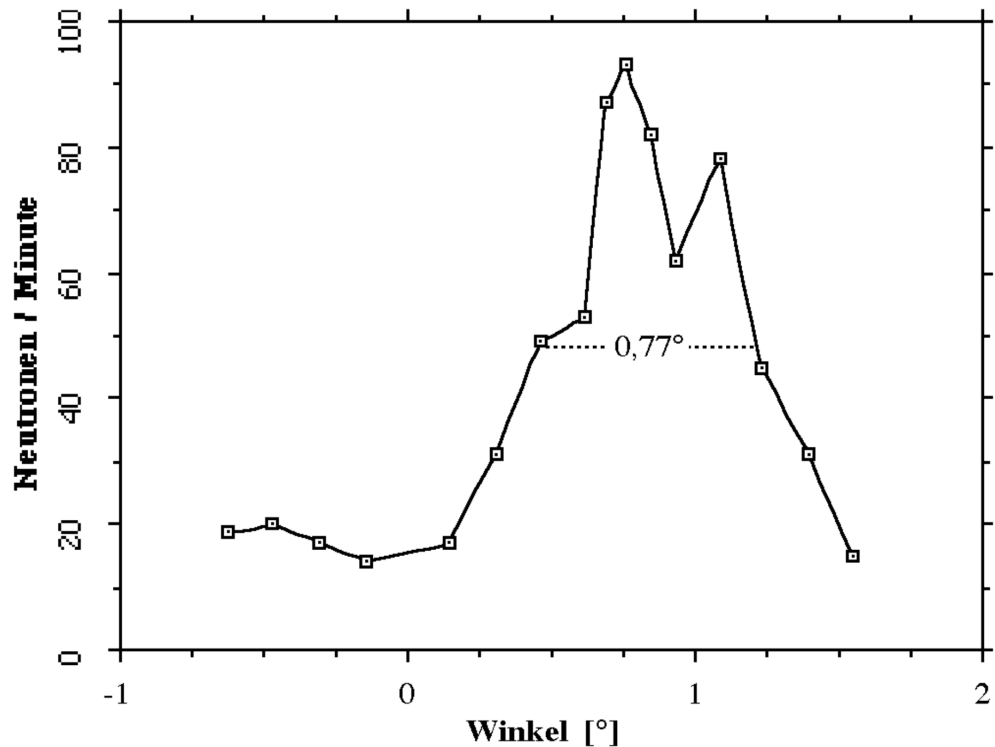

Abbildung 5.16:
Ausrichtung des Analysatorkristalls $K_2$. Die Halbwertsbreite der Kurve stimmt gut mit der geometrischen Abschätzung überein.

## 5.4.2.  Ausrichtung des Monochromatorkristalls

Nehmen wir in analoger Weise eine Kurve des Kristalls $K_1$ bei ausgerichtetem $K_2$ auf, so ergibt deren Breite sich aus der Unschärfe der Braggbedingung. Dafür betrachten wir eine kleine Winkelabweichung $\Delta\Theta$ aus der Rückstreugeometrie und entwickeln das so erhaltene Braggesetz um den Winkel $\pi/2$ bis zur zweiten Ordnung

$$\lambda = 2\,d\,\sin\!\left(\frac{\pi}{2} + \Delta\Theta\right)$$

$\Leftrightarrow$

$$\lambda = 2\,d\,\cos\!\left(\Delta\Theta\right)$$

$$\lambda = 2\,d\left(1 - \frac{(\Delta\Theta)^2}{2}\right) + O\left((\Delta\Theta)^4\right) \tag{5.5}$$

mit $\lambda_0 := 2d$ und $\Delta\lambda := \lambda_0 - \lambda$ gibt dies den gesuchten Ausdruck

$$\boxed{\frac{\Delta\lambda}{\lambda} = \frac{1}{2}\,(\Delta\Theta)^2} \tag{5.6}$$

Setzen wir zur Bedingung, daß die durch die Winkelabweichung gegebene Unschärfe von derselben Größe wie die Genauigkeit der Gitterkonstanten $\Delta d/d$ sein soll, also

$$\frac{1}{2}\,(\Delta\Theta)^2 = \frac{\Delta d}{d} \tag{5.7}$$

und drücken $\Delta d$ durch $\Delta E$ aus, so erhalten wir

$$\boxed{\Delta\Theta = \sqrt{\frac{\Delta E}{E}}} \quad . \tag{5.8}$$

Mit der theoretischen Auflösung des hier untersuchten GaAs[200]-Reflexes von $\Delta E/E = 3\cdot 10^{-6}$ bekommen wir als Abschätzung für die Winkelgenauigkeit $\Delta\Theta = 6'$. Die Messungen (Abbildung 5.17) liegen mit einer Halbwertsbreite von $\Delta\Theta \approx 8'$ in derselben Größenordnung, was bedeutet, daß der Kristall gut justiert ist.

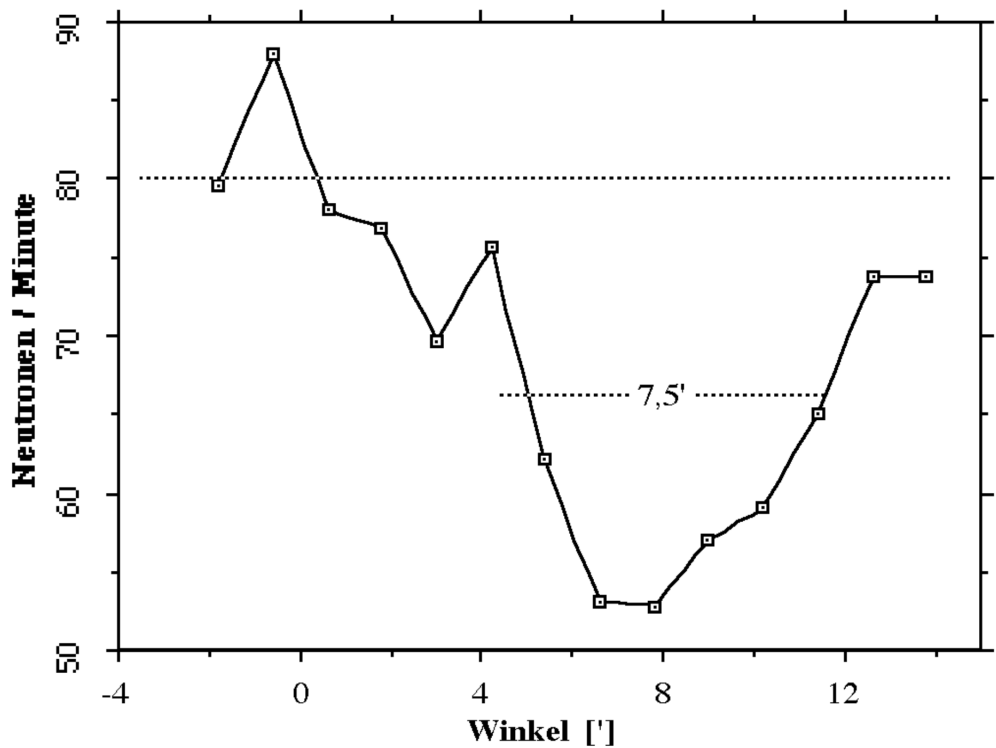

Abbildung 5.17:
Ausrichtung des Monochromatorkristalls K$_1$. Die Breite der Kurve liegt in
der Größe der Abschätzung aus der Unschärfe des Braggesetzes.

## 5.5.    Abschätzung der Temperaturstabilität

Wie schon in Kapitel 2.4 gezeigt, kann die thermische Ausdehnung
der Kristalle anstelle des Dopplereffekts zum Abtasten eines ge-
wünschten Wellenlängenintervalls herangezogen werden. Ander-
erseits bedeutet das aber auch, daß unkontrollierte thermische
Schwankungen unsere Messungen stören oder sogar zu nichte
machen. Hier soll eine Abschätzung für die thermische Stabilität
gegeben werden:
Während gleiche Temperaturschwankungen beider Kristalle nicht
ins Gewicht fallen, macht sich die Temperaturdifferenz zwischen
K$_1$ und K$_2$ empfindlichbemerkbar. Ändert diese sich während der
Messung, so bedeutet das, daß  die in Kapitel 2.2 beschriebene
Bragglinie sich auf der Dopplergeschwindigkeits- und damit auch
auf der Wellenlängenskala verschiebt. Sie wird ausgeschmiert und
somit breit und flach. Der maximal vertretbare Temperaturunter-
schied soll so abgeschätzt werden, daß sich das  Spektrum gerade
um seine eigene Linienbreite verschiebt.

Sei $\alpha(T)$ der thermische Ausdehnungskoeffizient des Kristalls, dann ergibt sich für die relative Änderung $\Delta d / d$ seiner Gitterkonstanten bei einer Temperaturänderung um $\Delta T$

$$\frac{\Delta d}{d} = \alpha(T)\,\Delta T \qquad\qquad (5.9)$$

Aufgelöst und mittels (2.5) in Energien ausgedrückt erhalten wir für die maximal zulässige Temperaturdifferenz

$$\boxed{\Delta T_{max} = \frac{1}{2\,\alpha(T)} \cdot \frac{\Delta E}{E}} \;. \qquad\qquad (5.10)$$

In der folgenden Tabelle sind die hier interessierenden Werte für den GaAs[200] Reflex zusammengestellt und mit Si[111] und Ge[111] verglichen.

|  | GaAs[200] | Si[111] | Ge[111] |
|---|---|---|---|
| $\alpha(300K)$ | $5{,}7{\cdot}10^{-6}$ | $2{,}3{\cdot}10^{-6}$ | $5{,}8{\cdot}10^{-6}$ |
| $\Delta E/E$ | $3{,}7{\cdot}10^{-6}$ | $4{,}24{\cdot}10^{-5}$ | $8{,}03{\cdot}10^{-5}$ |
| $\Delta T_{max}$ | $0{,}32$ K | $9{,}2$ K | $6{,}9$ K |

Bei GaAs erreicht dieser Wert bereits eine experimentell kritische Grenze, während man bei dem heute verwendeten Siliziumreflex weitgehend unempfindlich gegenüber Temperaturschwankungen im Labor ist.

Bei den Auflösungsmessungen waren keine Temperaturkontrollen vorgesehen. Es wurde jedoch darauf geachtet, daß nach Möglich-keit keine Luftströmungen z.B. vom Zerhacker, dem Doppleran-trieb, der Elekronikkühlung oder dem flüssigen Stickstoffvorrat des Berylliumfilters zu den Kristallen gelangen. Gegebenenfalls wurden Kunststoffolien aufgespannt.

## 5.6.    Eichung des Vielkanalanalysators

Die Spektren werden in direkter Abhängigkeit der Dopplerge-schwindigkeit des Analysatorkristalls aufgenommen, die dann in die gewünschten Größen wie Wellenlängen oder Energien umge-rechnet werden kann. Ein an der Kristallhalterung angebrachter induktiver Geschwindigkeitsaufnehmer liefert eine zur Momen-tangeschwindigkeit proportionale elektrische Spannung, die linear verstärkt und an den Analogeingang des Vielkanalanalysators gelegt wird. In unserem Fall ist der Linearverstärker auf sein Maximum fest eingestellt und damit bei allen Messungen gleich, sodaß nur einmal eine Eichung vorgenommen werden muß, diese aber von Zeit zu Zeit überprüft wird. Schwankungen sind nicht feststellbar. Hier soll die Breite eines Kanals des Vielkanalanalysators in Energieeinheiten bestimmt werden.

Die Eichung kann nach zwei Methoden vorgenommen werden, wobei die erste, herkömmliche Bestimmung der Kanalbreite ein Spezialfall der hier angewendeten zweiten Möglichkeit ist. Ausgangspunkt beider Methoden ist die Messung eines weißen Monitorspektrums, also eine von der augenblicklichen Dopplergeschwindigkeit unabhängige Zählrate. Dies kann man sehr leicht mit Hilfe eines Frequenzgenerators verwirklichen, indem wir an den Zählereingang des Vielkanalanalysator eine feste Frequenz legen.

Bei der ersten Methode läuft der Dopplerantrieb so langsam, d. h. die Geschwindigkeitsamplitude ist so klein, daß die ersten und letzten Kanäle des Vielkanalanalysators nie angesprochen werden. Die Zahl der angesprochenen Kanäle entspricht demnach der doppelten Geschwindigkeitsamplitude, woraus man leicht die Breite eines Kanals errechnen kann. Diese einfache Methode hat jedoch in unserem Meßaufbau den Nachteil, daß dazu die nötige Dopplerfrequenz unter 0,4 Hz liegen muß. Der hier benutzte, für das IN10C gebaute Dopplerantrieb ist nicht mehr für solch niedrige Frequenzen ausgelegt und läuft daher sehr ungleichmäßig. Dies verhindert eine genaue Bestimmung der Geschwindigkeitsamplitude.

Bei der zweiten Methode kann der Dopplerantrieb schneller laufen. Dabei wird der Analogeingang des Vielkanalanalysators übersteuert, und wir sehen nur den uns interessierenden Ausschnitt des Spektrums bei kleien Geschwindigkeiten. Die Elektronik hat den Vorteil, daß keine Zählerimpulse verlorengehen. Bei Über- oder Untersteuerung werden die ankommenden Ereignisse in den Kanälen mit der höchsten, bzw. niedrigsten Nummer registriert. Gemäß der Abbildung 5.18 messen wir das Spektrum innerhalb des Geschwindigkeitsintervalls von $V_1$ bis $V_2$. Dieses Intervall sei in K Kanäle aufgeteilt. Weiterhin werden alle Geschwindigkeiten $V < V_1$ dem Kanal $K_-$, alle mit

V > V$_2$ dem Kanal K$_+$ zugeordnet. Dann ist die Breite eines Kanals in Geschwindigkeitseinheiten durch

$$\delta V_K = \frac{V_2 - V_1}{K}$$

(5.11)

und auf der Energieskala durch

$$\delta E_K = 2\frac{E_0}{v_0} \cdot \frac{V_2 - V_1}{K}$$

(5.12)

gegeben. E$_0$ und v$_0$ bedeuten die Neutronenenergie bzw. -geschwindigkeit bei ruhendem Analysatorkristall.

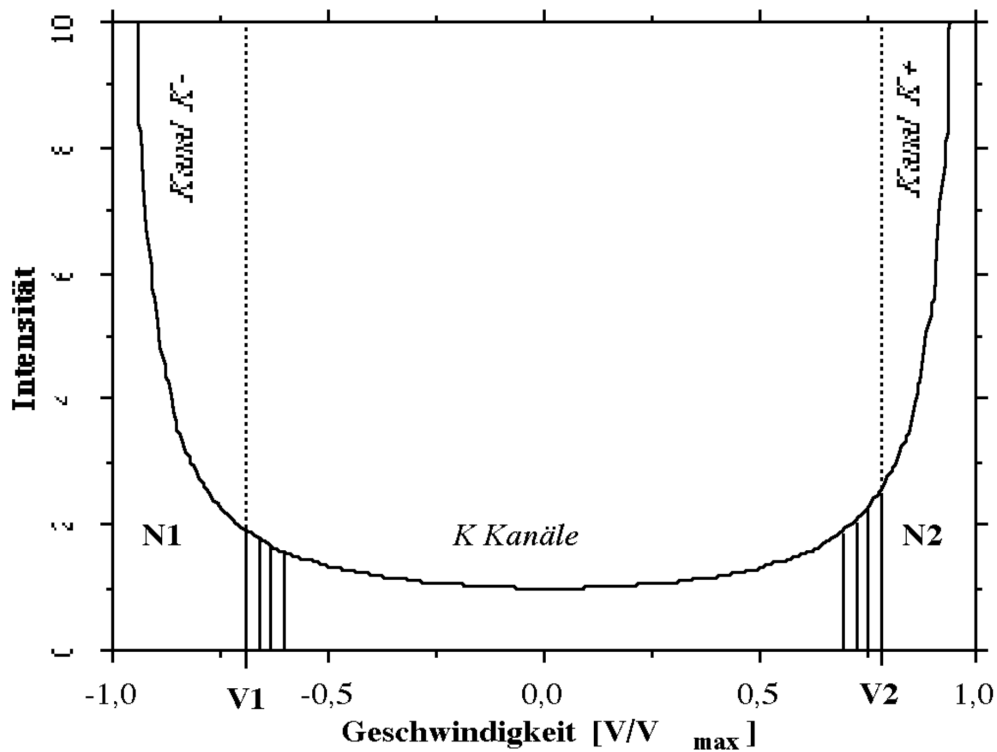

Abbildung 5.18:
Unterteilung des Monitorspektrums in die verschiedenen Kanäle. Das Intervall zwischen **V$_1$** und **V$_2$** ist in **K** Kanäle unterteilt. Die restlichen Ereignisse werden den Kanälen **K$_+$** und **K$_-$** zugeteilt.

Die Bestimmung von V$_1$ und V$_2$ geschieht durch Kenntnis der Verteilung $N$(V,t) wie sie in Kapitel 2.6 besprochen wurde: Seien N$_-$ und N$_+$ die Anzahl der Ereignisse der Kanäle K$_-$ bzw. K$_+$ sowie N die Gesamtzahl der Ereignisse des gesamten Spektrums, so können wir diese Zahlen mit den Flächenstücken unterhalb der Verteilungsfunktion $N$(V,t) identifizieren und nach den gesuchten

Größen auflösen (siehe Abbildung 5.18). Somit erhalten wir folgende Integrale:

$$N_1 = \int_{-V_{max}}^{V_1} N(V,t) \, dV$$

(5.13)

$$N_2 = \int_{V_2}^{V_{max}} N(V,t) \, dV$$

(5.14)

und wie in Kapitel 2.6 gezeigt

$$N = R \, t \; .$$

(5.15)

Das unbestimmte Integral

$$\int N(V,t) \, dV = \frac{R \, t}{\pi} \int \frac{1}{\sqrt{V_{max}^2 - V^2}} \, dV$$

(5.16)

ist elementar lösbar und ergibt

$$\int N(V,t) \, dV = \frac{R \, t}{\pi} \, asin\left(\frac{V}{V_{max}}\right)$$

(5.17)

Für $N_1$ und $N_2$ erhalten wir

$$N_1 = \frac{R \, t}{\pi} \left[ \frac{\pi}{2} + asin\left(\frac{V_1}{V_{max}}\right) \right]$$

$$N_2 = \frac{R \, t}{\pi} \left[ \frac{\pi}{2} - asin\left(\frac{V_2}{V_{max}}\right) \right]$$

(5.18)

oder Gleichung (5.15) eingesetzt und nach den gesuchten Größen $V_1$ und $V_2$ aufgelöst

$$V_1 = - \, V_{max} \, cos\left(\frac{N_1}{N} \, \pi\right)$$

(5.19)

$$V_2 = V_{max} \cos\left(\frac{N_2}{N}\pi\right)$$

Die zuerst angedeutete Methode erhalten wir hieraus als Spezial-fall, nämlich wenn wir $N_1 = N_2 = 0$ und K der Anzahl der angesprochenen Kanäle setzen.

Im Folgenden wird die Kanalbreite an zwei gemessenen Monitor-spektren ausgewertet. Sie unterscheiden sich in der Frequenz der Dopplermaschine und somit in der Geschwindigkeitsamplitude. Beiden liegt jedoch die gleiche Einstellung der Elektronik zu Grunde und sollten daher das gleiche Ergebnis liefern. Im ersten Spektrum (Abbildung 5.19) sehen wir, wie vorher diskutiert, nur den Ausschnitt zwischen $V_1$ und $V_2$ der Geschwindigkeitsskala, während im zweiten (Abbildung 5.20) die Dopplergeschwindigkeit verkleinert wurde, so daß das ganze Monitorspektrum sichtbar ist.

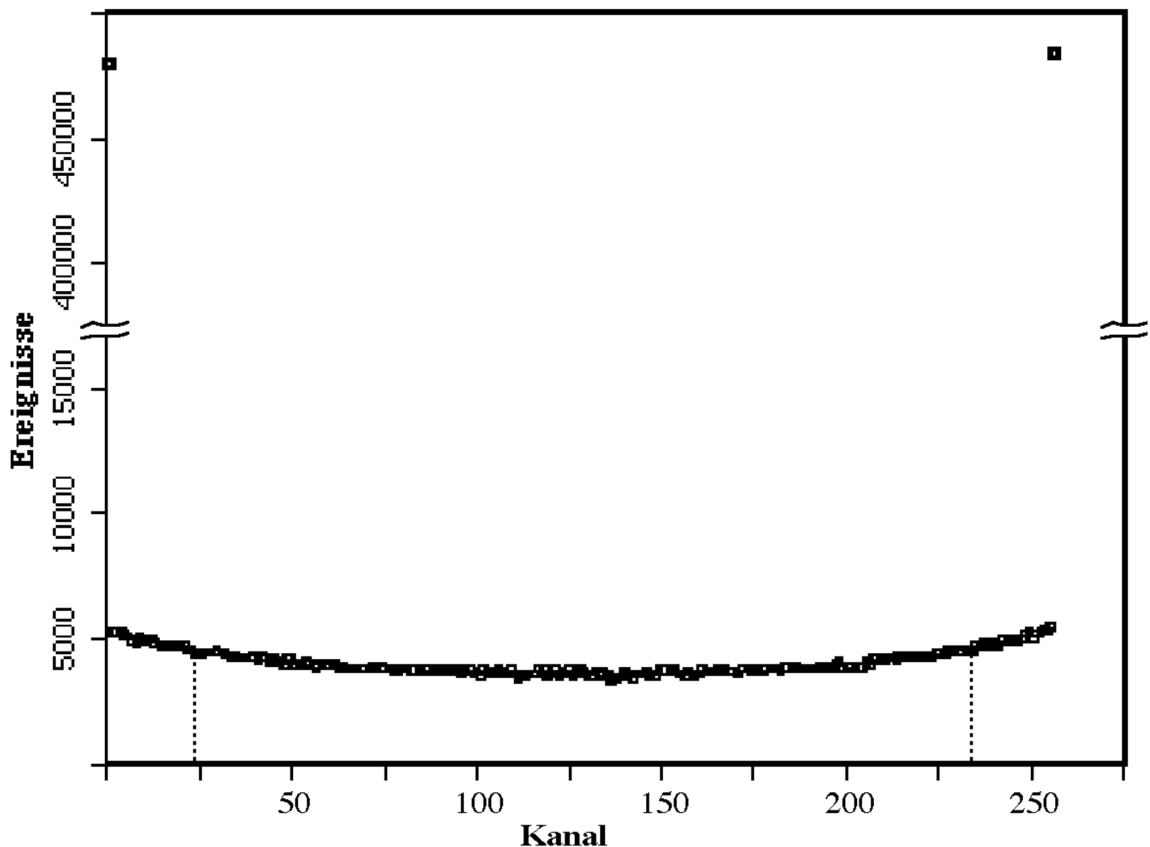

Abbildung 5.19:
Monitorspektrum #1. Die Geschwindigkeitsamplitude ist so groß, daß sie nicht mehr auf die Skala paßt. Bei Übersteuerung werden die Ereignisse im ersten und im letzten Kanal registriert (siehe Punkte bei Kanal 1 und 256).

Die gestrichelten Linien geben die Intervallgrenzen für dieAuswertung #1' an. Alle Kanäle, die außerhalb dieser Grenzen liegen, werden zu $N_1$ bzw. $N_2$ aufsummiert.

Der Hub der Dopplermaschine beträgt A = 2,5 cm. Er wird benötigt, um die Geschwindigkeitsamplitude $V_{max} = 2\pi\nu A$ zu bestimmen. In der nachfolgenden Tabelle sind die verschiedenen Daten zusammengestellt und für die GaAs[200]-Rückstreuwellenlänge von $\lambda_0 = 5,65$ Å ausgewertet. Die Energie der Neutronen geht mit $E_0 = 2,56 \cdot 10^6$ neV und deren Geschwindigkeit mit $v_0 = 700$ m/s in die Rechnungen ein.

| Größe | Spektrum #1 | Spektrum #1' | Spektrum #2 |
|---|---|---|---|
| **Frequenz $\nu$** | 0,700 Hz | 0,700 Hz | 0,372 Hz |
| **$V_{max}$** | $11,0 \frac{cm}{s}$ | $11,0 \frac{cm}{s}$ | $5,84 \frac{cm}{s}$ |
| **$N_1$** | 477213 | 589847 | 63086 |
| **$N_2$** | 474440 | 589620 | 64490 |
| **N** | 1997758 | 1997758 | 636308 |
| **K** | 254 | 208 | 176 |
| **$V_1$** | $-8,045 \frac{cm}{s}$ | $-6,598 \frac{cm}{s}$ | $-5,562 \frac{cm}{s}$ |
| **$V_2$** | $8,078 \frac{cm}{s}$ | $6,601 \frac{cm}{s}$ | $5,550 \frac{cm}{s}$ |
| **$\delta V_K$** | $0,634 \frac{mm}{s}$ | $0,634 \frac{mm}{s}$ | $0,631 \frac{mm}{s}$ |
| **$\delta E_K$** | 4,643 neV | 4,641 neV | 4,618 neV |

Das Spektrum #1' ist identisch mit#1. Der Unterschied liegt hier in der Wahl der Intervallgrenzen für die Auswertung. Während bei #1 nur der erste und der letzte Kanal zu $N_1$ bzw. $N_2$ beitragen sind bei #1' die ersten 24 Kanäle zu $N_1$ und die letzten 24 Kanäle zu $N_2$ aufsummiert. Stattdessen ist für K eine kleinere Zahl eingesetzt worden. Dies soll bestätigen, daß diese Methode zur Eichung der Skala anwendbar ist. Wären die Ergebnisse von #1 und #1' unterschiedlich, so hieße dies, daß wir nicht mit der aus der sinusförmigen Dopplerbewegung hervorgegangen

Verteilung $N$(V,t) rechnen könnten. Die Werte stimmen auf drei Stellen überein, was zeigt, daß diese Methode brauchbar ist. Vergleichen wir die Ergebnisse des Spektrums #1 mit dem Spektrum #2, so finden wir nur auf zwei Stellen Übereinkunft. Dies möchte ich auf die eingangs angesprochene Tatsache zurückführen, daß der Dopplerantrieb bei derart kleinen Geschwindigkeiten keineswegs gleichmäßig läuft. Die aus­geschmierten Kanten des Spektrums #2 um $-V_{max}$ und $V_{max}$ deuten ebenfalls auf eine relativ ungenau definierte Geschwindigkeits­amplitude hin.

Zur Eichung der Energieskala wird somit der Wert von

$$\delta E_K = 4{,}64 \text{ neV} \tag{5.20}$$

für die Breite eines Kanals vorgeschlagen.

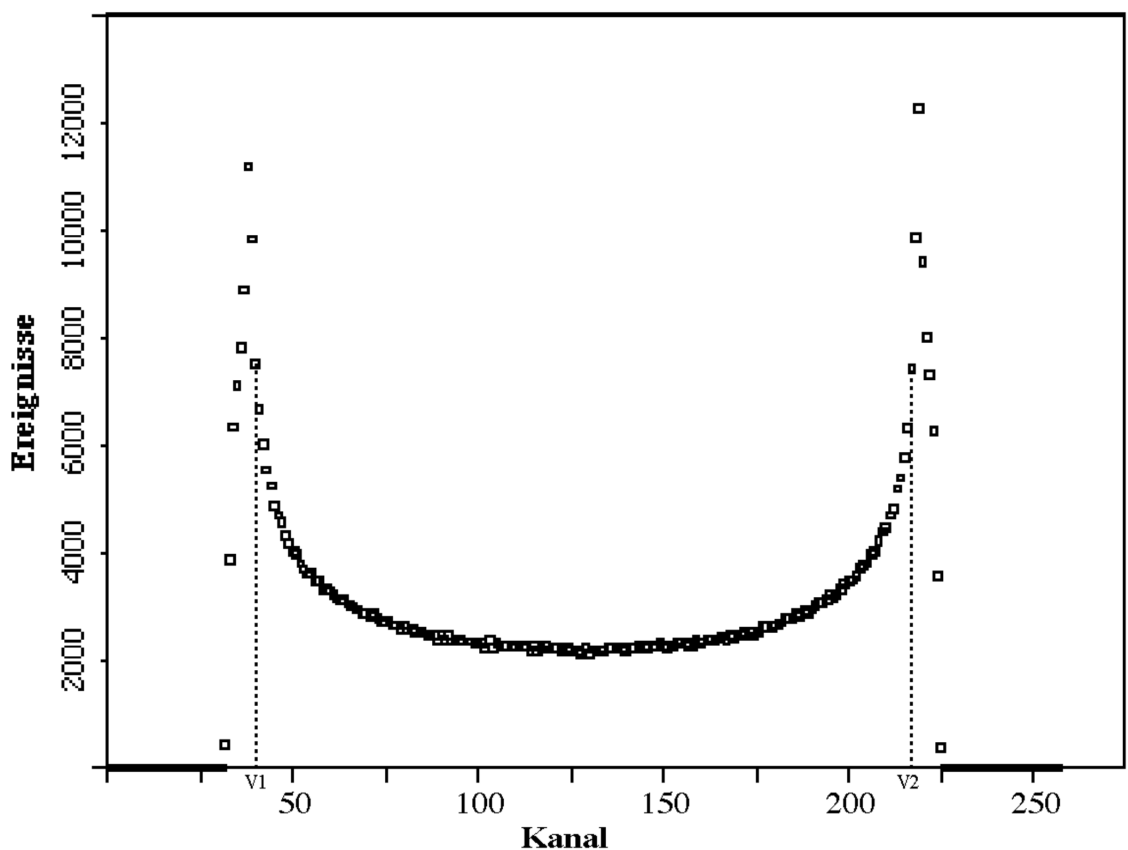

Abbildung 5.20:
Monitorspektrum #2. Die Geschwindigkeitsamplitude ist klein genug, daß das vollständige Spektrum auf die Skala paßt. Die gestricheltenLinien geben die Intervallgrenzen von $V_1$ bis $V_2$ an. Über alle Kanäle, die außerhalb dieser Grenzen liegen, wird zu $N_1$ bzw. $N_2$ aufsummiert.

# 6.     Messung und Auswertung

In diesem Kapitel soll eine stellvertretend Messungen des GaAs[200]-Reflexes gezeigt und diskutiert werden. Dabei wird auf die Auswertung der Bragglinie eingegangen und die so gewonnenen Daten mit den Ergebnissen der dynamischen Streutheorie verglichen.

## 6.1.     Gemessene Daten

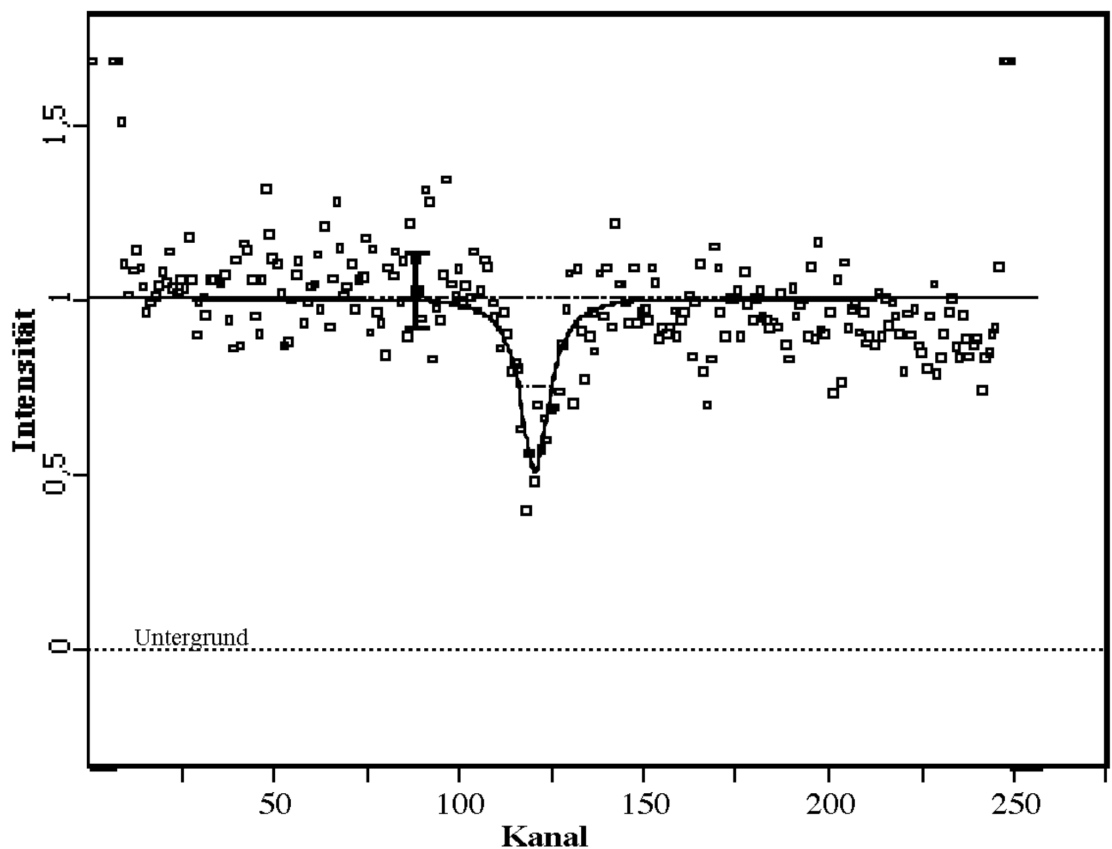

Abbildung 6.1:
Gemessene Bragglinie des GaAs[200]-Reflexes. Das Spektrum ist bereits auf ein Monitorspektrum normiert. Die Breite der angepaßtenLorentzkurve beträgt 43 neV. Einzelheiten sind im Text beschrieben. Ein repräsentativer Fehlerbalken ist links neben der Bragglinie eingezeichnet.

Ein typisches, bereits auf einen Monitor normiertes Spektrum wird in Abbildung 6.1 gezeigt. Die Justierung der Kristalle für diese Messung ist in den Abbildungen 5.16 und 5.17 wiedergegeben. Während der Meßzeit von 14 h wurden integral N = 62653 Neutronen gezählt, was eine Rate von

$$R = 75 \ \frac{n}{min} \qquad (6.1)$$

ergibt. Der Untergrund wurde bei abgedecktem Analysatorkristall zu

$$U = 18 \ \frac{n}{min} \qquad (6.2)$$

also

$$\frac{U}{R} = 0{,}24 \qquad (6.3)$$

bestimmt und von den Meßdaten subtrahiert.

Die Statistik ist noch nicht sonderlich gut, doch läßt sich die Bragg-linie sehr gut erkennen. Die Zahl n der Neutronen eines Kanals am Rande der Linie liegt typischerweise bei

$$n = 155 \ , \qquad (6.4)$$

woraus wir einen relativen Fehlerbalken dieses Kanals von

$$\pm \frac{1}{\sqrt{n}} = \pm 8\% \qquad (6.5)$$

erhalten.

Zur numerischen Auswertung wird nach der Methode der kleinsten Quadrate näherungsweise eine Lorentzkurve

$$f_b(x) = b_0 - b_2 \ \frac{1}{1 + \left( \dfrac{x - b_1}{b_3} \right)^2} \qquad (6.6)$$

den Meßpunkten angepaßt. Strenggenommen müßten wir mit der Faltung zweier Ewaldkurven rechnen, doch ist eine Lorentzfunktion innerhalb der vorgelegten Statistik vollkommen ausreichend. Die Fitparameter sind die $b_i$ ($i = 0 \ldots 3$), und x die Größe, in der wir die Abszisse messen, z.B. direkt in Kanalbreiten des Vielkanalanalysators oder mittels unserer Eichung (5.20) in Energieeineiten.

Um die Messung mit theoretisch gewonnenen Werten zu vergleichen, normieren wir Lorentzkurve und Meßwerte auf $b_0$, also

$$f(x) := \frac{f_b(x)}{b_0} = 1 - a_2 \, \frac{1}{1 + \left(\frac{x - a_1}{a_3}\right)^2} \, . \tag{6.7}$$

Die Parameter $a_i$ spiegeln dabei folgende Größen wider:
Die Linienposition $a_l$ und die volle Halbwertsbreite der Linie

$$\Delta E = 2 \, a_3 \tag{6.8}$$

werden beide in Energieeinheiten gemessen. Die Linientiefe

$$\tau = a_2 \tag{6.9}$$

ist dimensionslos, während die integrierte Reflektivität $R_E$ durch die Fläche des Loches mit

$$R_E = \frac{\pi}{2} \, \tau \, \Delta E \tag{6.10}$$

gegeben ist und gleichfalls die Dimension einer Energie trägt.
Die so erhaltenen Daten sind in der Tabelle 6.1 zusammengestellt.

| Tabelle 6.1 | | | |
|---|---|---|---|
| Größe | Zeichen | Theorie | Experiment |
| absolute Linienbreite (gefaltet) | $\Delta E$ | 13 neV | $43 \pm 5$ neV |
| relative Linien- breite | $\dfrac{\Delta E}{E_0}$ | $5,1 \cdot 10^{-6}$ | $(17 \pm 2) \cdot 10^{-6}$ |
| integrierte Reflektivität | $R_E$ | 13 neV | $33 \pm 5$ neV |
| Linientiefe | $\tau$ | 0,7262 | $0,49 \pm 0,05$ |

## 6.2.    Theoretische Werte

Die theoretischen Werte gehen aus der dynamischen Streutheorie, wie sie ursprünglich von Darwin und Ewald entwickelt wurde, hervor. (siehe Kapitel 3). Die für uns wichtigsten Größen der sogenannten Ewald-Lösung, die bei vernachlässigbarer Absorption der Strahlung im Kristall anwendbar ist, sind in Kapitel 3.4 zusammengestellt. Die gerechneten Werte für GaAs[200] sind in der Tabelle 6.1 neben den Meßdaten aufgelistet.

## 6.3. Diskussion

Die gemessenen Werte aus Tabelle 6.1 stimmen schlecht mit der Theorie überein. Die experimentelle Linie ist um einen Faktor 3,3 breiter als die theoretische, ihre integrierte Reflektivität beträgt das 2,5-fache der theoretischen Lösung während die Linientiefe auf das 0,67-fache reduziert ist. Letzteres ist sinnvoll, denn bei unkontrollierten Verschiebeungen der Linie während der Messungen wie z.B. Temperaturschwankungen oder mechanische Vibrationen muß die Halbwertsbreite der Linie größer werden und deren Tiefe entsprechend abnehmen. Die integrierte Reflektivität des Reflexes sollte sich in diesem Fall nur unwesentlich ändern.

In Zahlen ist die gemessene Linie um

$$\frac{\varepsilon}{E_0} := \left.\frac{\Delta E}{E_0}\right|_{\text{Experiment}} - \left.\frac{\Delta E}{E_0}\right|_{\text{Theorie}} = 11 \cdot 10^{-6} \tag{6.11}$$

verbreitert. Dies kann auf verschiedene Mechanismen zurückgeführt werden, die im Folgenden besprochen werden.

## 6.3.1. Temperaturschwankungen

Die Temperaturdifferenz der beiden Kristalle ist während der Meßzeit nicht konstant. Die dadurch hervorgerufene Linienverbreiterung ist durch

$$\left.\frac{\varepsilon}{E_0}\right|_T = 2\,\alpha(T)\,\Delta T \tag{6.12}$$

gegeben. Führten wir unsere gemessene Verbreiterung ausschließlich auf diesen Effekt zurück, so erhielten wir eine Temperaturdifferenz von

$$\Delta T = 1 \text{ K} \quad . \tag{6.13}$$

## 6.3.2. Vibrationen

Mechanische Vibrationen mit einer Geschwindigkeitsamplitude $V_V$, die vor allem durch den Motor des Dopplerantriebs hervorgerufen werden, verbreitern ebenfalls die Meßkurve um

$$\left.\frac{\varepsilon}{E_0}\right|_V = 2\,\frac{V_V}{v_0} \quad . \tag{6.14}$$

Würde unsere Messung ausschließlich durch diesen Mechanismus verbreitert, so bekämen wir eine Vibrationsamplitude von

$$V_V = 4\,\frac{mm}{s} \quad . \tag{6.15}$$

## 6.3.3. Kristallverzerrungen

Enthält der Kristall mechanische Spannungen, seien sie durch makroskopische Verbiegung oder mikroskopische Kristallbaufehler hervorgerufen, dann wird mittels

$$\left.\frac{\varepsilon}{E_0}\right|_S = 2\left\langle\frac{\Delta L}{L}\right\rangle_{Verzerrung} \tag{6.16}$$

ebenfalls die Linie verbreitert. Die spitze Klammer bedeutet die Mittelung über alle Verzerrungen.

## 6.3.4. Verkippung

Stehen die Kristalle nicht parallel zueinander sondern um einen Winkel $\beta$ gegeneinander verkippt, so ergibt dies eine Linienverbreiterung von

$$\left.\frac{\varepsilon}{E_0}\right|_\beta = 2\,\beta\,\delta \quad , \tag{6.17}$$

wobei $\delta$ die Winkeldivergenz des Neutronenstrahls ist [73H]. Diese ist am Strahlrohr des IN10C besonders groß, nämlich horizontal $2°$ und vertikal $4°$. Durch die Geometrie von Eintrittsblende und Detektorfenster kann jedoch nur maximal eine effektive Divergenz von $\delta = 2°$ in das Zählrohr gelangen. Eine

Linienverbreiterung von (6.11) ausschließlich durch Verkippung, bedeutete eine Fehljustierung von

$$\beta = 32 \text{ "} .\qquad\qquad\qquad\qquad (6.18)$$

## 6.3.5. Ursache der Fehlerquellen

Wir sehen, daß alle hier aufgeführten Möglichkeiten sehr empfindlich zur Linienverbreiterung beitragen können.

a)   Die Temperatur wurde nicht kontrolliert. Wie schon in Kapitel 5.5 angedeutet, wurde lediglich darauf geachtet, die Kristalle von störenden Luftströmungen fernzuhalten, ohne sie jedoch vollständig unterdrücken zu können. Der Raum, in dem sich die Kristalle befinden, war in Richtung Dopplerantrieb geöffnet. Leichte Temperaturschwankungen können auch durch Erwärmung der Dopplermaschine sowie den stark stromdurchflossenen Spulen der Schrittmotoren der Goniometer entstehen. Temperaturdifferenzen zwischen den Kristallen $K_1$ und $K_2$ einiger zehntel Kelvin sind daher nicht auszuschließen.

b)   Die Vibrationen des Geschwindigkeitgebers lassen sich sehr schwer kontrollieren. Verschiedene Teile des mechanischen Aufbaus werden mehr oder weniger zu Resonanzschwingungen angeregt, weshalb es unmöglich ist, alle Frequenzkomponenten absolut und phasengerecht mit dem induktiven Geschwindigkeitsaufnehmer zu erfassen.

c)   Innere Spannungen der Kristalle können nicht ausgeschlossen werden. Gemäß Angabe des Herstellers [5] haben die Kristalle eine mittlere Versetzungsdichte von

$$\rho = 2{,}6 \cdot 10^3 \ \frac{1}{cm^3} \ . \qquad\qquad\qquad (6.19)$$

d)   Makroskopische Verzerrungen durch Verbiegung der Kristalle, z.B. durch die mechanische Halterung sind ebenfalls möglich.

---

[5]   Mitsubishi Metal Corporation, Japan

e) Die Verkippung der Kristalle trägt wegen der großen Strahldivergenz ganz empfindlich zur Linienverbreiterung bei. Abbildung 5.17 zeigt die Einjustierung des Kristalls $K_1$ parallel zu $K_2$. Um die Linie mit der gemessenen Breite zu erhalten dürfen die Kristalle nur um eine halbe Bogenminute gegeneinander verkippt sein. Das bedetet, sie müssen auf $1/15$ der Breite der in Abbildung 5.17 dargestellten Kurve ausgerichtet sein. Dies entspricht einem Schritt der Motorsteuerung für die Goniometer von $0,01°$.

f) Außer dem Justierfehler tragen auch intrinsische Verkippungen zu dem Kippwinkel $\beta$ bei: Versetzungswände und andere Fehlstellen können Kleinwinkelverkippungen herbeiführen, die die gleichen Auswirkungen wie der Justierfehler auf die Linienbreite haben.

g) Ein weiterer Fehler, der Verkippungen gleichkommt ist eine Krümmung der Netzebenen über den ganzen Kristall hinweg, wie sie bereits bei der Kristallzucht entstehen können. In diesem Fall erhalten wir eine Mittelung über die an verschiedenen Flächenstücken des Kristalls entstehenden Bragglinien.

## 6.4. Einfluß der Kristallitgrößen

Die dynamische Streutheorie basiert auf der Interferenz der Neutronenwelle mit dem Kristallgitter. Haben wir es jedoch nicht mit einem Idealkristall, sondern mit einem aus vielen zueinander inkohärenten Kristalliten aufgebauten Kristall zu tun, dann trägt nur das Volumen jedes einzelnen Kristallits zur Interferenz bei. Die dynamisch gerechnete minimale Auflösung (3.21) behält nur ihre Gültigkeit, wenn die Abmessung eines Kristallits größer als eine typische Eindringtiefe

$$\Delta_0 = \frac{\pi \, V_z}{\lambda \left| F^b_{hkl} \right|}$$

$$(6.20)$$

der Welle in das Kristallvolumen ist. Sobald diese Bedingung nicht mehr erfüllt ist bekommen wir sehr breite Reflexionskurven für jeden einzelnen Kristallit. Die Reflektivität eines jeden Kristallits ist entsprechend klein, doch kann die Gesamtreflektivität des Kristalls durch inkohärente Summation über alle Kristallite größer als die eines Idealkristalls werden, vorausgesetzt der Kristall ist genügend dick. Dieser Übergangsbereich von der dynamischen zur sekundären Extinktionstheorie ist theoretisch nur unzulänglich beschrieben.

---

[6] In der Wellenmechanik haben wir immer das Unschärfetheorem: Werden die Abmessungen im Ortsraum klein, dann werden sie im Impulsraum groß.

Wir wollen mit Hilfe dieser Eindringtiefe eine grobe Abschätzung für die Versetzungsdichte erstellen, an der diese Grenze eintritt: Stellen wir uns einen Kristall vor, der aus würfelförmigen Kristalliten mit einer Kantenlänge gleich der Eindringtiefe $\Delta_0$ aufgebaut ist. Nehmen wir an, daß jedem Kristallit eine Versetzungslinie zugeordnet werden kann, dann bekommen wir eine Versetzungsdichte an der Oberfläche von

$$\rho_{\Delta_0} = \frac{1}{\Delta_0^2} \ . \tag{6.21}$$

Für den GaAs[200]-Reflex mit $\Delta_0 = 0{,}035$ cm erhalten wir

$$\rho_{\Delta_0} \approx 10^3 \tag{6.22}$$

also einen Wert, in dessen Größenordnung wir uns durchaus befinden. Die hier verwendeten Galliumarsenidkristalle haben die geringsten Versetzungsdichten, die heutzutage erhältlich sind.

# 7.    Zusammenfassung

Es wurde experimentell gezeigt, daß eine erhöhte Energieauflö­sung für Rückstreuspektrometer erzielt werden kann, wie dies zur Zeit am ILL im Rahmen der Definition eines zukünftigen Moderni­sierungsprogrammes vorgeschlagen wird [89L]. Die an GaAs[200] gemessene Bragglinie ist mit $\Delta E = (43\pm5)$ neV[7] um einen Faktor 3,5 schmäler als die theoretische Breite des Si[111]-Reflexes. Im Ver­gleich mit dem bestehenden Rückstreuspektrometer IN10 am ILL ist dies ein Faktor 7. Damit stellt diese Arbeit die bisher schmalste Linienbreitenmessung an einem Braggreflex dar. Der theoretisch mögliche Wert von $\Delta E = 13$ neV für den gemessenen Reflex ist al­lerdings noch nicht erreicht.

Um die Messung verbessern zu können, werden folgende Maßnamen zur Unterdrückung von Fehlerquellen vorgeschlagen:

1.)    Die Temperatur beider Kristalle muß während der Messung auf 1/10 Kelvin genau bestimmt werden. Kann sie nicht kon­stant gehalten werden, so ist die Kanalzuweisung des Vielka­nalanalysators auf die Temperaturdifferenz der beiden Kris­talle zu korrigieren.

2.)    Mechanische Vibrationen müssen besser unterdrückt werden. die hier verwendete Dopplermaschine des IN10C ist nicht zur Messung derart hoher Auflösungen konzipiert worden. Even­tuell kann ein Mößbauerantrieb als Geschwindigkeitsgeber eingesetzt werden. Dabei muß darauf geachtet werden, daß bei der Bewegung keine Verkippungen der Kristalle verursacht werden.

3.)    Die Divergenz des einfallenden Strahls kann wesentlich zur Linienverbreiterung beitragen. Eine gute Kollimation erleich­tert nicht nur die Justierung sondern vermindert in gleichem Maße die Linienverbreiterung durch intrinsische Verkippungen der Netzebenen.

---

[7]    volle Halbwertsbreite, nicht entfaltet

# 8.  Literaturangaben

# 9.    Danksagung

An dieser Stelle möchte ich mich ganzherzlich bei all denen bedanken, deren Beratung und praktische Hilfe zum Gelingen dieser Arbeit beigetragen hat.

Den Herren Professor Dr. W. Gläser und Professor Dr. A. Zeilinger danke ich für die Aufnahme an ihrem Institut.

Den Herren Professor Dr. D. Richter , sowie Dr. A. Magerl danke ich für ihren Einsatz zum Zustandekommen dieser Arbeit.
Herrn Dr. A. Magerl gebührt auch Dank für die hervorragende Betreuung, sowie für die zahlreichen Diskussionen zur Physik. Es sollen ihm auch seine persönlichen Hilfen, die über den Rahmen der Arbeit hinausgehen, anerkannt werden.

Herrn Dr. H. Rüfer der Firma Wacker-Chemitronic GmbH, Burghausen, Deutschland, sowie Herrn Dr. K. Sassa der Firma Mitsubishi Metal Corporation, Saitama, Japan sei für die freundliche Bereitstellung der GaAs-Kristalle gedankt.

Bei Herrn Dr. B. Frick bedanke ich mich für die Geduld am Meßplatz des IN10C.

Für die mechanische und elektronische Unterstützung danke ich den Damen und Herren Y. Blanc, R. Chevalier, P. Flores, P. Joubert-Bousson, H. Just, E. Kalmbach, M. Locatelli, H. Schwab und P. Thomas.

Meinen Eltern danke ich für die finanzielle Unterstützung und meiner Freundin L. Payan für ihre Hilfsbereitschaft und andauernde Geduld während der Arbeit.

Schließlich danke ich den zahlreichen Mitarbeitern des ILL sowie den Angehörigen des Lehrstuhls E21 für ihr freundliches Entgegenkommen sowohl in der Arbeit, als auch in der Freizeit.